%% file: main.tex
\documentclass[sigconf]{acmart}
\usepackage{enumitem}
\usepackage{soul}
\usepackage{xspace}
\usepackage{wrapfig}
\usepackage{subcaption}
\usepackage{xcolor}
\usepackage{lscape}
\usepackage{multirow}
\usepackage{verbatim}
\usepackage{framed}
\usepackage{float}

\setcopyright{none}
\copyrightyear{2025}
\acmYear{2025}
\acmConference[CHI '25]{CHI Conference on Human Factors in Computing Systems}{April 26-May 1, 2025}{Yokohama, Japan}
\acmDOI{10.1145/3706598.3713819}
\acmISBN{979-8-4007-1394-1/25/04}

\makeatletter
  \@printpermissiontrue
  \@printcopyrighttrue
  
  \renewcommand\@copyrightpermission{%
    \footnotesize
    This is the authors’ version of the work. It is posted here for your personal use. Not for redistribution. The definitive Version of Record was published in:
    \par
  }

  \renewcommand\@copyrightowner{%
  Copyright held by the owner/author(s).
  }
\makeatother

\definecolor{mygray}{gray}{0.9}
\newenvironment{promptbox}{%
  \MakeFramed {\FrameRestore}
  \raggedright
}%
{\endMakeFramed}

\begin{document}

\title{GPTCoach: Towards LLM-Based Physical Activity Coaching}

\author{Matthew J{\"o}rke}
\email{joerke@stanford.edu}
\orcid{0000-0003-2972-462X}
\affiliation{%
  \institution{Stanford University}
  \city{Stanford}
  \state{CA}
  \country{USA}
}

\author{Shardul Sapkota}
\email{sapkota@stanford.edu}
\orcid{0000-0002-0009-8672}
\affiliation{%
  \institution{Stanford University}
  \city{Stanford}
  \state{CA}
  \country{USA}
}

\author{Lyndsea Warkenthien}
\email{lyndsea@alumni.stanford.edu}
\orcid{0009-0005-7155-8356}
\affiliation{%
  \institution{Stanford University}
  \city{Stanford}
  \state{CA}
  \country{USA}
}

\author{Niklas Vainio}
\email{niklasv@stanford.edu}
\orcid{0009-0005-8490-9178}
\affiliation{%
  \institution{Stanford University}
  \city{Stanford}
  \state{CA}
  \country{USA}
}

\author{Paul Schmiedmayer}
\email{schmiedmayer@stanford.edu}
\orcid{0000-0002-8607-9148}
\affiliation{%
  \institution{Stanford University}
  \city{Stanford}
  \state{CA}
  \country{USA}
}

\author{Emma Brunskill}
\email{ebrun@cs.stanford.edu}
\orcid{0000-0002-3971-7127}
\affiliation{%
  \institution{Stanford University}
  \city{Stanford}
  \state{CA}
  \country{USA}
}

\author{James A. Landay}
\email{landay@stanford.edu}
\orcid{0000-0003-1520-8894}
\affiliation{%
  \institution{Stanford University}
  \city{Stanford}
  \state{CA}
  \country{USA}
}

\renewcommand{\shortauthors}{J{\"o}rke, et al.}

\begin{abstract}
  \input{sections/0_abstract}
\end{abstract}

\ccsdesc[500]{Human-centered computing~Empirical studies in HCI}
\ccsdesc[500]{Human-centered computing~Interactive systems and tools}
\ccsdesc[100]{Human-centered computing~Natural language interfaces}
\ccsdesc[100]{Computing methodologies~Natural language processing}
\keywords{Physical activity, health coaching, large language models (LLMs), personal informatics, conversational agents}

\begin{teaserfigure}
    \vspace{-0.8em}
    \includegraphics[width=\textwidth]{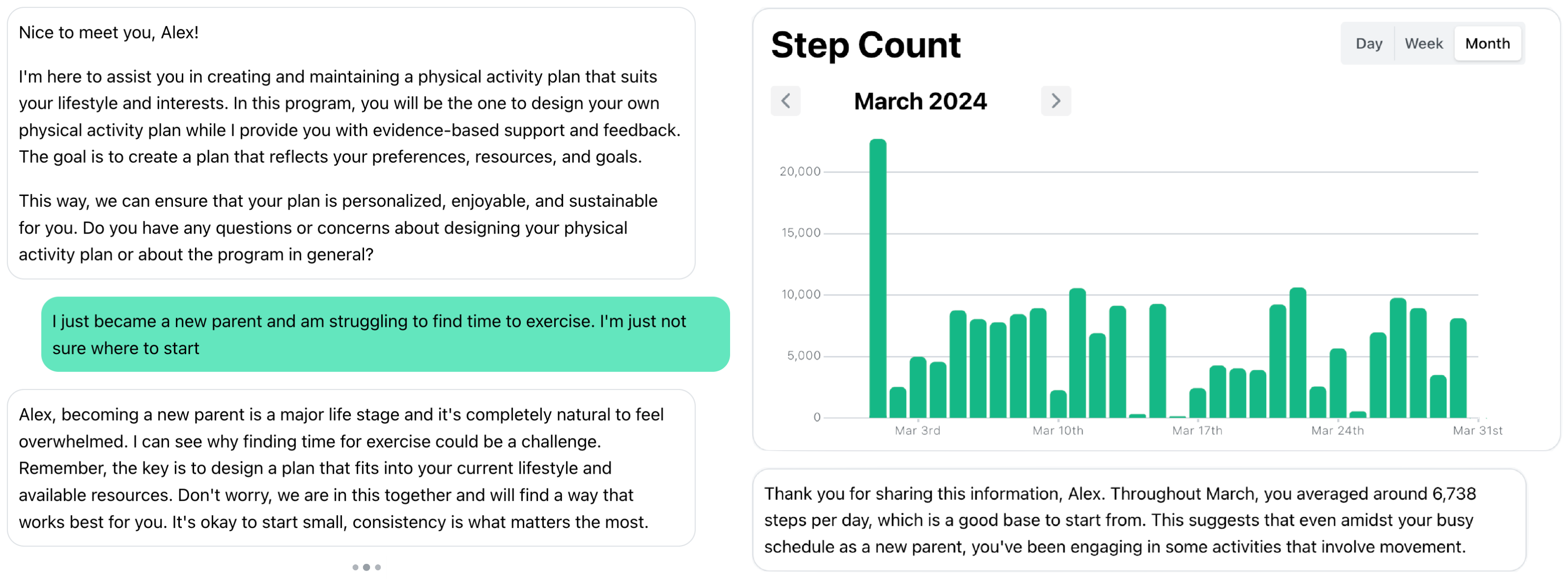}
    \caption{\textbf{\textsf{GPTCoach} is a health coaching chatbot designed to develop a physical activity plan that is tailored to the needs, abilities, and goals of a client.} \textsf{GPTCoach} implements the onboarding conversation from Active Choices~\cite{king2002stanford}, an evidence-based health coaching program, uses counseling strategies from motivational interviewing, and can query and visualize a user's health data from a wearable device through tool use. 
    On the left, we show an excerpt from an example conversation with \textsf{GPTCoach} that is representative of the conversation participants had in our lab study. 
    On the right, we show an interactive visualization displayed by \textsf{GPTCoach} at a later point in the conversation.}
  \label{fig:teaser}
  \Description{Example of a GPTCoach conversation and step count visualization. This figure consists of two parts: a text-based conversation and a bar chart. On the left, a chat interface shows a conversation between a user named Alex and GPTCoach. GPTCoach asks Alex about their physical activity goals and reassures them about the challenges of balancing exercise with parenting. Alex responds by sharing their struggle as a new parent to find time to exercise, and GPTCoach provides supportive feedback, emphasizing the importance of a personalized plan and small, consistent steps toward physical activity. On the right, a bar chart labeled “Step Count” for March 2024 shows the number of steps taken each day, ranging from 0 to 20,000 steps. The X-axis represents days in March, and the Y-axis represents the number of steps. A toggle at the top right allows switching between day, week, and month views, with the month currently selected. A brief message from GPTCoach below the chart mentions that Alex averaged around 6,738 steps per day in March and providing encouragement that they are still staying active amidst their busy schedule as a new parent.}
\end{teaserfigure}

\maketitle

\input{sections/1_introduction}
\input{sections/2_related_work}
\input{sections/3_formative_interviews}
\input{sections/4_system}
\input{sections/5_tech_probe_study}
\input{sections/6_results}
\input{sections/7_limitations}
\input{sections/8_discussion}
\input{sections/9_conclusion}

\begin{acks}
We thank the Stanford Institute for Human-Centered Artificial Intelligence (HAI) for partial funding support and the OpenAI researcher access program for providing API credits to support this research.
We extend our heartfelt gratitude to the Health Equity Action Research and Technology Solutions (HEARTS) lab for sharing their expertise and resources, particularly Maria Ines Campero, Prof. Abby C. King, and the Active Choices facilitators. 
We thank John Brelsford, Danielle Sessler, and the HETI Maine coding team for their consultation and support for MITI coding.
We thank all of the students who contributed to the design and development of our system, including Evelyn Hur, Bryant Jimenez, Niall Kehoe, Dhruv Naik, Evelyn Song, Romuald Thomas, Caroline Tran, and Anthony Xie, as well as the CS342 teaching staff for supervising our project team. 
Thank you to Andrea Cuadra, Ryan Louie, Omar Shaikh, and Danilo Symonette as well as the IxD research group for their continuous support and feedback throughout the project.
We are grateful for Beleicia Bullock, Alan Cheng, Elizabeth Childs, Cyan DeVeaux, Jane E, Ralf Herbrich, Michelle Lam, Lindsay Popowski,  Yujie Tao, Jordan Troutman, and Dora Zhao's contributions in recruitment, pilot testing, and feedback on paper drafts. 
We thank Michael S. Bernstein for providing feedback on inter-annotator agreement statistics.
Lastly, we thank all of our participants, without whom this work would not have been possible. 
\end{acks}

\bibliographystyle{ACM-Reference-Format}
\bibliography{references}

\newpage
\appendix
\onecolumn

\input{sections/appendix/A_formative}
\input{sections/appendix/B_system}

\input{sections/appendix/C_tech_probe}
\input{sections/appendix/D_mi_coding}
\input{sections/appendix/E_prompts}

\end{document}

%% file: sections/0_abstract.tex
Mobile health applications show promise for scalable physical activity promotion but are often insufficiently personalized. In contrast, health coaching offers highly personalized support but can be prohibitively expensive and inaccessible. This study draws inspiration from health coaching to explore how large language models (LLMs) might address personalization challenges in mobile health. We conduct formative interviews with 12 health professionals and 10 potential coaching recipients to develop design principles for an LLM-based health coach. We then built \textsf{GPTCoach}, a chatbot that implements the onboarding conversation from an evidence-based coaching program, uses conversational strategies from motivational interviewing, and incorporates wearable data to create personalized physical activity plans. In a lab study with 16 participants using three months of historical data, we find promising evidence that \textsf{GPTCoach} gathers rich qualitative information to offer personalized support, with users feeling comfortable sharing concerns. We conclude with implications for future research on LLM-based physical activity support.

%% file: sections/1_introduction.tex
\section{Introduction}
\label{sec:introduction}
Regular physical activity carries significant benefits to health~\cite{lieberman2021exercised, who2022}, yet over a quarter of the population worldwide~\cite{who2022} and nearly half of the population in the US~\cite{cdc2022} fail to meet standard recommendations for physical activity.
The gold standard for encouraging health behavior change is one-on-one interaction with a human health professional~\cite{bickmore2011reusable}. 
Health coaching is an effective and common form of such personalized support, but it is expensive, not widely accessible, and does not scale~\cite{bickmore2011reusable, mitchell2021automated}. 
Researchers have turned to mobile health technologies such as smartphone applications or wearable fitness trackers as a low-cost, scalable approach for promoting active lifestyles~\cite{hicks2023leveraging}. 
One promise of mobile health applications is to use data from activity trackers and other sensors to personalize health interventions and support. 
For example, nearly all modern smartphones include a pedometer, which could be leveraged to tailor interventions that encourage walking.
However, in comparison to support provided by health coaches, a key limitation of existing mobile health interventions is that they are insufficiently personalized to the \textit{qualitative} aspects of a person's context, such as goals, values, preferences, past experiences, life circumstances, time constraints, physical abilities, or access to resources \cite{murnane2020designing, rutjes2019beyond}.
These qualitative factors are generally the focus of coaching conversations~\cite{olsen2010health, wolever2013systematic, rutjes2019beyond, rutjes2022influence} and are crucial for effective support, but are challenging to standardize and quantify.

Recently, large language models (LLMs) have experienced rapid improvements in performance~\cite{ouyang2022training, bommasani2022opportunities}, presenting new opportunities to address personalization challenges in mobile health. Recent work has explored applications of LLMs in health and medicine, including health inference tasks~\cite{englhardt2023classification, kim2024health, liu2023large}, conversational agents for diagnosis~\cite{tu2024conversational, abbasian2023conversational}, and mental health counseling~\cite{chiu2024computational, shah2022modeling}. 
Inspired by these recent advances, we set out to explore the potential for LLMs to provide more personalized physical activity support than prior mobile health systems by drawing inspiration from human health coaching.
Specifically, LLMs offer the potential to integrate various sources of context, including qualitative context captured in natural language (e.g., goals, life circumstances, or past experiences) and quantitative context from self-tracking data.
Moreover, advances in conversational flexibility might allow a model to dynamically seek out information and adapt the structure and style of interaction in response to user input, much like a health coach. 

Despite this promise, several technical and design challenges remain. 
Due to the probabilistic nature of LLM outputs and the fickleness of prompting~\cite{zamfirescu2023herding}, an LLM-based health coach requires a fundamentally different system design than existing rule-based chatbots.
Moreover, off-the-shelf models do not natively support reasoning over raw sensor data as input~\cite{merrill2024language} and are instruction-tuned to answer questions, not engage in open-ended coaching conversations. 
A burgeoning interest in LLMs for behavioral health within both academic research and commercial products has focused on health question-answering or insight-extraction tasks~\cite{hegde2024infusing, ong2024advancing, vandelanotte2023increasing, englhardt2023classification, liu2023large, cosentino2024towards, merrill2024transforming, fang2024physiollm}. However, these concurrent efforts offer limited guidance on designing LLM agents that can seek out and incorporate nuanced qualitative information about a user's personal and environmental context.
Further, it remains unclear how to leverage these novel capabilities in ways that are endorsed by human health coaches and complements their coaching practice.

To navigate these challenges, we followed a human-centered process to design and evaluate \textsf{GPTCoach}, a GPT-4 based coaching chatbot for promoting physical activity. 
First, we conducted formative interviews with 22 participants, including 12 health professionals (health coaches, health educators, personal trainers, fitness instructors, and physical therapists) and 10 potential recipients of health coaching, ranging from highly sedentary individuals to athletes. 
Through a qualitative analysis of our formative interviews, we extracted three design principles that guided the design of \textsf{GPTCoach}: (1) follow a facilitative, non-prescriptive approach; (2) tailor information and advice using diverse source of context; and (3) adopt a supportive, non-judgmental tone.
\textsf{GPTCoach} implements the onboarding conversation from the Stanford Active Choices program~\cite{king2002stanford}, an evidence-based, clinically-validated health coaching program~\cite{wilcox2006results, wilcox2008active, king2007ongoing, king2014exercise, castro2011physical} developed by experts in behavioral medicine.
The onboarding conversation is one of the most important components of the Active Choices program and is crucial for effective support, setting the tone for the coaching relationship and requiring key skills such as building trust, maintaining a facilitative tone, and tailoring advice to a client's unique circumstances.
\textsf{GPTCoach} also uses strategies from motivational interviewing (MI), an established, evidence-based counseling approach~\cite{miller2023motivational, moyers2003assessing}, to engage in conversations that center the client's agency and motivation.
Moreover, \textsf{GPTCoach} can query and visualize health data from a wearable through tool use~\cite{schick2023toolformer}.
At the end of the conversation, \textsf{GPTCoach} generates a personalized physical activity plan.

In a single-session lab study with 16 participants using three months of their historical self-tracking data, we evaluate \textsf{GPTCoach}'s adherence to our design principles and participants' experiences interacting with \textsf{GPTCoach}.
We hired trained human coders to evaluate \textsf{GPTCoach}'s consistency with MI using the Motivational Interviewing Treatment Integrity (MITI) 4 code~\cite{moyers2016motivational} and additionally perform a counterfactual comparison to GPT-4.
Our single-session study design allowed us to evaluate \textsf{GPTCoach}'s ability to conduct a high-quality onboarding conversation under full researcher supervision, mitigating potential ethical risks associated with unpredictable LLM outputs while providing valuable insights to inform future work on multi-session coaching.

Through our formative interviews, system design, and lab study, we make three key contributions to the literature. First, we contribute a set of design principles for LLM-based physical activity coaching, emphasizing a non-prescriptive, non-judgmental approach that tailors advice to each client's unique circumstances. Second, we contribute the system design of \textsf{GPTCoach}, including a novel prompt chaining~\cite{wu2022ai} strategy to ensure appropriate adherence to the coaching program, use of MI strategies, and use of personal health data. Third, we offer insights and implications from our lab study evaluation of \textsf{GPTCoach}.

Our findings highlight the importance of considering the rich qualitative factors beyond sensor data that shape motivation, readiness for change, and the ability to take action. While concurrent work on LLMs for behavioral health has focused on extracting insights from wearable sensor data~\cite{merrill2024language, cosentino2024towards, fang2024physiollm}, our research demonstrates LLMs' potential to structure conversations in ways that centers the client's agency and fosters self-empowerment, which are critical aspects of successful health coaching programs~\cite{olsen2010health, wolever2013systematic}.
Survey and interview responses from our lab study indicate that participants perceived \textsf{GPTCoach}'s guidance to be highly personalized and actionable, as well as that they felt supported by and comfortable sharing concerns with \textsf{GPTCoach}.
Our MI coding results demonstrate that \textsf{GPTCoach} used MI-consistent or neutral behavior codes 93\% of the time and outperforms vanilla GPT-4 in MI-consistency.
However, \textsf{GPTCoach}'s ability to use sensor data was more variable, sometimes demonstrating the capacity to use data in ways that served conversations about change and other times failing to proactively incorporate data into its advice.
We conclude with a discussion of the implications for multi-session coaching, future LLM-based mobile health applications, and the risks of LLM-based health coaching.

%% file: sections/2_related_work.tex
\section{Related Work}
In this section, we summarize relevant literature on health coaching, conversational agents for health behavior change, as well as personal informatics and self-reflection on personal data.

\subsection{Health Coaching with Humans \& Conversational Agents}
\label{sec:related-work-health-coaching}
Health coaching is a popular and effective intervention for motivating health behavior change~\cite{olsen2010health, wolever2013systematic}. During health coaching, a coach establishes a partnership with their client and assists them in choosing healthy lifestyle behaviors by setting appropriate goals, maintaining accountability, and providing education and feedback~\cite{olsen2014health}. Most health coaching programs advocate for a client-centric and non-prescriptive approach, including collaborative goal-setting,  learning through self-discovery, fostering self-efficacy, and respecting client autonomy~\cite{wolever2013systematic}.
Many such programs draw from motivational interviewing~\cite{miller2023motivational}, an evidence-based counseling framework that provides guidance on how coaches and counselors can facilitate conversations to elicit motivation for behavior change.

Studies on the role of wearable and self-tracking data in health coaching have found that data can provide coaches with more objective reports on the client's behaviors, which can aid in providing personalized care~\cite{rutjes2019beyond, rutjes2022influence, chung2015more}. 
However, in a study examining how clients and coaches interpret data during coaching sessions, \citet{rutjes2022influence} found that data were not ``plug and play'' and needed to be contextualized through collaborative reflection to inform behavior change. 
Moreover, coaches may lack the time or necessary expertise to interpret clients' data~\cite{chung2015more}.

While highly effective, in-person health coaching is expensive, not always accessible, and does not scale~\cite{bickmore2011reusable, mitchell2021automated}. 
In contrast, \textit{automated health coaching}~\cite{mitchell2021automated} (also known as e-coaching~\cite{kamphorst2017coaching}) aims to simulate the health coaching experience using conversational agents. 
Early work by Bickmore et al. on 
health dialogue systems~\cite{bickmore2006health, bickmore2011reusable} has argued that automated systems can overcome the time and resource barriers affecting human counseling while effectively delivering evidence-based interventions at scale and low cost. In a meta-analysis of chatbots on lifestyle behaviors, \citet{singh2023systematic} found a significant, small-to-moderate effect of chatbots on motivating physical activity. 
Many coaching chatbots draw from or mirror principles and strategies from human health coaching, such as motivational interviewing~\cite{luo2021promoting, oh2021systematic, aggarwal2023artificial}, and some also make use of external data sources~\cite{mitchell2021automated, kocielnik2018reflection, king2020effects}.
Within HCI, several chatbot-based systems have been proposed for motivating physical activity~\cite{clavel2018wenner, kocielnik2018reflection, bickmore2005establishing, mollee2017evaluation}. 
For example, \citet{kocielnik2018reflection} developed an intent-based conversational system to explore mini-dialogues for triggering reflection on physical activity data.
\citet{mitchell2021automated} compared a scripted Wizard-of-Oz health coaching chatbot to a human counseling condition for patients with type-2 diabetes, finding that chatbots led to human-like experiences despite their rule-based nature. Human coaches were more skilled at expressing empathy and tailoring support, while the chatbot was more persistent and consistent. 

A limitation of all previously mentioned coaching chatbot systems is that they employ 
template-, rule-, or retrieval-based 
dialogue systems, which means that the system's response is always chosen from a pre-specified list of outputs.
This offers a higher degree of researcher control but falls short of the conversational flexibility and degree of personalization afforded by human coaches.
For example, in a review of conversational agents for physical activity promotion, \citet{luo2021promoting} report that the most common challenges were related to the agent's capabilities, such as only allowing multi-choice response options.
In contrast, LLM-based chatbots operate fundamentally differently by generating a unique response as a function of the conversation history. 
LLMs' responses mirror patterns learned in their internet-scale training data, which likely includes health blogs, medical guidelines, and coaching-related content, but may not always be reliable or up-to-date.
LLM-based chatbots have been explored in other domains of health and medicine, including educational support~\cite{baglivo2023exploring}, diagnosis~\cite{tu2024conversational}, and mental health~\cite{ma2023understanding, sharma2023human, chiu2024computational, hsu2023helping, chaszczewicz2024multi}. 
Emerging LLM-based approaches for physical activity coaching have explored prompting and re-ranking strategies to improve response quality~\cite{hegde2024infusing} or using LLMs for health question answering~\cite{ong2024advancing, vandelanotte2023increasing}. 
Recent work has also found that finetuned LLMs are capable of few-shot health tasks operating on raw self-tracking data~\cite{englhardt2023classification, liu2023large} and 
concurrent work has explored using LLMs for extracting insights from personal health data in the domains of physical activity and sleep~\cite{cosentino2024towards, merrill2024transforming, fang2024physiollm}.
Lastly, several commercially available LLM-based AI health coaches exist (e.g., WHOOP~\cite{whoopcoach} or
ONVY~\cite{onvy}), as well as the WHO's S.A.R.A.H~\cite{who2024sarah}. 
Unlike our approach, all of the aforementioned LLM-based systems focus on health question answering or insight extraction tasks and are thus not designed to seek out and incorporate nuanced qualitative information about a user's personal and environmental context. 
Informed by our expert interviews and the health coaching literature, we also place a greater focus on structuring open-ended conversations about change in ways that centers the client's agency and fosters self-empowerment.

\subsection{Personal Informatics \& Reflection on Personal Data}
The field of personal informatics designs systems to ``help people collect and reflect on personal information''~\cite{li2010stage}. 
Motivating physical activity is the most common application domain for personal informatics~\cite{epstein2020mapping}, with a long history of systems for motivating physical activity behavior change in HCI~\cite{consolvo2008activity, munson2012exploring, lin2006fish, anderson2007shakra}. 
Through reflection on personal data, personal informatics tools aim to generate new insights about behaviors and habits that can inform lasting behavior change~\cite{cho2022reflection, bentvelzen2022revisiting}. 
This has been referred to as the \textit{self-improvement hypothesis}~\cite{kersten2017personal}.
While the personal informatics literature has explored many design patterns for supporting reflection~\cite{bentvelzen2022revisiting}, many systems (and nearly all commercially available fitness trackers) make use of statistics and visualizations to support reflection~\cite{mcduff2012affectaura, choe2015characterizing, aseniero2020activity, huang2016field, thudt2015visual}.

Personal informatics tools are effective at fostering reflection and motivating behavior change~\cite{jin2022self}, but many practical challenges remain. 
First, data interpretation is challenging for non-experts~\cite{grammel2010information, rapp2016personal}. In a diary study with non-expert users, \citet{rapp2016personal} found that visualizations were ``neither meaningful nor tailored to the user needs'' and argued for greater personalization and contextualization of data.
Moreover, highly quantitative forms of feedback may be potentially harmful~\cite{murnane2020designing}, negatively impacting motivation~\cite{cohen2014psychology} and mindsets~\cite{crum2007mind}. 
In a review of self-tracking,~\citet{kersten2017personal} found that insights gained through reflection on personal data are frequently not actionable towards behavior change.
Many systems tacitly assume that reflection will occur naturally once data is processed and visualized~\cite{baumer2014reviewing}, while most theories of reflection highlight that reflection needs to be explicitly supported~\cite{slovak2017reflective, bentvelzen2021development}. 
Our work aims to better support reflection on personal data by using LLMs to contextualize a user's health data to their unique qualitative context. Moreover, \textsf{GPTCoach} is instructed to ask reflective questions and assist the user in translating insights into actionable plans for change.

%% file: sections/3_formative_interviews.tex
\section{Formative Interview Study}
\label{sec:formative-study}
Prior work has studied the role of self-tracking devices and data in the health coaching process~\cite{rutjes2019beyond, rutjes2022influence}, as well as user experiences interacting with rule-based conversational agents~\cite{mitchell2021automated}. However, it was unclear the degree to which these findings would transfer to LLM-based health coaching systems given the stark differences in capabilities.
Thus, we conducted formative semi-structured interviews with 22 participants, including 12 health experts and 10 potential recipients of health coaching.

The goal of our formative study was to better understand how LLMs can provide personalized physical activity support by drawing inspiration from human health coaching practices. While human and automated coaching share many similarities, they differ in their interaction medium and their respective capabilities. For example, as noted by \citet{mitchell2021automated}, automated coaches are always available and consistent, while human coaches can establish personal connection through embodied interaction and shared lived experience. 
Recognizing these differences, we solicited diverse perspectives on the affordances and limitations of LLMs for health coaching. This approach allows us to understand where LLM health coaches might (or might not) excel and which aspects of human coaching should (or should not) be replicated.

Specifically, we investigated the following research questions:
\begin{enumerate}[noitemsep]
    \item[\textbf{RQ1:}] What strategies do health experts use to help their clients overcome barriers to physical activity? What strategies do individuals use to help themselves overcome barriers to physical activity? \textit{Which of these strategies could LLMs employ and how?}
    \item[\textbf{RQ2:}] How do health experts make use of clients' self-tracking data to promote physical activity? How do individuals make use of their self-tracking data to promote physical activity? \textit{How might LLMs make use of self-tracking data to promote physical activity? }
\end{enumerate}

\subsection{Participants}
We recruited 22 participants from various sources, including university mailing lists, personal contacts, and former research participants. 12 participants were health experts with professions including health coach, health educator, personal trainer, fitness instructor, YMCA vice president, physical therapist, and behavioral scientist. 
Six of the health experts had previous experience as peer health counselors in a study that evaluated a rule-based chatbot~\cite{king2020effects}.
The 10 remaining non-expert participants were recruited from the general population to gain the perspective of potential recipients of health coaching.
This participant pool included an NCAA Divison 1 collegiate athlete, former athletes, avid self-trackers, physically active and inactive students, sedentary office workers, new parents, a program manager working on AI data quality, as well as active and inactive older adults.
Participant demographics are listed in Table~\ref{table:formative-demographics}.

\input{tables/formative_demographics}

\subsection{Procedure}
We conducted one-hour semi-structured interviews with participants, pre-approved by our university's institutional review board. Health experts were asked several questions about their relationships with clients, strategies for overcoming activity barriers, and the role of technology and data in their practice. Non-experts discussed their physical activity levels, barriers, motivation sources, goals, and use of data. In the last portion of our interview, participants were asked to ``imagine that you have access to an artificial intelligence chatbot that can help you improve your physical activity,'' and were asked to share feedback on several potential features. Further details are provided in Appendix \ref{appendix:formative}.

\subsection{Analysis}
\label{sec:formative-analysis}
We performed qualitative coding on our interview transcripts using thematic analysis~\cite{braun2012thematic}. 
Two authors coded two interviews collaboratively, then independently coded roughly 20\% of the interviews. 
One author coded the remaining 80\%, with periodic meetings to discuss codes and emerging themes, resolving any differences in interpretation through conversation. 

\subsection{Results}
Towards answering \textbf{RQ1}, we discuss the role of coaches as facilitators, educators, and supporters. Towards addressing \textbf{RQ2}, we then discuss the role of data as guiding, not driving, physical activity behavior change. 

\subsubsection{RQ1: The Role of Coaches as Facilitators, Educators, and Supporters}
\label{sec:formative-rq1}
We identified three major themes by analyzing the diverse strategies that coaches and individuals use to foster motivation and overcome barriers to behavior change. These themes center around coaches' roles in their interactions and relationships with clients. For each role, we discuss the implications for LLM coaching, including the affordances and limitations of LLMs in filling the role, as well as potential opportunities to go beyond the kinds of support provided by human coaches.

\noindentparagraph{\textbf{Facilitators:}}
Despite our health experts having different training, job titles, responsibilities, and client populations, they all described an approach that was facilitative, not prescriptive in nature. Experts used various language to describe this strategy, such as 
\textit{``You're not in the driver's seat, you're more in the passenger seat, providing maybe direction, steering the conversation one way or the other'' (P6).} Core to a facilitative strategy is that the client takes ownership of their behavior change journey. 
A facilitative approach decidedly does \textit{not} involve unsolicited advice or problem-solving: \textit{``No, definitely not giving them the solutions [...] We're not advice givers'' (P15).}

Experts cite several reasons for using facilitation as a strategy for overcoming barriers. First, several claimed that prescriptiveness does not work, as P12 reflected 
\textit{``Friends that are like, give me a diet, or give me a workout plan. They never follow it.''} Experts also noted that advising without comprehensive understanding and making assumptions could undermine credibility and trust.
Facilitation empowers clients and fosters self-efficacy so that clients can sustain healthy behaviors and learn to solve problems on their own. As P11 emphasized, their facilitation goal with their clients was \textit{``building that capacity so they can, so they can have the resiliency and tools needed to empower themselves.''}
Moreover, many coaches simply do not have time to adopt a prescriptive role: \textit{``I want you to build those habits, [...] I don't have the time or energy for it. And also, I don't think that long term, that's the best strategy for building intrinsic motivation.'' (P3).}

\noindentparagraph{Implications for LLM Coaching:}
Several experts independently cited motivational interviewing~\cite{miller2023motivational} as a guiding framework and many experts mentioned conversational strategies such as open-ended questions, reflective statements, affirmations, reframing, or advising with permission.
Surprisingly, health experts did not describe AI's behavior with facilitative terms,
such as P15 suggesting that an AI could help \textit{``come up with some solutions.''}
LLMs' conversational flexibility might allow them to structure facilitative conversations and implement strategies from frameworks like motivational interviewing, although this behavior generally conflicts with LLMs' instruction-following and questioning-answering objectives. 
This aligns with prior work in mental health support, which has found that LLM therapists prioritize providing solutions over asking questions~\cite{chiu2024computational}. 

\noindentparagraph{\textbf{Educators:}}
Experts refrain from giving unsolicited advice and spoke about applying their advanced knowledge to help clients overcome barriers.
As P8 noted, \textit{``my superpower is sifting through a lot of information and pulling together the patterns.''} 
This does not conflict with a facilitative approach; rather, information, education, and solutions should all be carefully tailored to a client's unique needs and situation: 
\textit{``What is motivating you right now? [...] And then trying to find the common threads with things that [...] I know about and can help'' (P3).}
Open-ended questions, reflective listening, asking for permission, and other strategies assist coaches in ensuring that they have gathered enough information to provide advice that is aligned with the client's needs, abilities, and resources: 
\textit{``The doctor said you need to be more physically active. So what I need you to do is go to a pool. [...] How do I know that you have access to a pool?'' (P19).}
Other kinds of education are targeted at reframing deep-seated beliefs about what counts as activity and why it is beneficial.

The process of tailoring advice is not an objective acquisition of information, as most experts also described engaging in interpretation:
\textit{``It's a mix of trying to provide what they necessarily want with what they need'' (P2).}
For example, experts and non-experts describe lack of time as an exceedingly common barrier to physical activity. While individuals differ greatly in their responsibilities and time commitments, experts noted that time barriers are also due to the perception of a lack of time.
Thus, coaches aiming to assist clients in overcoming a lack of time might suggest both time-management strategies and help clients reframe their perceived time barriers.

\noindentparagraph{Implications for LLM Coaching:} LLMs have the potential to perform well---perhaps better than human experts---at providing personalized physical activity information, given their broad, internet-scale knowledge and advanced question-answering capabilities. Several participants appreciated this possibility: \textit{``I think that is one of the most promising parts about generative AI, in my opinion, is being able to get the answer to somebody who doesn't necessarily know how to ask or how to look for it.'' (P4).}
However, participants were also quick to acknowledge the limitations of LLMs' knowledge. 
Many experts were concerned about the AI's sources of information and potential for hallucination. 
Others noted that AI might perform well for generic physical activity advice but could fail for highly specific activities.

\noindentparagraph{\textbf{Supporters:}} All of our health experts acknowledged the crucial role of personal connection and support, sharing comments like
\textit{``Just making everybody feel welcome. That's it. No matter who you are, where you're from, what your financial, social background is'' (P5).}
Health behavior change is a profoundly personal and emotional process. When asked to reflect on their barriers to physical activity, several of our non-expert participants shared traumatic experiences and insecurities, and experts mentioned that many of their clients lack confidence or have anxieties. The strongest differentiator between our most active and inactive participants revolved around identity---whether being active was core to who they were. Highly active participants used language that affirmed this identity,
\textit{``It's just in my DNA. It's just what I love to do'' (P5)}, while many of the most challenging barriers were rooted in identity conflict: 
\textit{``It's pretty depressing some days, you know, it's like I'm missing half of myself [...] I'm just a mom, and then I think back on those days when I did skate and compete pretty regularly'' (P10).} 

Health coaching is fundamentally relational---a partnership between a coach and client working towards a shared goal~\cite{olsen2010health}. 
Experts emphasized the importance of building rapport and trust with their clients as a way to reduce anxiety and fear: \textit{``once I meet with the clients and we kind of develop a little bit of rapport, it kind of takes a little bit of animosity from the weight room'' (P2).} 
Many coaches highlight positive affirmations, such as
\textit{``the strategies that I always use for this population is a lot of encouragement, a lot of celebration'' (P1).} 
These positive gestures can not only make clients more comfortable, but also encourage accountability and habit formation: 
\textit{``I think accountability is really important, having somebody at your corner and feeling that support'' (P6).}
Each coach supports dozens, if not hundreds, of clients, and there are limits to coaches' capacities:
\textit{``What I've had to learn is I can't help everybody, and not everybody can afford to see me, and I feel terrible about that'' (P8).}
Many experts emphasized the importance of building community to encourage mutual support and reduce reliance on the coach.

\noindentparagraph{Implications for LLM Coaching:} Expert and non-expert participants had mixed opinions on whether an AI can or should establish such personal connections. Most people liked the ability of an AI to hold them accountable with personalized and encouraging reminders. 
Many also liked the ability for an AI to adapt its persona, e.g., \textit{``depending on the day, I will need a different person. Sometimes I'm going to need that empathetic, but other times I'll need the strict regimen''~(P13).}
However, many hesitated to discuss highly personal, emotional topics with an AI, such as P10 disclosing \textit{``I guess that's a little too personal. [...] Are we here for exercise, or are we here for, like, mental counseling?''}
Some did not believe it was possible to form such a connection with an AI: \textit{``I don't think that my type of job, instructor wise, will ever be taken [...] Even though it will have all the information, it's not personal.'' (P5)}, while others were readily open to the idea.
Interestingly, a health coach who had previously worked on a rule-based chatbot mentioned that
\textit{``when you play video games, you immerse yourself in a world [...] these are older adults that have never been exposed to anything like this before, and I think they would develop a connection'' (P11).} Moreover, this chatbot had advantages over human coaches: \textit{``they had this avatar that they can go see once a week, was always there, like, didn't run late, you know, didn't judge them'' (P11).} 

In line with prior literature, our findings suggest that people can and will form personal connections with a chatbot (even if they deny it~\cite{reeves1996media}) and that this can be leveraged to create accountability and motivation toward physical activity behavior change with positive affirmations and encouragement. We expect that LLMs will offer marked improvements over rule-based chatbots. While LLMs can project empathy~\cite{cuadra2024illusion}, it is clear that LLMs \textbf{should not be designed to replace human connection}, but rather to foster motivation and bolster existing relationships with coaches and communities.

\subsubsection{RQ2: The Role of Data \& Technology as Guiders, not Drivers}
Many of our participants used wearable trackers and apps. 
In line with prior work in personal informatics~\cite{choe2014understanding, choe2017understanding}, participants cited many reasons for using data, including monitoring daily fluctuations, long-term trends, progress towards goals, or in-the-moment workout statistics. 
Participants appreciated wearable data as a more ``objective'' account of behavior, but were particularly frustrated when their behavior was not captured~\cite{consolvo2008activity}, like P5 mentioning \textit{``If I forget to put my watch on, my whole day is ruined. [...] the day just doesn't exist anymore.''}
Most health experts acknowledged the utility of self-tracking data 
and characterized data as most helpful in maintaining accountability and consistency for individuals already on the path to becoming active.
By analogy, one health educator asked us, 
\textit{``How would driving a car be a different experience if you had no gauges in front of you?'' (P18)} 

Despite this, several coaches did not analyze data with their clients except in high-performance athletic or clinical contexts. 
Many coaches do not have time to analyze data, \textit{``I can't scale that. I have, like, 20 clients'' (P3)}, and also acknowledged their biases, \textit{``Despite me having sort of more understanding of fitness science or exercise science broadly, I'm still falling into the same traps'' (P4).} Incorporating data into health coaching also presents additional challenges. Data showing a lack of progress can negatively impact motivation
and data often lacks important context:
\textit{``I feel like a lot of people, especially in fitness, think very quantitatively for everything, which is good to a certain extent, but, like, doesn't give you the whole picture'' (P1).}

\noindentparagraph{Implications for LLM Coaching:} Both expert and non-expert participants were generally excited by the prospect of an AI coach analyzing data.
Participants had a dominant view of AI as an information-synthesis machine, such as P5 stating \textit{``With AI, it has so much information within it. All you would have to do is tell them what your goals are, and it will obviously give you an answer.''} Participants wanted an AI coach to use their data to help them set goals and maintain accountability: 
\textit{``I'm not accountable to anyone other than myself, [...] but, if you had a chatbot going hey, you only got 7000 steps and now you're sitting on your butt'' (P7).} An AI coach could also help flexibly re-adjust goals over time to changes in motivation, life circumstances, or injuries.
Finally, many participants expressed privacy concerns when integrating health data with a chatbot.
Even when data is fully secured, experts noted that user perceptions of privacy and prior expectations play a critical role: \textit{``it's about what people think about what's going to happen. This is about people's perceptions. [...] Family and cultural dynamics that come into play also trust can also play a big role'' (P19).}

LLMs' use of wearable data is an opportunity to better support health behavior change. 
Conversational LLMs could enable personalized data analysis in ways tailored to the client, such as reflective sessions that examine historical data, establishing baselines for goal-setting, providing real-time support during or after workouts, or augmenting missing/inaccurate data with additional context.
To provide adequate data-driven support, 
an agent must integrate data with extensive qualitative information about a user's life and carefully present this information to foster motivation and empowerment.

%% file: tables/formative_demographics.tex
\begin{table*}[tb]
\small
\centering
\begin{tabular*}{\textwidth}{@{\extracolsep{\fill}} l p{15cm}}
\toprule
\textbf{Age} & Mean: 40.90, Median: 38, SD: 14.31, Min: 23, Max: 72 \\
\midrule
\textbf{Gender} & Female: 12, Male: 10 \\
\midrule
\textbf{Race/Ethnicity} & White: 10, Hispanic or Latino: 7, East Asian: 2, 
Southeast Asian: 2, South Asian: 1, African-American or Black: 1, 
Black English: 1 \\
\bottomrule
\end{tabular*}
\caption{\textbf{Summary of participant demographics in the formative study} $(N=22)$.}
\label{table:formative-demographics}
\end{table*}

%% file: sections/4_system.tex
\section{\textsf{GPTCoach}: Design \& Implementation}

Guided by our formative interviews, we designed \textsf{GPTCoach}, 
a chatbot system for physical activity coaching. 
In this section, we discuss our design process and system architecture. Our code is available at: \url{https://github.com/stanfordhci/gptcoach-chi2025}.

\subsection{Design Principles}
\label{sec:principles}
Our formative interviews revealed the role of coaches as facilitators, educators, and supporters, as well as insights into how LLMs might fill each of these roles. From these roles, we extracted three design principles for \textsf{GPTCoach}:

\begin{itemize}
    \item[{\textbf{DP-1:}}] \textbf{Follow a facilitative, non-prescriptive approach.}
    Health experts all described a facilitative approach to coaching in which clients drive their own behavior change journey. The chatbot should similarly stay ``in the passenger seat,'' empowering clients to make a change rather than prescribing what to do.
   
    \item[{\textbf{DP-2:}}] \textbf{Tailor information and advice using diverse sources of context.} Any information and advice provided by the chatbot should be tailored to a user's unique personal and environmental factors. The chatbot should integrate both qualitative and quantitative sources of context.

    \item[{\textbf{DP-3:}}] \textbf{Adopt a supportive, non-judgmental tone.}
    Health behavior change is a deeply personal journey and many clients face anxieties and fears around exercise. The chatbot should adopt an uplifting, non-judgmental tone to help clients feel comfortable and supported.
\end{itemize}

\subsection{The Active Choices Program}
\label{sec:active-choices}
We designed \textsf{GPTCoach} to implement the onboarding conversation of the Stanford Active Choices program~\cite{king2002stanford}, an evidence-based, clinically validated counseling program for physical activity promotion~\cite{wilcox2006results, wilcox2008active, king2007ongoing, king2014exercise, castro2011physical}.
Active Choices is grounded in behavior change theory, including the transtheoretical model~\cite{prochaska1997transtheoretical} and social cognitive theory~\cite{bandura1999social}.
During onboarding, coaches first introduce themselves and outline program expectations. They then discuss the client's past experiences, barriers to physical activity, health or injury concerns, and motivations. The session concludes with collaborative goal-setting based on the FITT (Frequency, Intensity, Time, and Type) framework along with advice (with permission, and if appropriate) for helping the client achieve their goal. Although other health-related topics may arise (e.g., nutrition, weight loss, sleep, and mental health), Active Choices coaches are encouraged to steer the discussion back to physical activity. We adopt a similar approach in \textsf{GPTCoach}, treating these topics as largely out of scope, but not entirely off-limits.

In the full Active Choice program, the onboarding conversation is followed by several shorter follow-up contacts that occur every few weeks over six months or more. In this work, we focus only on the onboarding conversation to evaluate \textsf{GPTCoach}'s adherence to design principles in a single-session lab study. Given the open-ended and highly personal nature of health coaching, the risks of LLM coaching are greater than traditional mobile health applications. This study design allows for full researcher supervision and intervention if necessary, minimizing potential risks from unpredictable outputs.
Although multi-session coaching is generally required for sustained behavior change, we note that the onboarding conversation is the longest and potentially most impactful session, setting the tone for the program and the relationship with the client. Successful onboarding conversations require key skills that are transferable to follow-up sessions, such as facilitating a supportive and non-judgmental conversation, seeking out important information about a client's background and motivation, and tailoring information and advice to a client's unique circumstances. Thus, a system capable of conducting high-quality onboarding conversations is likely to perform well in follow-up contacts. We explore how \textsf{GPTCoach} might be extended for follow-up sessions in Section \ref{sec:discussion}.

\subsection{Design Process}
LLMs present several design challenges due to their unpredictable outputs and the fickle nature of prompting~\cite{zamfirescu2023herding}. Thus, we designed \textsf{GPTCoach} as a \textit{technology probe}~\cite{hutchinson2003technology} to assess the viability of our approach and generate new ideas for design. 

We engaged in several design activities in creating \textsf{GPTCoach}.
We received training manuals from the Active Choices team along with descriptive statistics from a previous study on a rule-based chatbot~\cite{king2020effects}. 
During initial experiments, we utilized our own wearable data to develop a prototype resembling the final architecture, but lacking an additional prompt chain for tool calls (Section \ref{sec:prompt-chains}). This prototype was tested with two health experts and four non-experts in a pilot study. Feedback from participants highlighted a need for the model to more proactively query for data.
In response, we adjusted nearly all of our prompts and added the tool call prompt chain to explicitly check for opportunities to query data. Since our prompt chaining approach proved to be effective in our pilot studies, we did not explore finetuning or preference alignment~\cite{ouyang2022training} in this work, which would necessitate an additional dataset of expert annotations. 

\subsection{System Architecture}
\begin{figure*}
    \centering
    \includegraphics[width=\textwidth]{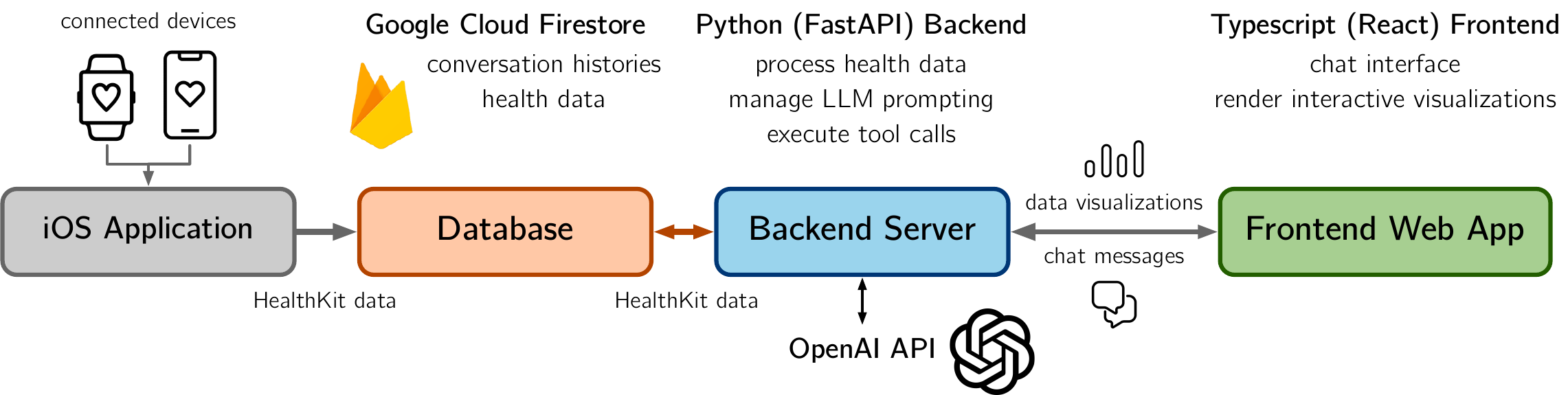}
    \caption{\textbf{Overview of \textsf{GPTCoach}'s System Architecture.} HealthKit data from connected devices (e.g., iPhones and wearables) are synced to our Google Cloud Firestore Database using an iOS application. A Python backend server handles several features: fetching health data from the database, aggregating and featurizing health data, handling LLM prompt management logic, interfacing with the OpenAI API to query GPT4, executing tool calls, generating data visualizations for the frontend, as well as sending and receiving chat messages from the frontend. The Typescript/React frontend web app hosts the chat interface and renders interactive data visualizations.}
    \label{fig:system-architecture}
    \Description{GPTCoach’s System Architecture Diagram. The diagram illustrates how data flows between the components. The iOS Application (left) receives HealthKit data from connected devices like smartwatches and smartphones, which is then sent to the database (middle-left). The backend Server (middle-right): The backend server interacts with the OpenAI API (below the server) and sends data visualizations and chat messages to the frontend (right).}
\end{figure*}

Our system's architecture (Figure~\ref{fig:system-architecture}) consists of four main components: 
(1) a \textit{database} containing raw health data and conversation histories; 
(2) an \textit{iOS application} that fetches three months of historical data using Apple's HealthKit API and uploads the data to our database; 
(3) a \textit{backend server} (Python), which handles all LLM logic and tool call execution, and 
(4) a \textit{frontend web interface} (Typescript/React)  that displays the chat interface and interactive data visualizations. 
Our system uses Google Cloud Firestore for our database and Google Cloud Run to host the backend and frontend. \textsf{GPTCoach} builds on the open-source Spezi ecosystem~\cite{schmiedmayer2024} for creating iOS-based digital health apps and uses GPT-4~\cite{achiam2023gpt} via the OpenAI Chat Completions API for LLM interactions. 
In the following sections, we discuss how our architecture supports a data pipeline that enables our chatbot to call tools that fetch personal health data, as well as several prompt chains~\cite{wu2022ai} to encourage adherence to the coaching program, adherence to motivational interviewing, and appropriate use of wearable data. 

\subsubsection{Data Pipeline}
Our data pipeline consists of a set of data source inputs (e.g., heart rate or step count). The exact set of data sources depends on the user's device(s) and permissions granted to our application. All possible data sources are listed in Appendix \ref{appendix:system-details}. We use the Spezi framework~\cite{schmiedmayer2024} (iOS/Swift) to read data from Apple HealthKit, encode it into the HL7 FHIR standard~\cite{HL7FHIR}, and upload it to Firebase. 
Though we make use of Apple HealthKit to fetch health data, our system is not limited to Apple devices. Any wearable device that syncs with HealthKit (e.g., Oura and WHOOP) can be used as input to our system. 

To provide our model access to information from wearable sensor data, we expose two tools to the LLM:
\begin{itemize}[leftmargin=*]
    \item \texttt{describe(data\_source, date, granularity)}:
    This function fetches all data within the given granularity (\texttt{`day|\\week|month'}) from the reference date and returns a natural language description containing aggregated summary statistics and a data source description. 
    \item \texttt{visualize(data\_source, date, granularity)}: 
    This function returns the same output as \texttt{describe}, but additionally sends a message to the frontend to display an interactive data visualization to the user (see Figure~\ref{fig:teaser}B).
\end{itemize}

The summary statistics and visualizations reported by the function call depend on the type of data source. For instance, count types, such as step count, active energy burned, or exercise time, are summed and visualized with a bar chart. Rate types, such as heart rate, oxygen saturation, and respiratory rate, are averaged and visualized with a line chart. Our system is extensible in that new types and visualizations can be added flexibly for new data sources.

A tool call is initiated by GPT-4, which determines when and how to trigger a function based on the input. The Chat Completions API generates a JSON output based on the function signature, specifying the input arguments and their types. When the model initiates a tool call, the backend fetches the data from the database, generates the aggregated summary statistics, and returns a summarized text description to the model. For example, our data pipeline would return the following text description to the LLM after generating the visualization in Figure~\ref{fig:teaser}:
{
\begin{quote}
\small 
\begin{verbatim} 
> describe(
    data_source_name="health.stepcount",
    start="2024-03-01", 
    end="2024-03-31", 
    granularity="day"
  ) 
           
  2024-03-01:00:00:00 to 2024-03-01:23:59:59:
   22728 steps from Apple Watch (108 entries)
  2024-03-02:00:00:00 to 2024-03-02:23:59:59:
   2528 steps from Apple Watch (21 entries)
  2024-03-03:00:00:00 to 2024-03-03:23:59:59:
   4987 steps from Apple Watch (13 entries)
  ...
  2024-03-29:00:00:00 to 2024-03-29:23:59:59:
   3510.00 steps from Apple Watch (60 entries)
\end{verbatim}
\end{quote}
}

\begin{figure*}
\centering
    \includegraphics[width=\textwidth]{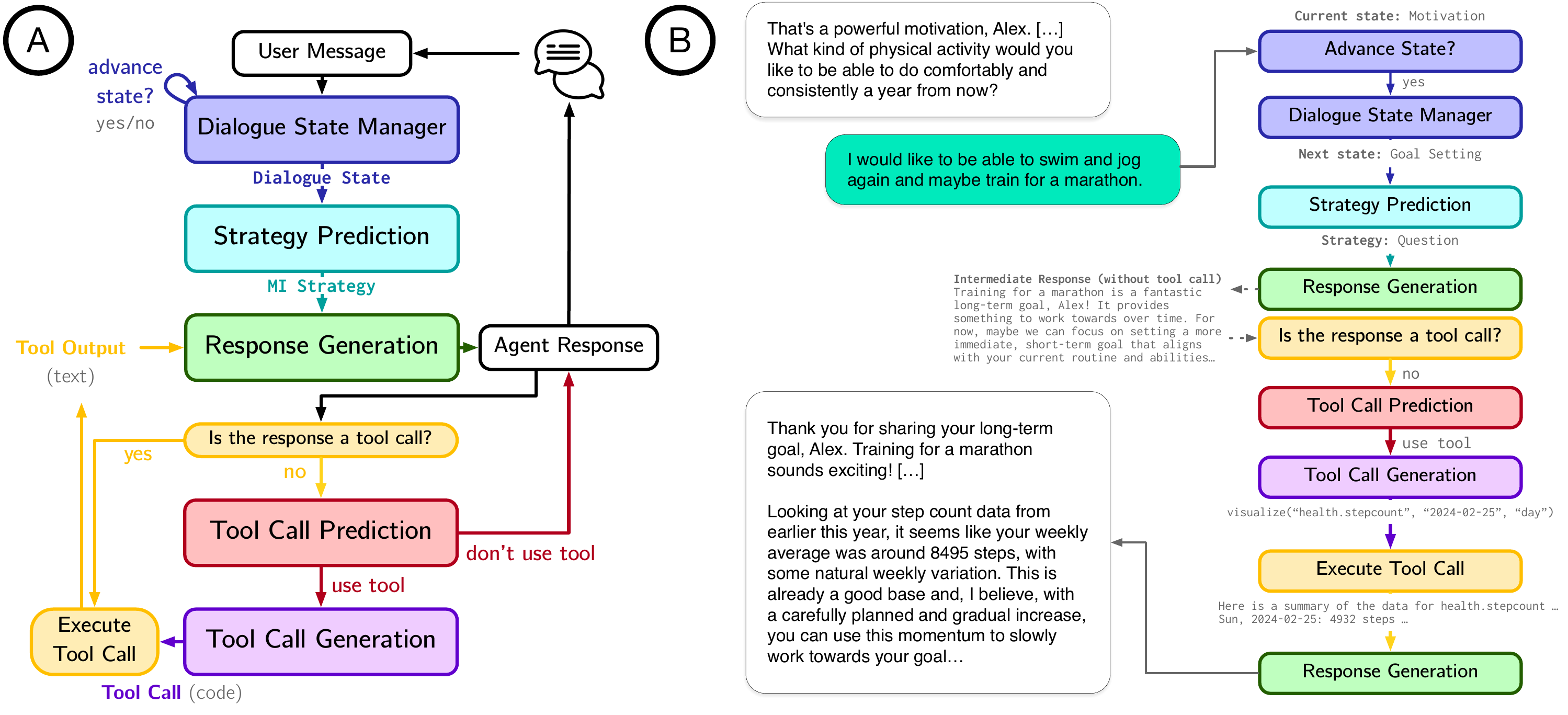}
    \caption{\textbf{Overview and a walkthrough of \textsf{GPTCoach}'s Prompt Chains.} (A) On the left, we show an overview of the prompt chains, in which the first chain manages the dialogue state, the second chain grounds the model's response in MI strategies, and the third chain determines whether the response should be augmented with health data. (B) On the right, we show the outputs of each prompt chain for an example exchange.}
    \label{fig:prompting-diagram}
    \Description{Overview and Walkthrough of GPTCoach’s Prompt Chains. This figure consists of two panels. In panel A (left), a flowchart illustrates the overall architecture of GPTCoach’s prompt chains using arrows that represent decision paths and information flows. The process begins when a user message is received. This message is processed by the dialogue state manager, which determines whether to advance the conversation state. The next step is strategy prediction, where an appropriate motivational interviewing (MI) strategy is selected. The selected strategy informs the response generation module, which creates the agent response. If the response does not already contain a tool, it enters tool call prediction to decide whether to use the tool or not. If a tool is required, the process moves to tool call generation, which creates a tool call (code) that is executed by the system. The tool output (text) is used to generate a new agent response. If a tool call is not required, the original agent response is returned. In panel B (right), the prompt chains’ intermediate outputs are shown for an example interaction between Alex and GPTCoach. Alex expresses the desire to swim and jog again and train for a marathon. The dialogue state manager decides to advance the state from Motivation to Goal Setting. The strategy prediction outputs the Question strategy. Response generation produces an intermediate response (without a tool call) providing encouragement and offering assistance in setting a short-term goal. Since this response does not contain a tool call, the message is passed to tool call prediction, which decides to call a tool. Tool call generation calls visualize(“health.stepcount”, “2024-02-25”, “day”) and tool call execution produces a summary of the data. Response generation is called again and produces a new message that offers encouragement and incorporates Alex’s step count baseline.}
    \vspace{-1em}
\end{figure*}

\subsubsection{Prompt Chains}
\label{sec:prompt-chains}
We encountered several limitations to vanilla prompting approaches\footnote{We define ``vanilla prompting'' as single-step output generation where a model $f$ samples a response $\mathbf{y} \sim f(\cdot | \mathbf{x})$ conditioned on a prompt $\mathbf{x}$. In contrast, a prompt chain relies on several intermediate generations $\mathbf{x}_t \sim f(\cdot | \mathbf{x}_1, ..., \mathbf{x}_{t-1})$ to produce the final response $\mathbf{y} \sim f(\cdot | \mathbf{x}_1, ..., \mathbf{x}_n)$.}. The model struggled to adhere to the onboarding session's structure, easily veered off-course, and had a strong tendency to give unsolicited advice. 
We provide further quantitative evidence of vanilla GPT-4's bias towards unsolicited advice in Section \ref{sec:counterfactual-results}.
To address these issues, we constructed three prompt chains~\cite{wu2022ai} to elicit our desired behaviors: a \textit{dialogue state} chain, a \textit{motivational interviewing} chain, and a \textit{tool call} chain. Upon receiving a user message, the conversation history sequentially passes through these chains (see Figure~\ref{fig:prompting-diagram}), each of which initiates a separate call to a GPT agent. We describe each prompt chain below and provide the full structure of all prompts in Appendix \ref{appendix:prompt-chains}.

\noindentparagraph{Dialogue State Chain.} The onboarding session consists of a series of topics that the coach discusses with the client. 
However, the agent needs to maintain flexibility throughout the conversation. 
For example, if a client asks an off-topic question, the coach should gently redirect the conversation back on topic. If a client gives an incomplete answer, the agent should politely follow up. Similarly, if a client mentions an injury early in the session, the agent should either reference this concern when asking about health or injury concerns or skip this question.

We organized the session into a linear sequence of dialogue states, each associated with a prompt. 
We created these prompts based on our formative interviews with health experts, as well as the Active Choices manual. 
Each prompt consists of a clear high-level task (e.g., \textit{``Your current task is to help your client set a physical activity goal.''}), along with state-specific subtasks (e.g., \textit{``First, help them set a short term goal, if they have not already identified one themselves.''}) and advice (e.g., \textit{``Connecting their short term goal to larger motivations can help them stay motivated.''}).
When the user sends a message, an external LLM agent classifies whether \textit{``the agent has successfully completed the following task,''} advancing to the next dialogue state if the task is complete. All dialogue state prompts are provided in Appendix \ref{appendix:prompts}, Figure~\ref{fig:dialogue-state-prompts}. 

\input{tables/mi-strategies}

\noindentparagraph{Motivational Interviewing Chain.}
While our dialogue state chain manages \textit{what} the model should talk about, it offers little guidance on \textit{how} it should say it. 
We selected 11 motivational interviewing strategies from the Motivational Interviewing Skills Code (MISC)~\cite{moyers2003assessing}, filtering out codes corresponding to undesired counselor behavior, merging some codes to reduce overlap, and adapting the examples to the domain of physical activity coaching. The 11 codes are listed in Table \ref{table:mi-strategies}. We use another prompt chain, inspired by~\cite{shaikh2023rehearsal, hsu2023helping, park2023thinking}, to ground the model's behavior in motivational interviewing strategies:
one agent selects an MI strategy conditioned on the current dialogue state and history, another then generates a response following this strategy. 

\noindentparagraph{Tool Use Chain.} 
If the response generation step does not call a tool, we prompt an additional tool call prediction agent to \textit{``determine whether the agent's response should be augmented with the user’s health data.''} If yes, we force the model to generate an output that calls the \texttt{visualize} function.

%% file: tables/mi-strategies.tex
\begin{table*}[tb]
\small
\centering
\begin{tabular*}{\textwidth}{@{\extracolsep{\fill}} l p{6cm} p{7.3cm}}
\toprule
\textbf{Strategy} & \textbf{Description} & \textbf{Example} \\
\midrule
\textsc{Advise with Permission} 
  & Give advice, make a suggestion, or offer a solution or possible 
    action, after gaining permission.
  & ``Would it be all right if I suggested something?'' \\
\midrule
\textsc{Affirm}
  & Say something positive or complimentary to the client.
  & ``You’re a very resourceful person.'' \\
\midrule
\textsc{Facilitate}
  & Simple utterances that function as ``keep going'' acknowledgments.
  & ``Hmm. Tell me more.'' \\
\midrule
\textsc{Filler}
  & Responses not categorizable elsewhere, such as pleasantries.
  & ``Good morning, John.'' \\
\midrule
\textsc{Giving Information}
  & Give information to the client, explain something, educate or provide 
    feedback or disclose personal information.
  & ``Your heart rate was higher during today’s workout.'' \\
\midrule
\textsc{Question}
  & Ask a question in order to gather information, understand, or 
    elicit the client's story.
  & ``How do you feel about that?'' \\
\midrule
\textsc{Raise Concern}
  & Point out a possible problem with a client's goal, plan, or intention.
  & ``I’m worried about your plan to decrease workout days.'' \\
\midrule
\textsc{Reflect}
  & A reflective listening statement in response to a client statement.
  & ``You’re looking for a more relaxed environment.'' \\
\midrule
\textsc{Reframe}
  & Suggest a different meaning for an experience expressed by the client, 
    placing it in a new light.
  & {Client: ``My husband is always nagging me about going to the gym.''
  \newline
  Counselor: ``It sounds like he's concerned about your health.''} \\
\midrule
\textsc{Support}
  & Generally sympathetic, compassionate, or understanding comments.
  & ``That must have been difficult.'' \\
\midrule
\textsc{Structure}
  & Give information about what’s going to happen directly to the client 
    throughout the course of treatment or within a study format.
  & ``What we normally do is start by asking about your physical activity.'' \\
\bottomrule
\end{tabular*}
\caption{\textbf{Motivational interviewing strategies used by \textsf{GPTCoach} in the motivational interviewing prompt chain}. The strategies were adapted from the Motivational Interviewing Skills Code (MISC)~\cite{moyers2003assessing}. The model first picks which of the 11 strategies above to use and then generates a response conditioned on the chosen strategy.}
\label{table:mi-strategies}
\end{table*}

%% file: sections/5_tech_probe_study.tex
\section{Evaluation Study}
\input{tables/probe_demographics}

We evaluate \textsf{GPTCoach} as a technology probe in a lab study with 16 participants. Our study was approved by our university's institutional review board.
As \citet{klasnja2011evaluate} have argued, assessing behavior change in a traditional sense (e.g., via a longitudinal RCT) is often an inappropriate metric for early-stage technologies in HCI research. Instead, as is common in HCI research on systems targeting health behavior change, we focus our evaluation on participants' experiences and \textsf{GPTCoach}'s adherence to our design principles.

\subsection{Participants}
\label{sec:tech-probe-participants}
We recruited 16 participants from the general population using a variety of sources, including university mailing lists, social media advertisements, former research participants, and personal connections. Participant demographics are listed in Table \ref{table:probe-demographics}. Participants were required to own an iPhone (as our system relies on Apple HealthKit) and nine also owned an Apple Watch. All participants were required to be fluent in English. Nearly all participants (15/16) expressed interest in increasing their physical activity in the near future (one participant responded ``I don’t know''), aligning with our study’s focus on behavior change. We discuss the limitations of our recruitment procedure and potential biases in Section \ref{sec:limitations}.

Participants were selected using a screening survey for a balanced sample across several dimensions, including 
age (21–71, Mean: 38.2, SD: 14.8), 
gender (10 female, 6 male), 
exercise stage of change~\cite{marcus1992self} from the transtheoretical model~\cite{prochaska1997transtheoretical} (1 precontemplation, 8 contemplation, 4 action, 3 maintenance), and 
levels of physical activity assessed via the International Physical Activity Questionnaire (IPAQ)~\cite{craig2003international} (5 low, 7 moderate, 4 high). 
The majority of participants (13/16) were employed for wages and varied in their level of education, including associate degrees (2), bachelor’s degrees (6), master’s degrees (7), and one doctorate. 
Two participants self-identified as having a disability.
Most participants (14/16) reported having previously interacted with an AI chatbot. Their knowledge of AI varied, with one participant identifying as a novice, the majority (10/16) having basic knowledge, three participants having intermediate knowledge, and two reporting advanced knowledge.

\subsection{Procedure}
\label{sec:tech-probe-procedure}
Participants interacted with \textsf{GPTCoach} in a one-hour study session, including both in-person (5/16) and remote sessions conducted on Zoom (11/16). We did not encounter notable differences between the two formats and report usability, advice quality, and demographics by format in Appendix \ref{appendix:zoom-vs-inperson}. Prior to the session, participants uploaded three months of historical data from HealthKit using our iOS application (Appendix \ref{appendix:ios}). All participants were informed that a chatbot would have access to their health data in our study's consent form and were reminded again at the beginning of the session.

At the beginning of the session, participants were informed that they would be interacting with a chatbot acting as a health coach and that this conversation would mirror an onboarding conversation in an established health coaching program. They were told to interact with the system as they usually would while thinking aloud. There was no training and no further instructions were provided. 
If questions arose about using the chat interface, they were answered; otherwise, participants were reminded to interact as they normally would.
The session concluded with open-ended questions from the researcher about their overall experience and a post-study survey.

\subsection{Analysis}
Our analysis uses several methods, including survey measures, qualitative coding using thematic analysis, computational analysis of model states and transcripts, and motivational interviewing coding by human experts.

\subsubsection{Survey Measures} 
We asked participants several 5-point Likert scale questions about their experience interacting with \textsf{GPTCoach} and the quality of its advice. In addition, we measured usability using a subset of the Subjective Assessment of Speech Interfaces~\cite{hone2000towards} (the same subset as \citet{mitchell2021automated} used  with two additional questions from the habitability and speed factors). All questions are provided in Appendix \ref{appendix:surveys}.

\subsubsection{Qualitative Coding}
To better understand participants' experiences with \textsf{GPTCoach} and \textsf{GPTCoach}'s adherence to our design principles, we qualitatively coded interviews (including conversation
transcripts, think-alouds, and post-study interviews) using thematic analysis~\cite{braun2012thematic}, using a similar process that was used in the formative studies (Section \ref{sec:formative-analysis}). 

\subsubsection{Dialogue Analysis} We report the overall frequency and temporal progression of \textsf{GPTCoach}'s internal dialogue states and MI strategies (as used in the prompt chains), along with an analysis of tool calls. We also report differences in message length.

\begin{figure*}[h!]
    \centering
    \includegraphics[width=\textwidth]{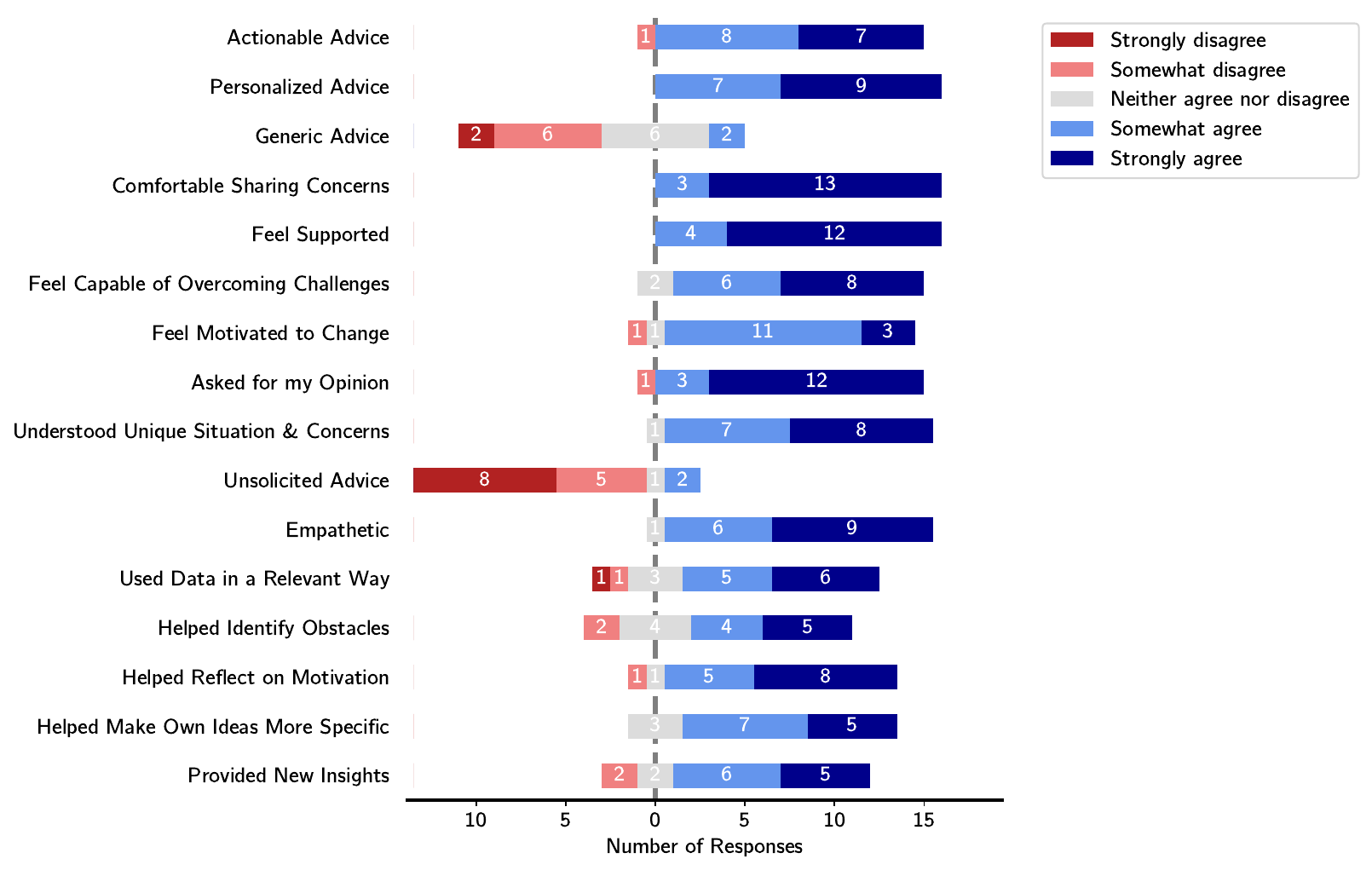}
    \caption{\textbf{Participant Responses to Survey Items on User Experience \& Quality of Advice.} Participants had an overwhelming positive, comfortable, and supportive experience interacting with \textsf{GPTCoach}. The advice they received was personalized, actionable, and not unsolicited. Full questions are provided in Appendix \ref{appendix:ux-questions}.}
    \label{fig:survey-response}
    \Description{Participant Responses to Survey Items on User Experience and Quality of Advice. A horizontal bar chart with color-coded segments representing the level of agreement with various survey items. The bars are segmented by five levels of agreement from strongly disagree (red) to strongly agree (dark blue).}
\end{figure*}

\subsubsection{Motivational Interviewing Coding}
\label{sec:miti-coding}
While each response is generated conditioned on a single MI strategy, the agent frequently makes use of several strategies within a single response. For example, this response was generated conditioned on the \textsc{Question} strategy, but the agent begins the response with \textsc{Giving Information}: \textit{``Starting and maintaining motivation for physical activity can indeed be challenging, and you're not alone in feeling this way. Now, have you ever had any health problems or injuries that have interfered with your ability to exercise?''} Moreover, \textsf{GPTCoach} may or may not implement a given MI strategy in a way that is consistent with MI principles.

To better assess \textsf{GPTCoach}'s consistency with MI, we performed motivational interviewing coding on all participant conversations with \textsf{GPTCoach}. 
As a first step, we performed LLM-based coding following Chiu et al's~\cite{chiu2024computational} prompt-based method for evaluating LLM psychotherapists, using the MISC coding scheme~\cite{moyers2003assessing} (full details provided in Appendix \ref{appendix:misc-coding}).
To more rigorously evaluate our system, we then partnered with an MI agency and hired six trained, experienced coders to perform behavior coding on all conversation transcripts according to the Motivational Interviewing Treatment Integrity (MITI) Code 4~\cite{moyers2016motivational}.
While we used MI strategies from MISC~\cite{moyers2003assessing} in designing our MI chain and initial LLM-based coding, we transitioned to MITI for human coding based on expert advice. 
MITI is a behavioral coding system derived from MISC that was specifically designed to measure MI fidelity with a simpler and more reliable codeset. MITI is commonly used in prior work~\cite{shah2022modeling, perez2016building, perez2017understanding, perez2019makes, welivita2022curating, welivita2023boosting, min-etal-2022-pair} and is used by the agency to evaluate human therapists. Each transcript was independently coded three times, with inter-rater reliability indicated by ICC scores ranging from $0.44$ to $0.98$ and an average (SD) ICC of $0.79$ ($0.17$), reflecting a high overall level of agreement \cite{cicchetti1981developing}. Full details on the annotation process and inter-rater reliability are provided in Appendix \ref{appendix:miti-coding}.

MITI defines 10 utterance-level \textit{behavior codes} that capture various counselor behaviors, which are highly analogous, but not exactly identical to the MISC codes listed in Table \ref{table:mi-strategies}.
Following \citet[p.~289]{miller2023motivational}, we consider the codes \textsc{Persuade} and \textsc{Confront} to be \textit{MI-inconsistent}. We consider \textsc{Giving Information} to be \textit{neutral} and the remaining 7 codes to be \textit{MI-consistent}. We report the frequency of each behavior code as well as the overall frequency of MI-consistent, neutral, and inconsistent codes. Moreover, we report a summary of qualitative feedback provided by the trained coders.
We refer to the MISC stragies (Table \ref{table:mi-strategies}) used in the \textsc{GPTCoach}'s motivational interviewing chain as \textit{internal} MI strategies and the MITI behavior codes annotated by human experts (Figure \ref{fig:external-misc-strategies}; Table \ref{tab:miti-code-table}) as \textit{external} MI strategies.

\subsubsection{Counterfactual Analysis.}
\label{sec:counterfactual}
We perform a counterfactual analysis to compare our model's behavior to GPT-4 with vanilla prompting.
We condition both models on the first five turns of each participant's conversation, covering program introduction, participants sharing their name and age, and ending with the agent asking for questions or concerns. For each of the 16 participants, we simulate 10 responses representing different barriers to physical activity based on coaching materials from collaborators in behavioral medicine. For each of the 160 simulated histories, we generate outputs using \textsf{GPTCoach} and using GPT-4 with only the system prompt and all prompt chains removed (which we refer to as ``vanilla GPT''). Subsequently, we evaluate all responses for MI consistency by having a single coder perform MITI behavior coding on each of the 320 total messages. The coder was the most experienced coder in our pool and was blind to condition. Full details on the counterfactual analysis along with illustrative examples are provided in Appendix~\ref{appendix:counterfactual}.

%% file: tables/probe_demographics.tex
\begin{table*}[tb]
\small
\centering
\begin{tabular*}{\textwidth}{@{\extracolsep{\fill}} l p{12cm}}
\toprule
\textbf{Age} & Mean: 38.2, Median: 32.5, SD: 14.8, Min: 21, Max: 71 \\
\midrule
\textbf{Gender} & Female: 10, Male: 6 \\
\midrule
\textbf{Race/Ethnicity} & White: 10, Hispanic or Latino: 2, Southeast Asian: 2, 
African-American or Black: 1, East Asian: 1, Middle Eastern: 1, South Asian: 1 \\
\midrule
\textbf{Education} & Associate: 2, Bachelor's: 6, Master's: 7, Doctorate: 1 \\
\midrule
\textbf{Stage of Change} & Precontemplation: 1, Contemplation: 8, Action: 4, Maintenance: 3 \\
\midrule
\textbf{Level of Activity (IPAQ)} & Low: 5, Moderate: 7, High: 4 \\
\midrule
\textbf{AI Knowledge} & Novice: 1, Basic: 10, Intermediate: 3, Advanced: 2 \\
\bottomrule
\end{tabular*}
\caption{\textbf{Summary of participant demographics in the technology probe evaluation study} $(N=16)$.}
\label{table:probe-demographics}
\end{table*}

%% file: sections/6_results.tex
\section{Results}
In this section, we report on the result of our survey measures, qualitative analysis, computational analysis of conversation transcripts, and MI coding evaluation.

\subsection{Survey Measures}
\subsubsection{User Experience \& Quality of Advice.} 
\label{sec:ux-and-advice}
Figure \ref{fig:survey-response} illustrates that participants had an overwhelmingly positive experience with \textsf{GPTCoach}.
On average, they felt supported (4.8/5; average response from \textit{1: Strongly disagree} to \textit{5: Strongly agree}), comfortable sharing concerns (4.8/5), capable of overcoming challenge (4.4/5), and motivated to change (4.0/5). The advice provided was rated as personalized (4.6/5), actionable (4.3/5), not unsolicited (1.8/5), and (to a lesser extent) not generic (2.5/5) The chatbot was empathetic (4.5/5), and helped them reflect on motivation (4.3/5). To a lesser extent, the chatbot used data in a relevant way (3.9/5), helped identify obstacles (3.8/5), helped make their own ideas more specific (3.9/5), and provided new insights (3.7/5). 

\subsubsection{Usability.} 
\label{sec:usability}
Among the subset of 12 questions selected from SASSI~\cite{hone2000towards} (see Appendix \ref{appendix:sassi}), our usability evaluations yielded an aggregated score of 49.4/60 (82.4\%). 
The questions are grouped into six factors, each representing a different aspect of usability. Each factor includes a different number of subquestions, scored on a 5-point Likert scale, resulting in varying maximum scores (5, 10, or 15).
Scores by factor (with reverse scoring; higher is better), were 7.7/10 for \textsc{Response Accuracy} (whether the system is accurate and does what the user expects), 14.1/15 for \textsc{Likeability} (whether the system is useful, the system is friendly, and it is clear how to send messages), 9.1/10 for \textsc{Cognitive Demand} (whether they felt confident or tense using the system), 7.6/10 for \textsc{Annoyance} (whether the interaction was repetitive or boring), 7.3/10 for \textsc{Habitability} (whether they always knew what to say to the system and knew what the system was doing), and 3.7/5 for \textsc{Speed} (whether the interaction was fast). This indicates that the system is likable and makes participants feel confident, with room for improvement in hability and speed.

\subsection{Qualitative Coding}
Next, we describe the results of our qualitative coding of participant interviews, reporting on \textsf{GPTCoach}'s ability to adhere to each of our three design principles.
\subsubsection{DP-1: Facilitation \& Non-Prescriptiveness}
We found that participants readily recognized facilitative and non-prescriptive qualities in their interaction with the system. When describing their overall experience with the chatbot, P10 said \textit{``it sort of met me where I was at. [...] It first asked about sort of some contextual things before prescribing anything,''} while P5 shared \textit{``I like how the system says, can you share? You know, it's not commanding.''}
Many of the participants recognized that a non-prescriptive approach ensures that advice is appropriate, such as P11:
\textit{``if you don't know what I do for work, or if you don't know what my daily schedule looks like and what things make me happy and what things I want to avoid, then it's like, you can't actually curate a good plan for me.''}
Many of the same participants appreciated having intentional time to reflect, such as
P5 sharing that \textit{``It made me really think about exercise and how positive it can be [...] When it brought up, who do you do this for? You know, what motivates you? It really, like, touched my heart a little bit.''}

However, many participants were not accustomed to a computer adopting a facilitative role. P4 shared \textit{``You can tell it's powered by a lot of positive recommendations and positive language and inclusive language. I think it is probably the most questions that I have ever been asked by a chatbot.''} On the other hand, P2 felt like the chatbot was asking too many questions: \textit{``it seems a little rapid fire, maybe just a lot of stuff. Yeah, so it's a little overwhelming.''} This was likely impacted by GPT-4's tendency towards verbosity, where it would sometimes ask multiple questions in one message. 
Several participants simultaneously expressed that they appreciated the facilitative tone, while also wanting more prescriptive advice.
P1 addressed this tension directly, sharing that \textit{``when I face obstacles, and this is going to be the hard part, because here I am asking for someone to be like, `Hey, you didn't do this,' like a taskmaster, but then, you know, in the moment, you're already feeling like, oh, I didn't do enough, [...] in that case, you would want this kindness.''}

\begin{figure*}[t!]
    \centering
    \includegraphics[width=0.9\textwidth]{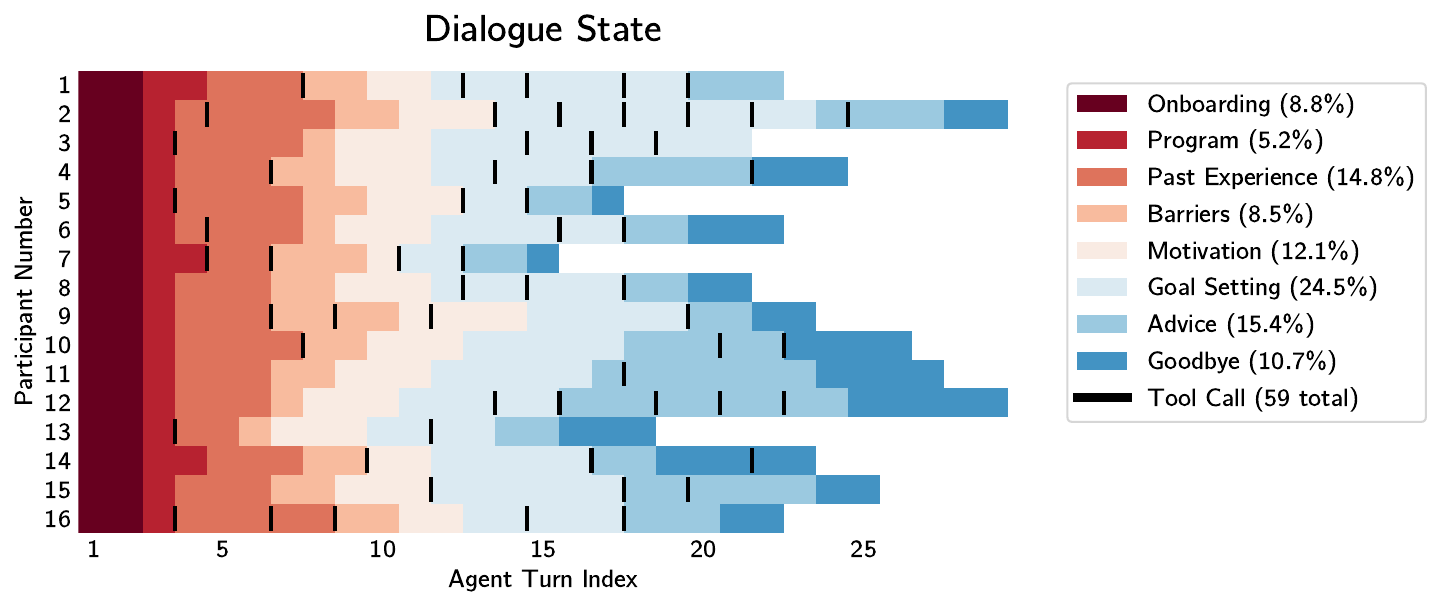}
    \caption{\textbf{Progression of \textsf{GPTCoach}'s Dialogue States By Turn Index.} We find that \textsf{GPTCoach} adaptively allocates more conversational turns for gathering information about past experiences, barriers, motivation. \textsf{GPTCoach} allocates the most turns for the goal-setting state. Tools calls are appropriately called during past experience, goal setting, and advice.}
    \label{fig:dialogue-state-progression}
    \Description{Progression of GPTCoach’s Dialogue States. This figure presents a heatmap showing the progression of dialogue states across 16 participants, with the Y-axis representing participant number (1-16) and the X-axis representing the agent turn index (1-25). The colors in the heatmap indicate the different dialogue states GPTCoach transitions through during the conversation. Tool calls are displayed as black dashes on top of the heatmap.}
\end{figure*}

\begin{figure*}
    \centering
    \includegraphics[width=0.9\textwidth]{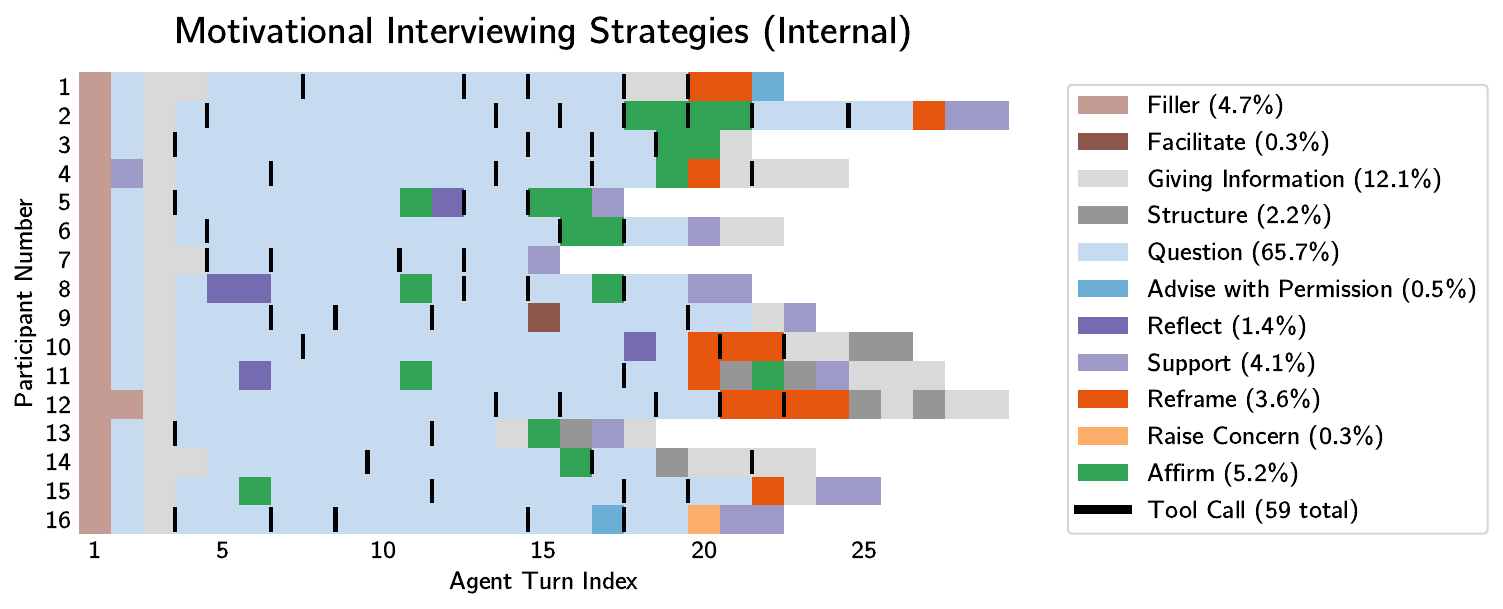}
    \caption{\textbf{Progression of \textsf{GPTCoach}'s Internal MI Strategies By Turn Index.} We find that most of the conversation is spent asking questions and that \textsc{Question}, \textsc{Reflect}, and \textsc{Affirm} precede 
    \textsc{Advise with Permission} and \textsc{Giving Information}. Questions and reflections preceding advice is more aligned with high-quality counselor behavior~\cite{perez2019makes}.
    }
    \label{fig:internal-strategies}
    \Description{Progression of GPTCoach’s Internal Motivational Interviewing Strategies. This figure presents a heatmap showing the progression of internal MI strategies across 16 participants, with the Y-axis representing participant number (1-16) and the X-axis representing the agent turn index (1-25). Each block in the heatmap is color-coded to represent one of the MI strategies used by GPTCoach. Tool calls are displayed as black dashes on top of the heatmap.}
\end{figure*}

\subsubsection{DP-2: Tailoring Advice and Integrating Context}
\label{sec:dp2-tailoring}
Participants overwhelmingly expressed that the system accurately understood their preferences, constraints, and concerns, drawing attention to the personalized nature of its advice. For example, P7 shared that 
\textit{``I really liked that it was accurate, that it was like my personal thing and not just abstract pictures before and after from the Internet of people who are not related,''}
while P13 expressed that \textit{``it felt like it was responding directly to the information I gave it, which is good because I think that's not always the case with the chatbots.''}
Participants built a mental model of the system's capabilities over time, learning that the system could take their context into account through explicit acknowledgments, such as P12 sharing \textit{``There was acknowledgment, [...] with my shoulder, acknowledging previous injury.''} 
These acknowledgments were particularly important as part of building trust in the system. For instance, P3 remarked early on that \textit{``usually systems don't take [back pain] into account. so, like, I already don't have trust,''}
later building more trust in the system as it appropriately acknowledged their concerns.
P7 initially was not sure if the chatbot would be able to understand their lack of motivation, noting that
\textit{``it's asking from the ideal world where robots don't have laziness, they have answers right away and they [are] always ready to work.''}

In comparing our system to existing tools, participants highlighted its ability to capture additional, important context beyond sensor data. For instance, P10 shared that \textit{``I like it just because I think of like, exercise and physical activity is not really like much of like a numbers game. [...] There's a lot more factors,''} while P12 noted that \textit{``it was the kind of the personal touches [...] Different things that were not part of the Apple Fitness questions.''} In reflecting on prior experiences with self-tracking, P4 shared that \textit{``I've tried to use, like, MyFitnessPal from a food tracking standpoint. [...] While I can use the data, sometimes the emotional side of my brain gets more over than the logical side. So I would say, in comparison, I think this was easy.''}

While \textsf{GPTCoach} consistently provided personalized and non-prescriptive guidance, participant impressions of advice specificity were more mixed. Some participants expressed a desire for more specific advice at the end of the conversation, particularly after having shared so much information about themselves. For example, 
P12 shared that they wanted \textit{``maybe more detail, you know, kind of digging into some of my exercise data and maybe some of my feelings on how things are going.''} 
Our chatbot 
would not always probe further for specific days and times, or which specific routines to do in each session. 
Several participants explicitly prompted the chatbot to be more specific, and many were satisfied with the resulting output but wanted the chatbot to have been more proactive.
At the same time, other participants, particularly those who were older and/or less physically active, were highly satisfied with the higher-level advice they received. For example, P4 noted, \textit{“When I see it written out like this, there's no reason I shouldn't do it.”} P2 found the advice not only specific, but even more actionable than that of human coaches, stating that \textit{“I think a lot of the conversations that I had with the [human] health coach were maybe not as specific [...] and that was one thing I kind of didn't like about and why I didn't continue that health coaching.”} These mixed results suggest that participants have differing needs and expectations regarding the level of specificity in advice. While some participants appreciated granular details and structured plans, others preferred more flexible, high-level guidance.
Crucially, none of the participants who wanted more specific advice expressed receiving any advice that was not useful or accurate.

Participants generally appreciated and were often impressed by the system's use of data. 
In its best moments, the system could fetch data at appropriate times and weave it into the conversation in ways that empowered the client. For example, in response to a participant sharing that they've been \textit{``feeling down/depressed if I don't exercise,''} the system responded with \textit{``Thank you for sharing that, [name]. It takes courage to discuss these frustrations, and I appreciate your openness. From your previous data, it seems you've been engaging in quite a diverse and healthy mix of activities in the past month. Look, you've managed a total of 35 workouts with varying duration! That's a wonderful achievement.''}
Showing interactive visualizations were also prompts for reflection, such as P12 remarking \textit{``Apparently on the weekends, I don't do much. I sit on the couch and watch TV,''}
or
P14 appreciated the ability to ask questions about data, sharing that \textit{``I cannot ask any questions about the data that Apple Watch is, like, collecting, but I can ask questions from the chatbot.''}

However, in other moments, the system's use of data was more variable. Sometimes, the system would display a chart without acknowledging it in a follow-up messsage, leading to confusion.
Other times, data discrepancies could sidetrack the conversation and break trust in the system, further complicated by the fact that the system tended to treat data as fact without asking the participant to confirm its accuracy or relevance. Interestingly, P8 told the chatbot that they did not use their phone to track steps, after which the system ignored their step count data. 
Moreover, several participants expected more of the system's data analysis capabilities than was currently supported by our tool calls. These expectations may have been shaped by general impressions of computers' advantages over humans in data analysis, e.g., \textit{``the real person, like therapist, they don't have time to read my data [...] it's a program that can read all of this like multiple data and the real person can't'' (P7).} Some participants wanted more granular analysis of existing data, 
while others wanted additional data sources from other wearables not supported by our system.

\subsubsection{DP-3: Supportive \& Non-Judgmental Tone}
Our strongest findings came from participants' impressions of the system's \textbf{positive, supportive, and empowering tone}. Participants felt comfortable, supported, and safe when speaking with the chatbot:
\textit{``I think the way the system answered, it didn't make me feel like I was being judged.'' (P2)}; 
\textit{``I think I've noticed that I feel some sort of, like, psychological safety with chatting with the chat bot.'' (P3)}.
Many compared the experience to interacting with a human coach, such as P6 noting that  \textit{``I had a really positive experience. [...] it felt like I was, like, chatting with a human, honestly, or like a coach'' (P6).}
Some participants shared that they preferred this interaction to prior interactions with humans.
P7 told us that they felt more safe, noting that \textit{``and here it's like, no gender, no body shape, nothing. [...] Yeah, comfortable, less judgmental. With the real person, you're always comparing.''}
P5 shared that \textit{``I had a doctor who was, now, I was probably only, like, 20 pounds overweight at the time, and she told me I was morbidly obese, and she told me I needed to exercise. Like, I mean, it was, like, a ridiculous amount. [...] and I stopped going to her because I just felt so threatened.''} Later, P5 shared that our system \textit{``doesn't feel threatening at all.''} 
For some, the positive tone was too much, such as P10 telling us \textit{``It is really good to be affirmed. But we've been affirmed quite a few times.''}
For others, the conversation felt decidedly neutral,  
\textit{``It did not emit any emotion out of me either way. [...] I was just having a conversation with a computer program as far as I can tell'' (P12).} Most importantly, none of the participants reported a negative or judgmental experience interacting with the chatbot.

\subsection{Dialogue Analysis}
Next, we report on a computational analysis of participants' conversations with \textsf{GPTCoach}. We analyze \textsf{GPTCoach}'s internal dialogue states, MI strategies, and tool calls, as well as a comparison of \textsf{GPTCoach}'s message length.

\subsubsection{Dialogue States: How does \textsf{GPTCoach} structure conversations?} As shown in Figure \ref{fig:dialogue-state-progression}, we find that dialogue states follow a similar progression across all participants, with individual variation in the number of turns allocated to each state. The chatbot spends no more than 1-2 turns during onboarding or describing the program, but adaptively allocates more turns when gathering information about past experiences, barriers, or motivation. Overall, we notice that 28.5\% of all messages were spent in goal setting, with an average of 5.6 agent responses during this state. This was followed by 15.4\% of the total conversations in the advice state (an average of 3.5 messages) and 14.8\% of the conversations in the past experiences state (an average of 3.4 agent responses). The messages between past experiences and motivation states account for a total of 35.4\% of the total conversations. 
This behavior is more aligned with high-quality counselor behavior \cite{perez2019makes}, while low-quality counselors focus on problem-solving before collecting sufficient context from the client \cite{cochran2015heart, chiu2024computational}. 

We also see that the chatbot almost exclusively performs tool calls between past experience and advice dialogue states. Of the 59 total tool calls across all users, 35.6\% of the tool calls were in the goal setting state, 30.5\% in the advice state, and 13.6\% in the past experience state. This indicates that tool calls are mostly called at appropriate conversation states: collecting information about users' past experiences, grounding goal setting in concrete historical baselines, or giving tailored advice.

\begin{figure}
    \centering
    \includegraphics[width=\columnwidth]{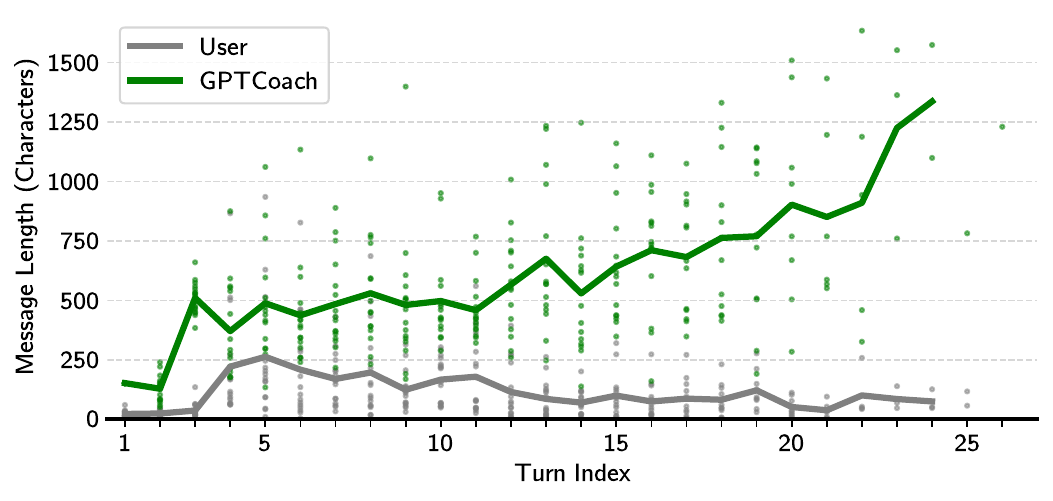}
    \caption{\textbf{Comparing the Length of \textsf{GPTCoach} and User Responses.} We find that \textsf{GPTCoach}'s responses are 2-3+ times as long as the user's responses, suggesting that the system could be improved with shorter responses.}
    \label{fig:turn-length}
    \Description{Comparing the length of GPTCoach and user responses. A line graph comparing message length between GPTCoach and users across conversation turns. The X-axis represents the turn index and the Y-axis represents the message length in characters, ranging from 0 to 1500. The line shows that GPTCoach’s responses are consistently longer than the user’s, with GPTCoach responses fluctuating between 250 and 1500 characters and increasing in length towards the later turns of the conversation. In contrast, the line representing user responses remains much shorter throughout the interaction, mostly ranging between 100 and 300 characters and decreasing as the conversation progresses.}
\end{figure}

\begin{figure*}[t!]
    \centering
    \includegraphics[width=\textwidth]{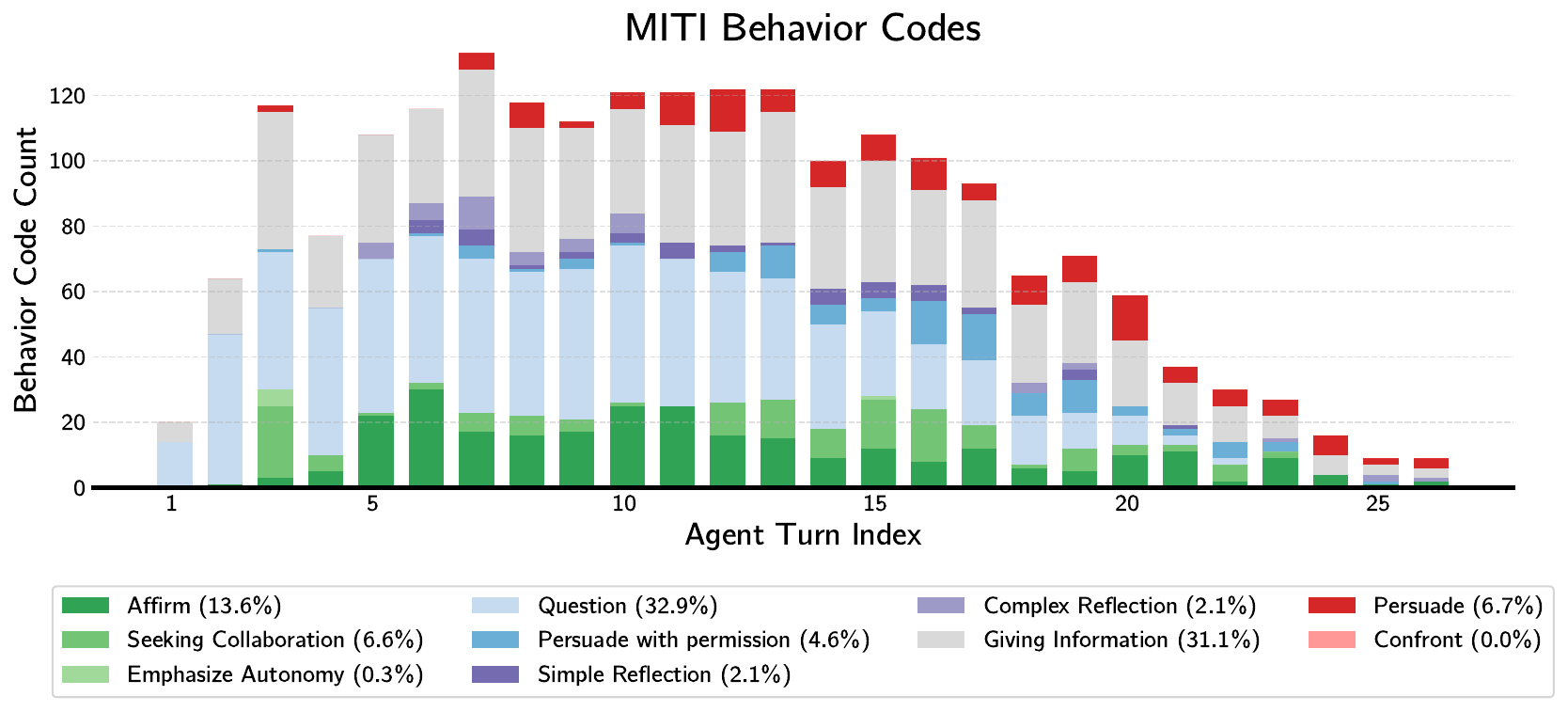} 
    \caption{
    \textbf{Distribution of \textsf{GPTCoach}'s External MITI Behavior Codes by Turn Index.} We find that \textsf{GPTCoach} most frequently engages in \textsc{Affirm}, \textsc{Question}, and \textsc{Giving Information}, which is aligned with MI. However, \textsf{GPTCoach} still engages in \textsc{Advise without Permission}, which is not aligned with MI. MITI behavior code defintions are provided in Appendix \ref{appendix:tech-probe} Table \ref{tab:miti-code-table}.}
    \label{fig:external-strategies}
    \Description{Distribution of GPTCoach’s External MITI Behavior Codes by Turn Index. A stacked bar chart shows the distribution of external MITI behavior codes, as coded by human experts, employed by GPTCoach over the course of a conversation, with the Y-axis representing the strategy count (ranging from 0-120) and the X-axis representing the agent turn index (1-25). Each color in the stacked bars corresponds to a specific MI strategy. This figure highlights that GPTCoach frequently uses strategies such as Affirm, Question, and Giving Information, which align with MI principles. However, it also occasionally uses Persuade, which is less aligned with these principles.}
\end{figure*}

\begin{figure*}[t!]
    \centering
    \includegraphics[width=\textwidth]{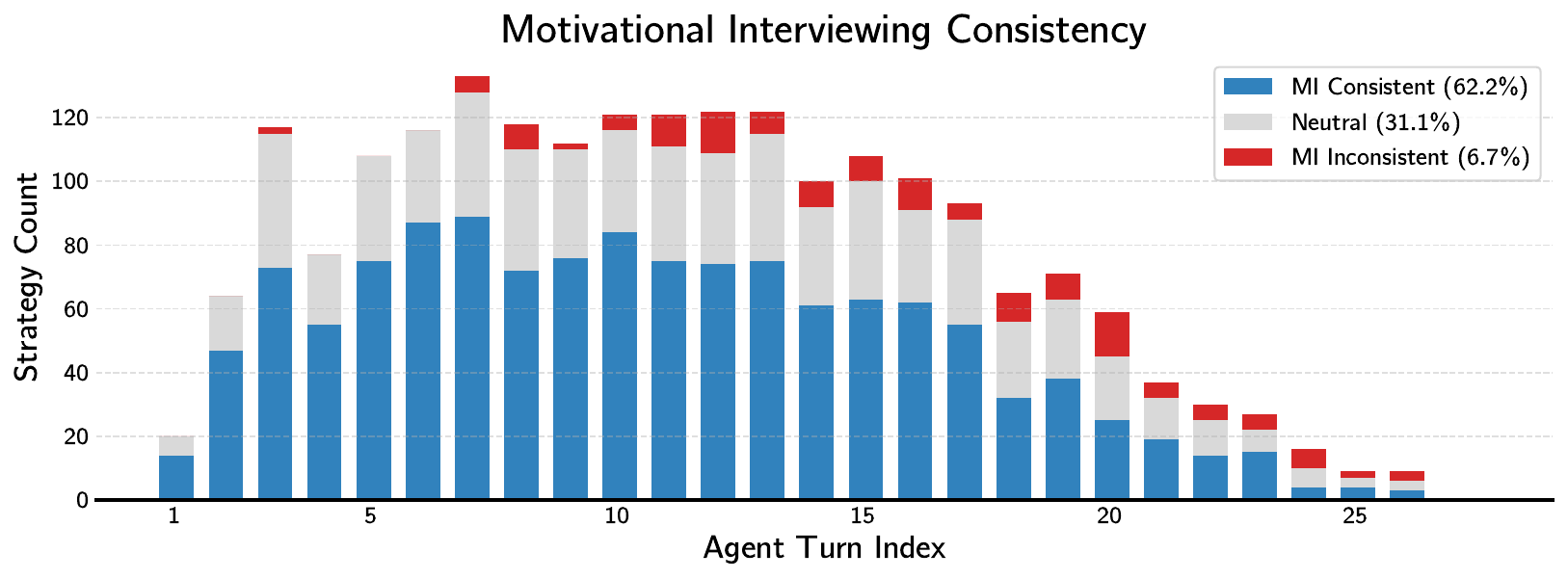}
    \caption{\textbf{\textsf{GPTCoach}'s MI Consistency by Turn Index.} This figure aggregates MITI behavior codes from Fig. \ref{fig:external-strategies} into MI-consistent, MI-inconsistent, and neutral codes~\cite{miller2023motivational}. We find that only 6.7\% of \textsf{GPTCoach}'s total codes are inconsistent with MI principles.}
    \label{fig:mi-consistency}
    \Description{GPTCoach’s MI Consistency by Turn Index. A stacked bar chart shows the proportion of MI-consistent, neutral, and MI-inconsistent MITI behavior codes over the course of a conversation, with the Y-axis representing the strategy count (ranging from 0-120) and the X-axis representing the agent turn index (1-25). MI-consistent strategies are the most frequent, especially during the early and middle turns. MI-inconsistent strategies, although present, account for a smaller portion, peaking occasionally throughout the conversation.}
\end{figure*}

\subsubsection{Response Length: How long are \textsf{GPTCoach}'s responses?} 
\label{sec:response-length}
We analyze the chatbot's utterance length and compare them to the user's utterance length. As seen in Figure \ref{fig:turn-length}, agent responses are 2-3 times the length of user responses, which increases as the conversation advances. Despite prompts to keep responses brief, the chatbot struggled to limit its verbosity, which is indicative of low-quality therapist behavior \cite{perez2019makes} and points to areas for future work.

\subsubsection{MI Strategies: How does the chatbot use motivational interviewing strategies?} 
\label{sec:mi-adherence}
We first look at the distribution of \textit{internal} MISC strategies that the chatbot uses to ground its responses in the MI chain. 
As shown in Figure \ref{fig:internal-strategies}, most of \textsf{GPTCoach}'s responses are \textsc{Questions}, covering 65.7\% of the entire conversation across all participants. This was followed by 12.1\% of chatbot responses grounded in \textsc{Giving Information} and 5.2\% in \textsc{Affirm}. 
We find that \textsc{Question}, \textsc{Reflect}, and \textsc{Affirm} generally precede \textsc{Advise with Permission} and \textsc{Giving Information}, which is more aligned with high-quality counselor behavior~\cite{perez2019makes}.

\subsection{Motivational Interviewing Coding: How well does the chatbot adhere to Motivational Interviewing?}
\label{sec:mi-consistency}
We next examine the distribution of \textit{external} MITI behavior codes present in \textsf{GPTCoach}'s utterances, as coded by human MI experts. 
On average, each agent response contained 2.3 different external strategies.
As shown in Figure \ref{fig:external-strategies}, the most frequent strategies are \textsc{Question} (32.9\% of all codes), \textsc{Giving Information} (31.1\%), and \textsc{Affirm} (13.6\%).
However, despite our effort to discourage unsolicited advice, we also encounter undesirable, MI-inconsistent codes like \textsc{Persuade} across 6.7\% of all codes.
MI-inconsistent codes tend to occur later in the conversation, as the agent shifts from asking questions to providing information and giving advice, some of which can be unsolicited.
Aggregating these strategies into MI consistent (\textsc{Affirm, Emphasize Autonomy, Persuade With Permission, Question, Seeking Collaboration, Simple Reflection, Complex Reflection}), inconsistent (\textsc{Persuade, Confront}), and neutral codes (\textsc{Giving Information}), we see in Figure \ref{fig:mi-consistency} that only 6.7\% of total codes corresponding to our chatbot's responses are inconsistent with MI principles. 

\begin{figure*}
    \centering
    \begin{subfigure}{0.48\textwidth}
      \centering
        \includegraphics[width=\textwidth]{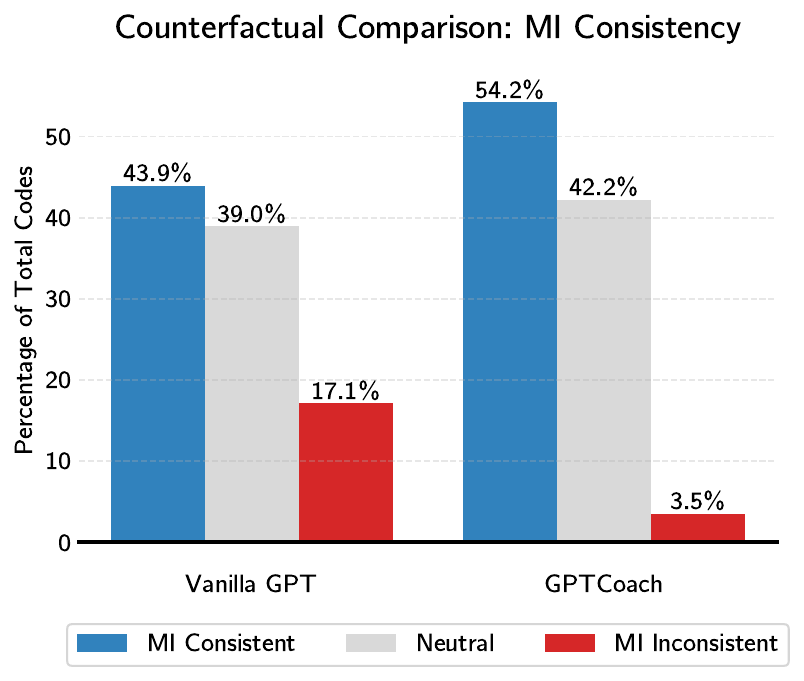}
        \caption{
        \textbf{Comparing the Overall Frequency of MI-constistent, MI-inconsistent, and Neutral Strategies in Our Counterfactual Analysis.} Percentages represent the proportion of strategies that are MI-consistent/MI-inconsistent/neutral strategies among all MI strategy codes for a given agent. We find that \textsf{GPTCoach}'s responses contain more MI-consistent and fewer MI-inconsistent responses than vanilla GPT. 
        \newline}
        \label{fig:counterfactual-mi-consistency}
    \end{subfigure}
    \hfill
    \begin{subfigure}{0.48\textwidth}
      \centering
        \includegraphics[width=\textwidth]{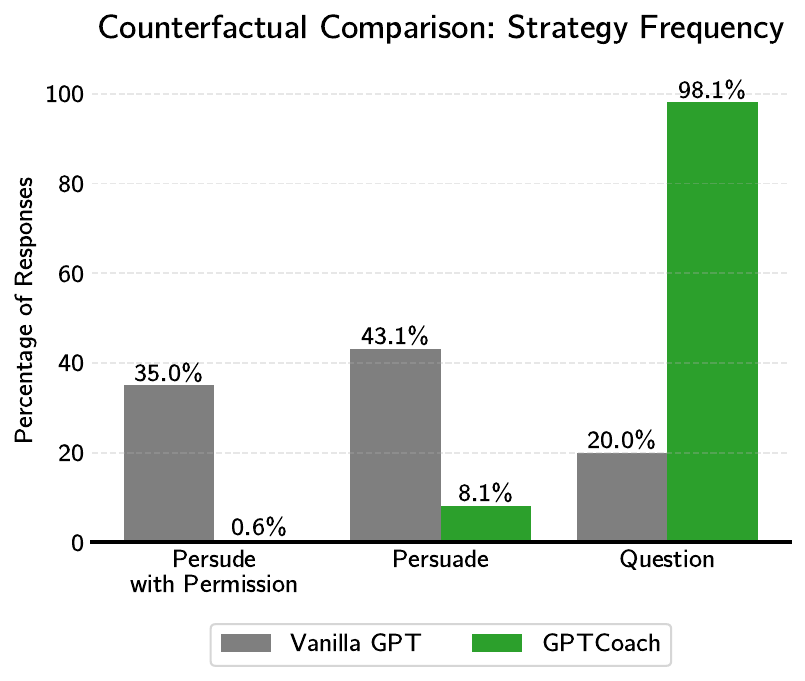}
        \caption{
        \textbf{Comparing the Frequency of MI Strategies by Response in Our Counterfactual Analysis.} These three strategies showed the largest differences in frequency across the two agents. Percentages represent the proportion messages that contain a given MI strategy across all of a given agent's counterfactual responses. We find that vanilla GPT uses persuasion (i.e., advice or solutions) far more frequently (and often without permission), while \textsf{GPTCoach} gives much less advice and asks far more questions.
        }
        \label{fig:counterfactual-strategy-frequency}
    \end{subfigure}
    \caption{\textbf{Counterfactual Comparison Between \textsf{GPTCoach} and Vanilla GPT-4.} The vanilla GPT condition includes the same system prompt, but with all prompt chains ablated. Illustrative examples are provided in Appendix \ref{appendix:illustrative-examples}.}
    \label{fig:counterfactual}
    \Description{Counterfactual comparison between GPTCoach and vanilla GPT-4. This figure consists of two bar charts. In panel A (left), the bar chart compares how the two models adhere to Motivational Interviewing (MI) principles. The Y-axis shows the percentage of total codes and the X-axis shows the two models being compared: Vanilla GPT and GPTCoach. The individual bars show the percentage of MI-consistent, neutral, and MI-inconsistent strategies for each agent. This comparison shows that GPTCoach has more MI-consistent responses and fewer MI-inconsistent responses compared to Vanilla GPT. In panel B (right), the bar chart compares the frequency of three MI strategies in each model’s responses. The Y-axis represents the percentage of responses that contain a particular strategy, while the X-axis shows three strategies: Advise with Permission, Advise without Permission, and Open Question. The individual bars show the percentage of responses containing the respective strategy for each model. This comparison demonstrates that Vanilla GPT frequently provides advice without permission, while GPTCoach uses Open Questions much more frequently, in line with MI principles.}
\end{figure*}

\subsubsection{Qualitative Feedback from MITI Coders}
\label{sec:miti-coding-qualitative}
In addition to behavior coding, we also received qualitative feedback from the coding team through video-call discussions and a final feedback document. 
The coders noted that \textsf{GPTCoach} demonstrated a respectful and non-confrontational approach, emphasizing client strengths and autonomy rather than focusing on their difficulties, which is in alignment with MI principles and fosters a more open relationship. 
However, they also highlighted several areas for improvement. One experienced coder remarked that \textsf{GPTCoach}'s MITI scores were similar to those of human counselors in training, who grasp basic building blocks but have yet to master MI.
Specifically, \textsf{GPTCoach} tended to offer motivational statements or advice in response to ``change talk'' (client statements in favor of behavior change) rather than leveraging these moments to further elicit client motivation and explore the benefits of change. 
While some good reflections were present, they were underutilized, with \textsf{GPTCoach} often defaulting to providing information or asking questions instead of letting the client ``hear their own thoughts and emotions in the voice of another.''
Skilled MI practitioners would use more reflections (particularly complex reflections), give less information, and avoid persuasion or advice altogether. Lastly, as confirmed by our analyses (Section \ref{sec:response-length}), \textsf{GPTCoach} tended to speak more than the client, which is contrary to MI's client-centric approach.

\subsection{Counterfactual Analysis: How does \textsf{GPTCoach} compare to vanilla GPT-4?}
\label{sec:counterfactual-results}
We report on our counterfactual analysis comparing \textsf{GPTCoach} to GPT-4 using only the system prompt, with all prompt chains removed. 
As shown in Figure \ref{fig:counterfactual-mi-consistency}, 54.2\% of \textsf{GPTCoach}'s codes are MI-consistent, compared to 43.9\% of vanilla GPT-4's codes. Meanwhile, 3.5\% of \textsf{GPTCoach}'s codes are MI-inconsistent, compared to 17.1\% of vanilla GPT-4's codes. We found that the top three codes driving this difference in MI-consistency are \textsc{Persuade with Permission}, \textsc{Persuade} (without permission), and \textsc{ Question}, matching our intuitions. In Figure \ref{fig:counterfactual-strategy-frequency}, we plot the percentage of messages that contain each of these codes. We find that GPT-4 contains \textsc{Persuade} in 43.1\% of its responses, compared to 8.1\% of \textsf{GPTCoach}'s responses. Meanwhile, \textsf{GPTCoach} contains \textsc{Question} in 98.1\% of its responses, compared to 20.0\% of vanilla GPT-4's. Lastly, we also find that vanilla GPT-4 contains \textsc{Persuade with Permission} in 35.0\% of its responses, compared to 0.6\% of \textsf{GPTCoach}'s replies.
Vanilla GPT-4 has a stronger bias towards persuasion and giving advice, frequently without permission, while \textsf{GPTCoach} is more inclined to ask  questions before jumping to giving advice. 

\subsection{LLM-Based Motivational Interviewing Coding}
We briefly report on our LLM-based coding results, which leveraged Chiu et al.'s~\cite{chiu2024computational} prompt-based method for evaluating LLM psychotherapists and used the MISC coding scheme~\cite{moyers2003assessing}. 
While there are differences between the MITI and MISC codeset, our LLM-based coding results are highly analogous to our human coding results, indicating that \textsf{GPTCoach} is largely consistent with MI, with room for improvement in advanced MI skills. 
We find that 54.1\% of codes are MI-consistent, 30.0\% are neutral, and 15.9\% are MI-inconsistent. The most commonly used strategies are \textsc{Affirm} (18.8\%), \textsc{Open Question} (18.0\%), and \textsc{Giving Information} (16.4\%). In the counterfactual analysis using LLM-based coding, 62.6\% of \textsf{GPTCoach}’s codes are MI-consistent, compared to 55.0\% of vanilla GPT-4’s codes. Meanwhile, 14.5\% of \textsf{GPTCoach}’ codes are MI-inconsistent, compared to 22.9\% of vanilla GPT-4’s codes. Given the growing interest in LLM-based MI coding in HCI and NLP~\cite{min-etal-2022-pair, perez2016building, perez2017understanding, perez2019makes, shah2022modeling, welivita2022curating, welivita2023boosting}, these findings suggest that prompt-based methods show promise for automated MI coding.
Full details on LLM-based coding are provided in Appendix \ref{appendix:misc-coding}.

%% file: sections/7_limitations.tex
\section{Limitations}
\label{sec:limitations}
A primary limitation of our current study is that we did not examine long term use of \textsf{GPTCoach} and thus could not evaluate its ability to sustain physical activity behavior change. As discussed in Section \ref{sec:active-choices}, we believe that many of our technical innovations and learnings will transfer to the multi-session setting.
However, we emphasize that we opted for a single-session study design in this formative work such that researchers could supervise all interactions, thereby mitigating the risk of harm from unpredictable outputs. 
We believe that additional work is needed to ensure that \textsf{GPTCoach} is safe to use without researcher oversight and we discuss ethical considerations and possible risk mitigation strategies in Section \ref{sec:risks}.

We encountered several challenges with current models that may have limited our chatbot's efficacy, such as difficulties adhering to the coaching program, effectively employing motivational interviewing strategies, and using tool calls at the appropriate times. In Section \ref{sec:llm-training-eval-implications}, we discuss how this behavior may arise from instruction-tuning and reinforcement learning from human feedback (RLHF)~\cite{ouyang2022training}.
We were able to address many of these challenges through our prompt chaining architecture, which greatly improved MI-consistency compared to a baseline GPT-4. However, qualitative feedback from MI experts indicated that the system still fell short of achieving the performance level of skilled human coaches and future work could explore how to further improve adherence to MI. Moreover, our final system was not always consistent in utilizing data proactively.
It is possible that this could be improved with finetuning or with more sophisted data analysis functionality, such as multi-agent architectures or code generation~\cite{merrill2024transforming}.  
Despite these challenges in incorporating data, participants’ ratings of advice quality and personalization were not noticeably affected (see Section~\ref{sec:ux-and-advice} and Figure~\ref{fig:survey-response}). This could suggest that personalization based on qualitative context is more impactful to user experience than data-driven personalization. Alternatively, it may reflect the difficulty of achieving meaningful personalization based on quantitative data, mirroring challenges faced by human coaches in incorporating data~\cite{rutjes2019beyond, rutjes2022influence}.

Lastly, we acknowledge limitations in the demographics of our participant pool. While we aimed for a diverse sample across age, gender, race/ethnicity, physical activity levels, and stages of behavior change—including older adults, individuals from minority groups, non-native English speakers, individuals with chronic health conditions, and neurodivergent individuals—our participant pool reflects some socioeconomic bias. Participants were all based in the US and required to own iPhones, as our system relied on Apple HealthKit. iPhones ownership is associated with higher income and education~\cite{jamalova2019comparative}, and all participants had some post-secondary education. While we believe that LLM coaching can increase access to physical activity support, future work that critically engages with socioeconomic class is needed to ensure the technology meets its potential for benefit~\cite{kouaho2024socioeconomic}.

%% file: sections/8_discussion.tex
\section{Discussion}
\label{sec:discussion}
Our formative interviews and technology probe study demonstrate promising evidence that LLMs can collect rich qualitative context about people's unique life circumstances and use this context to personalize its physical activity support. Moreover, \textsf{GPTCoach}'s conversational flexibility allowed it to acquire this information in a non-judgmental, supportive manner, reflecting many of the positive qualities of human coaching. 
We now discuss our findings, including implications for mobile health interactions beyond conversation, considerations for extending \textsf{GPTCoach} to multi-session coaching, implications for LLM training and evaluation, as well as the ethics and privacy risks of LLM-based health coaching. 

\subsection{Implications for Future Mobile Health Systems}
\label{sec:future-mobile-health-implications}
While conversational interaction proved highly effective for eliciting rich qualitative information, we also believe that LLM agents could make use of qualitative context to facilitate a range of interactions for mobile health applications beyond simple chat interfaces.
For instance, a mixed-initiative~\cite{horvitz1999principles} system might couple a chat interaction with ambient displays~\cite{consolvo2008activity, lane2012bewell, murnane2020designing}, adjusting the display in response to both quantitative sensor data and qualitative information acquired through conversation. 
Such a system might also allow users to set more flexible goals, which could be adapted and renegotiated as users progress or encounter new obstacles~\cite{munson2012exploring}. 
By integrating push notifications and calendar entries, an agent could tailor actitivy plans, schedule reminders, and readjust plans based on natural language input (e.g., \textit{``I can't go on a run this Thursday because I need to pick up my kids''}). 
In fact, many participants suggested several such additions, like personalized reminders, scheduling support, guidance during life changes or injuries, and real-time assistance. 
These interactions could be implemented by seeding a controller LLM agent with qualitative context and augmenting its capabilities through tools (e.g., functions that control UI elements or schedule push notifications).

Participants also requested different personas, such as a regimented ‘trainer’ when they needed an extra push and an empathetic ‘counselor’ when they were feeling down. Such a persona could enhance engagement and adherence through narrative~\cite{murnane2020designing} or be represented as an embodied agent~\cite{bickmore2005establishing, king2020effects} to increase the emotional and relational connection. 
Participants had mixed opinions on whether users should choose the personas or have the agent adapt, raising interesting open questions for design. Moreover, it is unclear whether increased emotional connection with an LLM-based coach should be encouraged, as this might lead to a dependence on the coach, conflicting with client empowerment and self-efficacy.

While our study did not incorporate any nudging or just-in-time adaptive interventions~\cite{nahum2018just}, the qualitative context elicited during an onboarding conversation could also help drive adaptive experimention algorithms. 
For instance, an LLM agent could help the user set a specific and measureable behavior change goal, which could in turn be optimized by an adaptive experimentation algorithm~\cite{mandyam2023adaptive}. Qualitative information about a user's preferences, abilities, and constraints could help warm-start an algorithm with better priors over optimal intervention timing and content.
When new obstacles arise or life circumstances change, LLMs could help detect a distribution shift to better adapt the underlying learning algorithm.

A natural extension of our current system would provide users with continuous access to \textsf{GPTCoach} over multiple coaching sessions. After the onboarding conversation, the Active Choices program consists of several follow-up conversations, scheduled every few weeks. Our prompt chaining architecture could be adapted to conduct follow-up conversations using modified dialogue state prompts. Qualitative feedback elicited in follow-up sessions could be used by the agent to adjust exercise plans, provide strategies for overcoming barriers, make adjustments to the client's goals, and provide encouragement. When coupled with additional features described above (e.g., push notifications, calendar entries, and ambient displays), the agent could also serve as an accountability tool, scheduling reminders for planned workouts or helping clients monitor progress towards their goal. 

\subsection{Implications for LLM Training \& Evaluation}
\label{sec:llm-training-eval-implications}
Instruction-tuning and reinforcement learning from human feedback (RLHF)~\cite{ouyang2022training}
optimize models for single-turn question answering, 
which prior work suggests can bias the model's behavior towards problem-solving and advice-giving~\cite{chiu2024computational, shaikh2023rehearsal}.
While this behavior is sensible for ``helpful and harmless''~\cite{bai2022training, ouyang2022training} assistants, it runs contrary to foundational principles of frameworks like motivational interviewing~\cite{miller2023motivational}---conversations that empower clients to change are not served by problem-solving and advice-giving.
Instruction-tuning also biases the model to call tools only when prompted (e.g., \textit{``visualize last month's step count"}), rather than proactively incorporating relevant data. 

It is possible that steerability towards facilitative behaviors, as well as effective integration of context, will improve with model scale, particularly in light of a growing interest in multi-turn, information seeking objectives~\cite{zhou2024archer, andukuri2024star}. 
With advancements in multimodal learning, future foundation models for sensor data~\cite{abbaspourazad2023large} might jointly encode sensor data and text, enabling multimodal understanding without representing sensor data as text.
However, as argued by \citet{ma2024evaluating} in the domain of LGBTQ+ mental health support, task-specific models may be more effective for handling sensitive subjects rather than repurposing general foundation models. This necessitates large-scale training and evaluation datasets for health behavior change, which are time-consuming to create, but could enable rigorous, evidence-based health coaching models. 
While one could feasibly collect a dataset of existing health coaching conversations, care is needed because human and automated coaching differ significantly and it may not be ideal to mirror human coaching exactly~\cite{miller2023motivational}.
Such datasets could also include rigorous evaluations of ethics and safety risks, which we discuss in the following section.

\subsection{Risks of LLM-Based Health Coaching}
\label{sec:risks}

While LLMs present several promising opportunities for health coaching, these new capabilities also introduce new risks that must be addressed for the technology to realize its potential for benefit. Below, we discuss several of these risks along with possible mitigating strategies. We focus on ethical concerns and risks that are specific to LLM-based health coaching, not general behavior change or mobile health applications. 

\subsubsection{Privacy.} The collection and use of personal health data raises significant privacy implications. 
Regarding technical security measures in our system, 
we used HIPAA-compliant storage with Google Firebase, encryption using HTTPS, and privacy access controls in our iOS app (Appendix \ref{appendix:ios}).
Our system leveraged the OpenAI API, which does not use API data for training and does not retain data after 30 days\footnote{\url{https://openai.com/enterprise-privacy}}.
Although participants consented to sharing their health data, a preferable alternative would be to use self-hosted or on-device LLMs. Current on-device models were not performant enough at the time of our study, but their performance may improve with growing interest in small models and on-device inference~\cite{abdin2024phi, gemma2024, alizadeh2023llm}.

We take a human-centric stance towards privacy, arguing that technical data security measures are necessary but insufficient to ensure privacy.
For example, a user may consent to collect body temperature but be unaware this data can be used to expose pregnancy status.
While our system did provide access controls, it provided little support for understanding the benefits and risks of sharing various data sources. 
In future work, an LLM agent might offer the potential to educate users about privacy decisions. On the other hand, the use of a conversational agent may unintentionally influence users to disclose more information than intended.

\subsubsection{Bias \& Equity} 
LLMs have documented biases that can negatively impact marginalized groups~\cite{blodgett2020language, kumar2022language, weidinger2021ethical}, such as perpetuating harmful stereotypes, performing poorly with certain subgroups, or withholding empathy in response to identity disclosures~\cite{cuadra2024illusion}.
Many of our participants readily shared aspects of their identities or health conditions with our chatbot.
Researchers with training in computing ethics supervised all interactions with \textsf{GPTCoach} and analyzed study transcripts, while participants were asked about negative aspects of their interaction in post-study interviews.
We did not encounter nor did participants report any instances of bias or potentially harmful responses, though we were limited by our small sample size. A study at larger scale and/or without researcher supervision would likely require systematic redteaming~\cite{ganguli2022red}. 
Our study provided insights into factors that such a systematic investigation should account for, including (but not limited to) gender, race and ethnicity, access to resources (e.g., financial barriers or lack of housing), motor impairments and disability, neurodivergence, and cultural differences. 
In addition, while \textsf{GPTCoach} focuses on providing support for physical activity only, physical activity is often entangled with other sensitive topics such as nutrition and diet, weight loss, body image, eating disorders, mental health, substance use, sleep disorders, and medical concerns. Though these topics can be considered out of scope, an LLM agent should be able to acknowledge the concerns, point the client to relevant resources, and gently redirect the conversation back to physical activity without causing harm. \textsf{GPTCoach} exhibited preliminary evidence of being able to appropriately navigate conversations related to weight loss, nutrition, and mental health, but further systematic validation of our system's robustness is warranted.

While no method can guarantee safe and unbiased outputs with certainty, we are encouraged by recent work on redteaming, guard models, and evaluation benchmarks~\cite{mazeika2024harmbench, inan2023llama, chiu2024culturalteaming, zeng2024air, chang2024red}. We believe that many of these methods can also be leveraged for risk mitigation in the domain of health coaching. In addition, we draw particular attention to LLM-based mental health support, which has received considerable attention in the literature and shares many of the same risks as health coaching~\cite{de2023benefits}. 

\subsubsection{Hallucination \& Factual Accuracy.}
Even the most performant language models are known to confidently ``hallucinate'' false information~\cite{maynez2020faithfulness, weidinger2021ethical}. This is of utmost concern when offering health-related advice. While our model was based on an evidence-based coaching program that did include guidelines for giving advice to common barriers, it did not have access to external knowledge bases via retrieval~\cite{shuster-etal-2021-retrieval-augmentation, lewis2020rag}. We encountered two instances of the model justifying its responses with reputable sources (the Center for Disease Control and American College of Sports Medicine) that were not included in its prompts. While both instances were innocuous and happened to contain factually accurate information that matched the source, we find this behavior concerning: not only can guidelines change, but it points to a risk of hallucinating reputable sources for factually inaccurate information. 

One popular mitigation strategy for hallucination is retrieval augmented generation~\cite{shuster-etal-2021-retrieval-augmentation}. For example, \citet{merrill2024transforming} provide their PHIA agent with web search to retrieve relevant health information from reliable sources. 
As with bias, we also believe that a study conducted at larger scale and/or without researcher oversight should include systematic evaluations of factual accuracy and evidence-based advice.
One such example is described by \citet{cosentino2024towards}, who evaluated their PH-LLM model on the National Strength and Conditioning Association's Certified Strength and Conditioning Specialists exam and received a passing score (88\%). Beyond static benchmarks, it may also be necessary to evaluate a model's ability to provide accurate information in the context of a personalized health coaching conversation.

Future work on \textsf{GPTCoach} will be guided by the risk areas identified above, with a focus on implementing effective mitigation strategies for privacy, bias, and factual accuracy, such as enhanced data security measures, auditing and redteaming for bias and factual accuracy, or technical approaches such as finetuning or retrieval-augmented generation. This is of particular importance for studies in which participants interact with an LLM agent without researcher oversight. 

%% file: sections/9_conclusion.tex
\section{Conclusion}
This work draws inspiration from health coaching to explore the potential for personalized, LLM-based physical activity support. 
Through formative interviews with 12 health professionals and 10 potential recipients of health coaching, we identify three key design principles for LLM-based health coaching that emphasize the importance of a non-prescriptive approach, the integration of both qualitative and quantitative context, and a non-judgmental tone. These principles center the client's agency and motivation, establishing an alternative to the question-answering and advice-giving objectives common in prior work and commerical products.
Through the design and evaluation of \textsf{GPTCoach} in a lab study with 16 participants, we demonstrate that LLMs can effectively implement the onboarding conversation of an evidence-based physical activity coaching program. We find promising evidence that participants felt comfortable sharing concerns with and supported by \textsf{GPTCoach}, as well as that \textsf{GPTCoach}'s advice was personalized and actionable. 
By integrating qualitative context from conversational interaction with quantitative data from wearable devices, \textsf{GPTCoach} represents a promising step towards more personalized and effective mobile health interventions.
We discuss our work's implications for future mobile health applications, how \textsf{GPTCoach} might be extended to multi-session coaching, implications for LLM training and evaluation, as well as the risks of LLM-based health coaching. 

%% file: sections/appendix/A_formative.tex
\section{Formative Study Details}
\label{appendix:formative}
We conducted a one-hour semi-structured interview with each participant. 
For health experts, we first asked a series of questions guided by the following structure:
\begin{itemize}
    \item What does a typical day look like for you? 
    \item Why do you do the work that you do?
    \item How did you approach your relationship with clients?
    \item What types of clients did you typically work with? 
    \item What are some of the common challenges that your clients face?
    \item How do you help people overcome those barriers? Do you have any common strategies or techniques?
    \item How do you adapt your strategies to cater to different people’s needs and circumstances?
    \item Are there any strategies you tried in the past but no longer use?
    \item Do you use any technology to assist you during your own exercise? 
    \item Do your clients use any digital technologies?
    \item How do you think technology changes your relationship with clients?
    \item How do you feel about the growing role of technology within the health/wellness space? 
\end{itemize}
\noindent
For non-experts, we used the following set of questions to guide our interview:
\begin{itemize}
    \item How would you define the term ``physical activity''? What about ``exercise''? What about ``fitness''?
    \item What kinds of activities do you do for exercise? (if applicable)
    \item How many days per week do you exercise in a typical week? (if applicable)
    \item Where do you exercise? (if applicable)
    \item Do you try to get people to be active with you? (if applicable)
    \item Have your levels of physical activity changed over time?
    \item If possible, can you tell us about a time in your life when you were particularly active? 
    \item If possible, can you tell us about a time in your life when you were particularly inactive? 
    \item What helps you stay motivated?
    \item What are some challenges you face in being active? Have you tried anything to overcome these challenge?
    \item Are you currently interested in getting more physical activity? If so, do you have any goals?
    \item Do you use any technology to assist you during your exercise? Why?
    \item Were there any technologies you tried before, but no longer use? Why? 
\end{itemize}

In the last portion of our interview, both groups of participants were asked to \textit{``Imagine that you have access to an artificial intelligence chatbot that can help you improve your physical activity. The chatbot has access to information from a fitness tracker (such as an Apple Watch or Fitbit) as well as suggestions and strategies for improving your exercise. We’re going to be asking you questions about a number of different features this hypothetical technology might offer.''} 
We then asked participants to consider each of the following features, one at a time:
\begin{itemize}
    \item Imagine that this chatbot is able to help you set effective goals
    \item Imagine that this chatbot is able to help keep you accountable towards those goals. 
    \item Imagine that this chatbot is able to help you formulate a training plan for an upcoming event (e.g., a competition, or a wedding). 
    \item Imagine that this chatbot is able to help you adjust your exercise routine in response to an injury.
    \item Imagine that this chatbot is able to help you visualize data from your fitness tracker and provide you with personalized insights. 
    \item Imagine that this chatbot can adopt different personas, such as an empathetic wellbeing counselor or a regimented personal trainer. 
    \item Imagine that this chatbot has access to your personal journal and offers insights into your mental wellbeing, personal relationships, and barriers to getting physical activity.
\end{itemize}
For each potential feature, we asked participants which aspects they liked, disliked, or had concerns about. 

The research team synthesized these features through several research activities. First, we reviewed relevant literature on human health coaching and coaching manuals provided to us by our collaborators. We then individually prepared short storyboards grounded in the health coaching literature and our own explorations using GPT-4 to interpret our personal data. We presented our storyboards to an external group of researchers and used this feedback to synthesize a set of core interactions. These interactions were presented to our collaborators and in our research group for additional feedback. 

%% file: sections/appendix/B_system.tex
\newpage
\section{\textsf{GPTCoach}: Implementation Details}
\label{appendix:system-details}

\begin{figure}[H]
    \centering
    \begin{subfigure}{0.5\textwidth}
      \centering
        \includegraphics[width=0.7\textwidth]{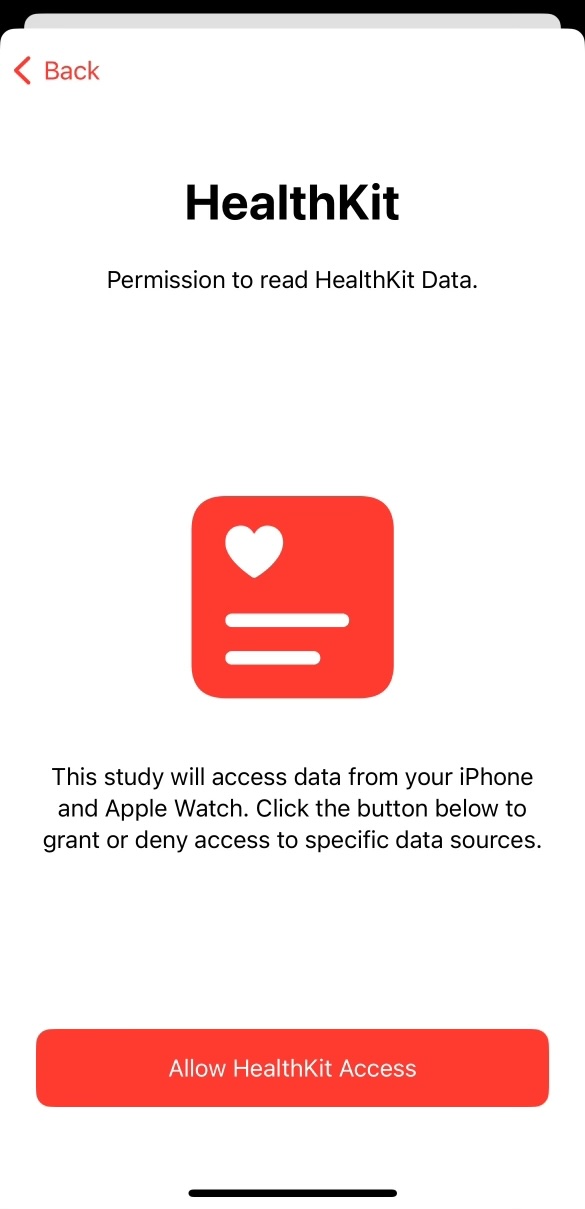}
        \caption{HealthKit permissions screen}
        \label{fig:healthkitupload-onboarding}
    \end{subfigure}%
    \begin{subfigure}{0.5\textwidth}
      \centering
        \includegraphics[width=0.8\textwidth]{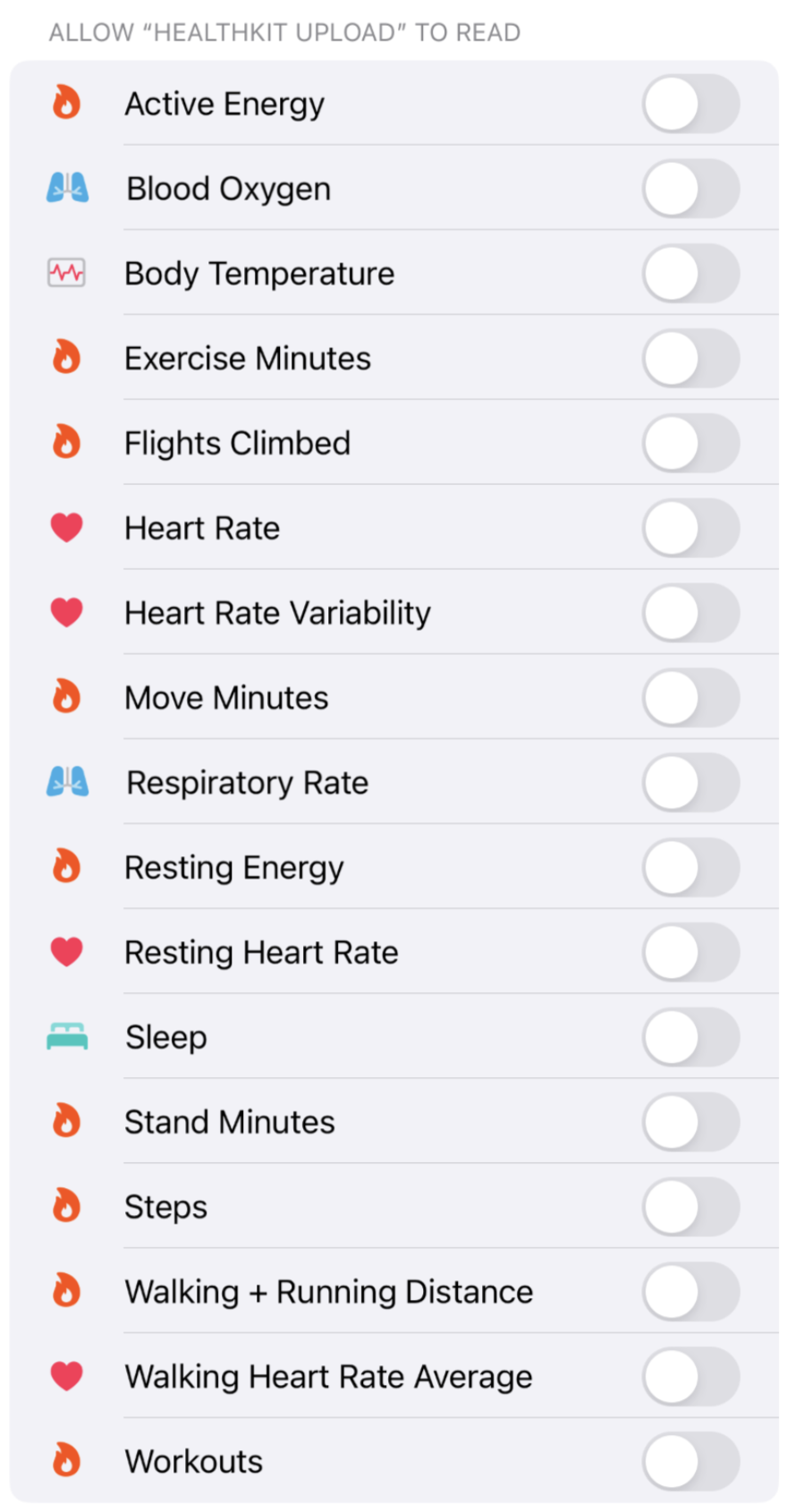}
        \caption{All data sources available in our system}
        \label{fig:healthkitupload-privacy-controls}
    \end{subfigure}
    \label{fig:ghealthkitupload}
    \caption{HealthKit permissions screen and data sources from our iOS application}
    \Description{HealthKit permissions screen and data sources from our iOS application. The figure consists of two screenshots. Fig 11A (left) shows a screenshot of the HealthKit permissions page, which displays a large red icon with a heart and the text “This study will access data from your iPhone and Apple Watch. Click the button below to grant or deny access to specific data sources/” A button labeled "Allow HealthKit Access" is centered at the bottom. Fig 11B (right) shows a screenshot of the privacy permissions page, with a list of toggle switches for different types of health data that can be accessed via HealthKit. The data sources include active energy, blood oxygen, body temperature, exercise minutes, flights climbed, heart rate, heart rate variability, move minutes, respiratory rate, resting energy, resting heart rate, sleep, stand minutes, steps, walking and running distance, walking heart rate average, and workouts. All toggles are shown in the off position.}
\end{figure} 

\subsection{iOS Application}
\label{appendix:ios}
To fetch users' historical data from the Apple HealthKit API and upload it to our Firestore database, we developed an iOS application using the Spezi open source framework~\cite{schmiedmayer2024}. 
Prior to installing our application, participants signed a consent form that informed them that they would be uploading three months of their health data and that they would be interacting with a chatbot system that has access to this data. Selected participants were also reminded of this in email communications prior to scheduling a study session. Participants were also informed that they could deny access to whichever data source they did not feel comfortable sharing. 

As part of the app's onboarding, users are shown a permissions screen (Figure \ref{fig:healthkitupload-onboarding}) which requests access to their HealthKit data. 
Upon granting permissions to read HealthKit data, participants were shown the default iOS HealthKit permissions screen, which contains granular toggles for individual data sources (Figure \ref{fig:healthkitupload-privacy-controls}). Users who did not own an Apple Watch were able to share active energy, basal energy, flights climbed, step count, and walking + running distance. Participants who had other fitness trackers that sync with HealthKit (e.g., Oura or WHOOP) were also able to upload their data through our application.

\subsection{Prompt Chains}
\label{appendix:prompt-chains}
In this section, we provide the structure for each of our prompt chains. 
LLMs are known to have poor performance when following instructions with long contexts. Given that instruction following is highest when relevant information is either at the beginning or end of the context \cite{liu2024lost}, we re-iterate important task instructions relevant to the current prompt chain to the agent at the end of the context window as an assistant message. The general structure of our prompts thus includes: a 1) system prompt, 2) dialogue history, and 3) an agent prompt. 
Full prompt texts are all provided in Appendix \ref{appendix:prompts}.

\subsubsection{Dialogue State Chain} Upon receiving a new user message, the dialogue state chain first classifies whether or not to advance to the next dialogue state. This dialogue state classifier uses the following prompt structure:

\begin{table}[H]
    \begin{tabular}{ll}
    \toprule
    \textbf{Role} & \textbf{Prompt} \\
    \midrule
    System: & Dialogue State Classification Instructions (Fig.~\ref{fig:dialogue-state-classification-prompt})  \\
    \midrule
    \multicolumn{2}{c}{--- Dialogue History --- } \\
    \midrule
    Agent Prompt: & Dialogue State Classifier Agent Instructions (Fig.~\ref{fig:dialogue-state-classification-agent-prompt}) \\
    
    \bottomrule    
    \end{tabular}
    \vspace{10 pt}
            \caption{Prompt structure for dialogue state classification} \label{table:prompt-structure-dialogue-state-classification}

\end{table}

\subsubsection{Motivational Interviewing Chain}
After deciding on the dialogue state, the motivational interviewing chain decides how to ground the model's responses to the 11 Motivational Interviewing strategies. This motivational interviewing chain uses the following prompt structure:

\begin{table}[H]
    \centering
    \begin{tabular}{p{0.12\textwidth} p{0.4\textwidth}}
    \toprule
        \textbf{Role} & \textbf{Prompt} \\
    \midrule
        System: & 
        System Instructions (Fig.~\ref{fig:system-prompt}) \newline
        + Dialogue State Prompt (Fig.~\ref{fig:dialogue-state-prompts}) \newline
        + Predict Strategy Instructions (Fig.~\ref{fig:strategy-prediction-instructions-prompt}) \newline
        + MI Interviewing Strategies (Fig.~\ref{fig:strategy-prediction-description-prompt}) \\
    \midrule
        \multicolumn{2}{c}{--- Dialogue History --- } \\
    \midrule
        Agent Prompt: & Strategy Prediction Agent Instructions (Fig.~\ref{fig:strategy-prediction-agent-prompt}) \\
    \bottomrule
    \end{tabular}
    \caption{Prompt structure for motivational interviewing strategy prediction.}
    \label{table:prompt-structure-strategy-prediction}
\end{table}

\subsubsection{Response Generation}

After the MI strategy prediction, our prompt chain then uses this strategy to predict the response from the LLM. This response generation step uses the following prompt structure:

\begin{table}[H]
    \begin{tabular}{p{0.12\textwidth} p{0.4\textwidth}}
    \toprule
    \textbf{Role} & \textbf{Prompt} \\
    \midrule
    
    System: & 
        System Instructions (Fig.~\ref{fig:system-prompt}) \newline
        + Dialogue State Prompt (Fig.~\ref{fig:dialogue-state-prompts}) \newline 
        + Generate Response Instructions (Fig.~\ref{fig:response-generation-instruction-prompts}) \newline 
        +  MI Interviewing Strategies (Fig.\ref{fig:strategy-prediction-description-prompt}) \newline 
        + Few-Shot Tool Call Examples (Fig.~\ref{fig:response-generation-fewshot-tool-call-examples-prompts})  \\
    \midrule
    \multicolumn{2}{c}{--- Dialogue History --- } \\
    \midrule
    Agent Prompt: & Response Generation Agent Instructions (Fig.~\ref{fig:response-generation-agent-prompts}) \\
    
    \bottomrule    
    \end{tabular}
    \vspace{10 pt}
    \caption{Prompt structure for response generation.} 
    \label{table:prompt-structure-response-generation}
\end{table}

\subsubsection{Tool Call Prediction}
We introduce a tool call prediction prompt chain to improve the timeliness of fetching and visualizing a user's health data. If the response generation step does not call a tool, using the output from the previous resposne generation step, we use an external LLM to predict whether tool call is appropriate to augment the conversation between the user and \textsf{GPTCoach}. This Tool Call Prediction chain uses the following prompt structure:

\begin{table}[H]
    \begin{tabular}{p{0.12\textwidth} p{0.4\textwidth}}
    \toprule
    \textbf{Role} & \textbf{Prompt} \\
    \midrule
    System: & 
        System Instructions (Fig.~\ref{fig:system-prompt}) \newline
        + Dialogue State Prompt (Fig.~\ref{fig:dialogue-state-prompts}) \newline
        + Tool Call Prediction Instructions (Fig.~\ref{fig:tool-call-prediction-instruction-prompts}) \newline
        + Few-Shot Tool Call Examples (Fig.~\ref{fig:response-generation-fewshot-tool-call-examples-prompts})
    \\
    \midrule
    \multicolumn{2}{c}{--- Dialogue History + Agent Response --- } \\
    \midrule
    Agent Prompt: & Tool Call Prediction Agent Instructions (Fig.~\ref{fig:tool-call-prediction-agent-prompt}) \\
    
    \bottomrule    
    \end{tabular}
    \vspace{10 pt}
            \caption{Prompt structure for tool call prediction.} 
            \label{table:prompt-structure-tool-call-prediction}

\end{table}

\subsubsection{Tool Call Generation}

If the response generation step did not have a tool call and if the tool call prediction step determined that tool call was appropriate in the conversation, we introduced another agent to determine which tool call is appropriate in the conversation. This Tool Call Generation chain uses the following prompt structure:

\begin{table}[H]
    \begin{tabular}{p{0.12\textwidth} p{0.4\textwidth}}
    \toprule
    \textbf{Role} & \textbf{Prompt} \\
    \midrule
    
    System: & 
        System Instructions (Fig.~\ref{fig:system-prompt}) \newline
        + Dialogue State Prompt (Fig.~\ref{fig:dialogue-state-prompts}) \newline
        + Tool Call Generation Instructions (Fig.~\ref{fig:tool-call-generation-instruction-prompts}) \newline
        + Few-Shot Tool Call Examples (Fig.~\ref{fig:response-generation-fewshot-tool-call-examples-prompts})  \\
    \midrule
    \multicolumn{2}{c}{--- Dialogue History + Agent Response --- } \\
    \midrule
    Agent Prompt: & Tool Call Prediction Agent Instructions (Fig.~\ref{fig:tool-call-generation-agent-prompt}) \\
    
    \bottomrule    
    \end{tabular}
    \vspace{10 pt}
            \caption{Prompt structure for tool call generation.} 
            \label{table:prompt-structure-tool-call-generation}

\end{table}

%% file: sections/appendix/C_tech_probe.tex
\newpage
\section{Evaluation Study Details}
\label{appendix:tech-probe}
In this section, we provide the survey measures for our evaluation study along with additional details for our automated motivational interviewing coding and counterfactual analyses. We also provide randomly sampled examples of our counterfactual analyses.

\subsection{Survey Measures}
\label{appendix:surveys}
\subsubsection{User Experience \& Quality of Advice}
\label{appendix:ux-questions}
We asked participants the following survey items after interacting with \textsf{GPTCoach}. Each question was rated on a 5-point Likert scale, from 1: Strongly disagree to 5: Strongly agree.

\begin{table}[h!]
\small
\begin{tabular}{l}
    \toprule
    The chatbot’s advice was actionable.\\ 
    The chatbot’s advice was personalized.\\ 
    The chatbot’s advice was generic.\\ 
    I felt comfortable sharing my concerns with the chatbot.\\ 
    I felt supported by the chatbot.\\ 
    The chatbot made me feel capable of overcoming challenges.\\ 
    The chatbot made me feel more motivated to change.\\ 
    The chatbot asked me for my opinion about what activities I would like to do.\\ 
    The chatbot understood my unique situation and concerns.\\ 
    The chatbot gave me unsolicited advice.\\ 
    The chatbot was empathetic.\\ 
    The chatbot used my data in a way that was relevant.\\ 
    The chatbot helped me identify obstacles to engaging in physical activity.\\  
    The chatbot helped me reflect on what motivates me to be physically active.\\ 
    The chatbot helped make my own ideas about how to increase my physical activity more specific.\\ 
    Interacting with the chatbot provided me with new insights about my physical activity.\\
    \bottomrule
\end{tabular}
\caption{User Experience \& Quality of Advice Questions}
\end{table}

\subsubsection{Subjective Assessment of Speech Systems Interfaces (SASSI)}
\label{appendix:sassi}
We measured usability using a subset of the Subjective Assessment of Speech System Interfaces~\cite{hone2000towards}. 
We use the same subset as \citet{mitchell2021automated} with two additional questions from the habitability and speed factors. 
Each question was rated on a 5-point Likert scale, from 1: Strongly disagree to 5: Strongly agree.

\begin{table}[h!]
\small
\begin{tabular}{ll}
    \toprule
    \textbf{Subscale} & \textbf{Question} \\
    \midrule
    \textsc{Response Accuracy} & The system is accurate \\ \midrule
    \textsc{Response Accuracy} & The system didn’t always do what I wanted\\ \midrule
    \textsc{Likeability} & The system is useful\\ \midrule
    \textsc{Likeability} & The system is friendly\\ \midrule
    \textsc{Likeability} & It is clear how to send messages to the system\\ \midrule
    \textsc{Cognitive Demand} & I felt confident using the system\\ \midrule
    \textsc{Cognitive Demand} & I felt tense using the system\\ \midrule
    \textsc{Annoyance} & The interaction with the system is repetitive\\ \midrule
    \textsc{Annoyance} & The interaction with the system is boring\\ \midrule
    \textsc{Habitability} & I always knew what to say to the system\\ \midrule
    \textsc{Habitability} & I was not always sure what the system was doing\\ \midrule
    \textsc{Speed} & The interaction with the system is fast \\
    \bottomrule
\end{tabular}
\caption{Usability Questions (SASSI~\cite{hone2000towards})}
\end{table}

\subsection{Remote vs. In-Person Participants}
\label{appendix:zoom-vs-inperson}
In this section, we report demographics, usability, and advice quality conditioned on remote (11 participants) and in-person (5) participant subgroups. We did not observe notable differences across the two groups, both in our quantative data (reported here) or in our qualitative feedback. In Table \ref{table:probe-demographics-zoom-vs-inperson} below, we report demographics within each subgroup.

The SASSI~\cite{hone2000towards} usability evaluation showed similar overall scores for Zoom (48.8/60, 81.3\%) and In-Person (50.8/60, 84.7\%) participants. Scores by factor for Zoom vs. In-Person were: Response Accuracy (7.6 vs. 7.8), Likeability (13.7 vs. 14.8), Cognitive Demand (8.9 vs. 9.6), Annoyance (7.5 vs. 7.6), Habitability (7.4 vs. 7.2), and Response Time (3.6 vs. 3.8). These results suggest that in-person participants reported marginally higher usability.

The advice quality evaluations also showed minimal differences between Zoom and In-Person participants. Scores by question for Zoom vs. In-Person were: Actionable (4.2 vs. 4.6), Personalized (4.5 vs. 4.6), Generic (2.5 vs. 2.6), Share Concerns (4.7 vs. 5.0), Supported (4.7 vs. 4.8), Capable (4.3 vs. 4.6), Motivated (3.9 vs. 4.2), Opinion Considered (4.8 vs. 4.2), Understood as Unique (4.4 vs. 4.6), Not Unsolicited (2.0 vs. 1.4), Empathetic (4.5 vs. 4.4), Relevant Data (3.9 vs. 3.8), Identify Obstacles (4.0 vs. 3.4), Reflect Motivation (4.3 vs. 4.4), Made Own Ideas Specific (4.1 vs. 4.2), and New Insights (4.3 vs. 3.2).

\input{sections/appendix/zoom-vs-inperson-demographics-table}

%% file: sections/appendix/zoom-vs-inperson-demographics-table.tex
\begin{table}[h!]
\small
\begin{tabular*}{\linewidth}{@{\extracolsep{\fill}} p{3cm} p{1.2cm} l}
\toprule
    \textbf{Age} & \textit{Zoom} & Mean: 39.4, Median: 33, SD: 16.0, Min: 21, Max: 71 \\
                 & \textit{In-Person} & Mean: 35.6, Median: 27, SD: 13.2, Min: 27, Max: 57 \\
    \midrule
    \textbf{Gender} & \textit{Zoom} & Female: 6, Male: 5 \\
                    & \textit{In-Person} & Female: 4, Male: 1 \\
    \midrule
    \textbf{Race/Ethnicity} & \textit{Zoom} & White: 8, East Asian: 1, Hispanic or Latino: 1, Middle Eastern: 1, Southeast Asian: 2, South Asian: 1 \\
                            & \textit{In-Person} & White: 3, Hispanic or Latino: 1, African-American or Black: 1 \\
    \midrule
    \textbf{Education} & \textit{Zoom} & Associate: 2, Bachelor's: 4, Master's: 4, Doctorate: 1 \\
                       & \textit{In-Person} & Bachelor's: 2, Master's: 3 \\
    \midrule
    \textbf{Stage of Change} & \textit{Zoom} & Precontemplation: 1, Contemplation: 6, Action: 2, Maintenance: 2 \\
                             & \textit{In-Person} &  Contemplation: 2, Action: 1, Maintenance: 2 \\
    \midrule
    \textbf{Level of Activity (IPAQ)} & \textit{Zoom} & Low: 4, Moderate: 4, High: 3 \\
                                      & \textit{In-Person} & Low: 1, Moderate: 3, High: 1 \\
    \midrule
    \textbf{AI Knowledge} & \textit{Zoom} & Basic: 7, Intermediate: 3, Advanced: 1 \\
                          & \textit{In-Person} & Novice: 1, Basic: 3, Advanced: 1 \\
\bottomrule
\end{tabular*}
\vspace{0.1em}
\caption{\textbf{Participant demographics by format (Zoom vs. In-Person) in the technology probe evaluation study $(N=16)$}}
\label{table:probe-demographics-zoom-vs-inperson}
\vspace{-2em}
\end{table}

%% file: sections/appendix/D_mi_coding.tex
\section{Motivational Interviewing Coding}
In this section, we report on two motivational interviewing coding analysis. First, we report on our paper's primary analysis, in which human MI experts coded transcripts according to the MITI coding scheme. We then report on an analogous LLM-based coding analysis using the MISC coding scheme.

\subsection{MITI Coding (Human Experts)}
\label{appendix:miti-coding}
We hired trained coders to code transcripts according to the Motivational Interviewing Treatment Integrity (MITI) 4 Code~\cite{moyers2016motivational}. We partnered with an agency that provides a wide range of MI training services, including MITI coding, and had previous experience coding chatbot transcripts. All of the coders were trained in MITI coding and had extensive coding experience.

\subsubsection{Behavior Codes}
In MITI coding, each ``volley'' (i.e., a turn or message) is partitioned into utterances and each utterance receives a unique behavior code. All MITI behaivor codes are listed below in Table \ref{tab:miti-code-table}. The codes assigned to each volley are the deduplicated set of codes aggregated across all utterances contained in the volley (e.g., if a volley contains two \textsc{Questions} utterances and one \textsc{Complex Reflection} utterance, that volley will receive the aggregate label: \textsc{Complex Reflection, Question}). Additional rules for aggregating utterance-level codes (e.g., a \textsc{Simple Reflection} and \textsc{Complex Reflection} occuring is the same volley is coded only as \textsc{Complex Reflection}) are described in the MITI coding manual.\footnote{\url{https://casaa.unm.edu/assets/docs/miti4_21.pdf}}

\input{sections/appendix/miti-code-table}

\subsubsection{Inter-Rater Reliability}
Each of the 16 transcripts were independently coded by three different coders. Coders were aware that the transcripts were from AI coaching conversations because \textsf{GPTCoach} discloses that it is a chatbot. In particular, transcripts 1-8 were coded by coders 1-3, and transcripts 9-16 were coded by coders 4-6. After all transcripts had been coded once, all coders met together to review their scores and recoded segments with substantial disagreement.

Following best practices for computing inter-rater reliability (IRR)~\cite{hallgren2012computing} and Moyer et al.'s~\cite{moyers2016motivational} original reliability study, we compute IRR using a one-way, random effects, absolute agreement, average-measures intraclass correlation (ICC) for transcript-level behavior counts. That is, we compare agreement among the integer-valued total sum of all codes in a transcript for a given category. 
We use ICC because these data are ordinal and we have more than two coders. We use a one-way, random effects ICC because each coder did not code all 16 transcripts. ICC values for each measure are provided in Table \ref{tab:irr-table}, with an average (SD) ICC of 0.79 (0.17). Following \cite{cicchetti1981developing} (0.00–0.40 = poor, 0.40–0.59 = fair, 0.60–0.74 = good, and 0.75–1.00 = excellent), this corresponds to excellent agreement on average. In addition, we report absolute agreement and the maximum difference in annotator scores for each code. For a given behavior code, absolute agreement describes the percentage of transcripts where all three coders had the exact same behavior count, and maximum difference describes the maximum difference in counts among the three raters.

\input{sections/appendix/irr-table}

\subsubsection{Counterfactual Analysis}
\label{appendix:counterfactual}
We perform a counterfactual analysis to compare \textsf{GPTCoach}'s behavior to vanilla GPT-4\footnote{All experiments were run with \texttt{gpt-4-0613} with temperature 1.}.
We condition on the first five turns of each participant's true conversation history with \textsf{GPTCoach} from our evaluation study.
The first five turns of the conversation are highly consistent across participants, including an introduction to the program, the participant sharing their name and age, and all ending with the agent asking whether they have any questions or concerns. 
We focus our evaluation on early stages of the conversation since we found that early interactions greatly shape the quality of the resulting conversation. 

We simulate 10 different user responses to the agent's question, each of which correspond to a different barrier to physical activity. We source the barrier categories from the coaching materials we received from our collaborators. 
The barriers are listed in Table \ref{appendix:barrier-messages} below.

\begin{table}[H]
\small
\begin{tabular}{l p{0.7\linewidth}}
    \toprule
    \textbf{Barrier} & \textbf{User Message} \\
    \midrule
    Feeling discomfort & I haven't exercised in a while and I'm worried about being sore and feeling pain from not having exercised in a long time. \\ \midrule
    Feeling unmotivated & On most days, it's hard for me to find the motivation to exercise. \\ \midrule
    No energy & I feel too tired to exercise when I get back from work. I just don't have the energy to exercise. \\ \midrule
    No time & I just feel like I don't have time to exercise. Between work and my family responsibilities, I'm always so busy. \\ \midrule
    Feeling sick & I had been sick for the last few weeks and that has thrown me off track. Now it feels hard to get back into my routine. \\ \midrule
    Feeling stressed & I've been feeling quite stressed lately and that's been preventing me from getting exercise. \\ \midrule
    Feeling ashamed & I’ve never really exercised before and I worry about what others might think of me. I feel ashamed. \\ \midrule
    Feeling unsafe & I don’t feel safe going for a workout outside in my neighborhood, which makes it hard to stay active. \\ \midrule
    Feeling unsupported or alone & I don't have anyone I can exercise with together. I feel like I’m doing this alone. \\ \midrule
    Weather & It's difficult to get exercise because it's really cold and dark outside this time of year. \\ 
    \bottomrule
\end{tabular}
\caption{Simulated user messages based on common barriers to physical activity}
\label{appendix:barrier-messages}
\end{table}

\noindent
This set of 16 true histories and 10 simulated user messages yields a test set of 160 simulated histories.
For each simulated history, we sample an output from \textsf{GPTCoach} and from GPT-4 using only the system instruction prompt (Fig. \ref{fig:system-prompt}), ignoring all prompt chains. 

All 320 responses were coded for MITI behavior codes by a single coder, who was the most experienced coder from our pool of six coders. The coder was blind to condition.

\newpage
\subsubsection{Illustrative Examples}
\label{appendix:illustrative-examples}
We randomly sample five outputs and provide them as illustrative examples below. Participant names and ages have been redacted.

\begin{table}[H]
\small
\caption*{\textbf{Example 1}}
\vspace{-1em}
\begin{tabular}{p{0.13\linewidth} p{0.8\linewidth}}
    \toprule
    \textbf{Role} & \textbf{Message} \\
    \midrule
    Agent: & Hello, it's wonderful to meet you! I'm a health coaching chatbot and am excited that you're here to start this journey with me. How are you doing today? \\
    User: & I'm doing well. Thanks for asking. \\
    Agent: & That’s great to hear! May I know your name and age? \\
    User: & My name's \{NAME\}. I'm \{AGE\} years old \\
    Agent: & Welcome to the program, \{NAME\}! It's great to know that you're taking this positive initiative for your health. We will be creating a physical activity plan tailored just for you. This plan will be designed based on your unique interests, preferences and the resources you have access to. My role is to assist you throughout this process, providing you with the necessary information to make your plan effective and sustainable. \newline \newline
    Does the concept make sense to you, and do you have any questions or concerns so far? Please feel free to discuss anything that's on your mind. \\
    \midrule
    User (simulated): & It's difficult to get exercise because it's really cold and dark outside this time of year. \\
    \midrule
    Vanilla GPT: & I totally understand that it's challenging to keep up with an exercise routine during the cold and dark months. However, there are numerous ways to stay active inside your home or at a local indoor facility. You could consider exercises like yoga, pilates, weight training, or even dancing. There are great resources online that guide you through these home workouts. 
    \newline \newline
    Also, remember, our goal is to make exercise enjoyable for you, so I encourage you to think about activities that you personally enjoy. Let's work together to find a solution that fits your lifestyle and preference. \\
    GPTCoach: & I understand that the weather can indeed make it challenging to go outside for physical activities, \{NAME\}. If it's okay, may I please ask what types of physical activities you have engaged in the past and for how long you would typically engage in these activities? Your past experiences can help us tailor a plan that fits your preferences and circumstances.\\
    \bottomrule
\end{tabular}
\end{table}

\begin{table}[H]
\small
\caption*{\textbf{Example 2}}
\vspace{-1em}
\begin{tabular}{p{0.13\linewidth} p{0.8\linewidth}}
    \toprule
    \textbf{Role} & \textbf{Message} \\
    \midrule
    Agent: & Hello, it's wonderful to meet you! I'm a health coaching chatbot and am excited that you're here to start this journey with me. How are you doing today? \\
    User: & I'm doing well \\
    Agent: & That's great to hear! Could you please share with me your name and age? \\
    User: & \{NAME\}, age \{AGE\} \\
    Agent: & Nice to meet you, \{NAME\}! As we embark on this journey together, it will be your task to design your own physical activity plan. This approach can be really rewarding because it is tailored specifically to your preferences, interests, and resources. I'm here to assist and provide evidence-based insights to help you determine the specifics and effectiveness of your plan. Can I assure you understand this approach? Do you have any questions or concerns to start with? \\
    \midrule
    User (simulated): & I just feel like I don't have time to exercise. Between work and my family responsibilities, I'm always so busy. \\
    \midrule
    Vanilla GPT: & I completely understand, \{NAME\}, and it's a common concern. Balancing time between responsibilities and personal health can be a real challenge. What if we start by identifying some short, manageable physical activities that can easily be incorporated into your daily routine? Small steps can lead to big changes. Also, activities involving family can be enjoyable and beneficial for everyone. Please remember to always consult with your healthcare provider before starting any physical activity program. What are your thoughts about this?
    \newline \newline
    Also, remember, our goal is to make exercise enjoyable for you, so I encourage you to think about activities that you personally enjoy. Let's work together to find a solution that fits your lifestyle and preference. \\
    GPTCoach: & I completely understand, \{NAME\}, it can indeed be challenging to fit in exercise in a busy schedule, but it's definitely not impossible. To better assist you, it would be very helpful to understand a bit more about your past experiences with physical activity. 

    To begin, could you tell me what types of physical activities you've done in the past and for approximately how long? For example, something like ``I used to jog for about 30 minutes in the mornings''.\\
    \bottomrule
\end{tabular}
\end{table}

\begin{table}[H]
\small
\caption*{\textbf{Example 3}}
\vspace{-1em}
\begin{tabular}{p{0.13\linewidth} p{0.8\linewidth}}
    \toprule
    \textbf{Role} & \textbf{Message} \\
    \midrule
    Agent: & Hello, it's wonderful to meet you! I'm a health coaching chatbot and am excited that you're here to start this journey with me. How are you doing today? \\
    User: & I'm doing fine, thanks for asking \\
    Agent: & That's great to hear! To get us started, can I have your name and your age, please? \\
    User: & \{NAME\} and \{AGE\} \\
    Agent: & Great to meet you, \{NAME\}! As we begin, I'd like to clarify a few things. Throughout this program, you'll have the opportunity to create your own physical activity plan. This means your plan will be based on your preferences, interests, and available resources. I'm here to assist you in figuring out the specifics of that plan.
    \newline \newline
    How does that sound, \{NAME\}? Please feel free to let me know if you have any questions or concerns before we proceed. \\
    \midrule
    User (simulated): & I just feel like I don't have time to exercise. Between work and my family responsibilities, I'm always so busy  \\
    \midrule
    Vanilla GPT: & \{NAME\}, your feelings are completely valid and many people can relate to the struggle of balancing work, family, and personal health. It's fantastic that you're starting to prioritize your health, though. What we could do is explore ways on how you might fit small, manageable blocks of activity into your day which doesn't take much extra time from your busy schedule. Could this approach be helpful for you, \{NAME\}? \\
    GPTCoach: & I understand, \{NAME\}, balancing work, family, and personal time can be challenging. You're definitely not alone in feeling that time is a barrier to physical activity. But remember, any physical activity is better than none and it all adds up. 
    \newline \newline
    In order to best assist you, could you please share a little more about your past experiences with physical activity? What types of activities have you previously engaged in, and for how long typically? This will give us a nice starting point to work from.\\
    \bottomrule
\end{tabular}
\end{table}

\begin{table}[H]
\small
\caption*{\textbf{Example 4}}
\vspace{-1em}
\begin{tabular}{p{0.13\linewidth} p{0.8\linewidth}}
    \toprule
    \textbf{Role} & \textbf{Message} \\
    \midrule
    Agent: & Hello, it's wonderful to meet you! I'm a health coaching chatbot and am excited that you're here to start this journey with me. How are you doing today? \\
    User: & I'm doing great! \\
    Agent: & Fantastic to hear that! Could you please tell me your name and how old you are? \\
    User: & \{NAME\} and I'm \{AGE\} \\
    Agent: & Nice to meet you, \{NAME\}! Given you're now part of our program, together we'll design a physical activity plan uniquely tailored to you. This plan will reflect your particular preferences, interests, and available resources.
    \newline \newline
    That being said, this will be a collaborative process where your input is not only valued but essential. I'm here to guide and assist you in creating your plan and achieving your goals. I hope I've made our approach clear.
    \newline \newline
    Before we move on, \{NAME\}, do you have any questions or concerns about designing your own physical activity plan? \\
    \midrule
    User (simulated): & On most days, it's hard for me to find the motivation to exercise. \\
    \midrule
    Vanilla GPT: & I completely understand, \{NAME\}, and it's perfectly normal to feel this way. Increasing your motivation for physical activity can be a journey in itself. What's most important is to find activities that you truly enjoy and look forward to. That's why our collaboration in designing your physical activity plan will be crucial. 
    \newline \newline
    If your lack of motivation persists and begins to affect other areas of your life too, it can be helpful to speak with a mental health professional. They're equipped with the right tools to provide additional support. \\
    GPTCoach: & I see, \{NAME\}. It's not uncommon to encounter days when motivation to exercise can feel out of reach. Many of us have been there. Would it be okay for us to explore your past experiences with physical activity? This can give us some valuable insights to better understand your journey so far. Could we start by knowing what types of activities you did in the past, and approximately how long you sustained them?\\
    \bottomrule
\end{tabular}
\end{table}

\begin{table}[H]
\small
\caption*{\textbf{Example 5}}
\vspace{-1em}
\begin{tabular}{p{0.13\linewidth} p{0.8\linewidth}}
    \toprule
    \textbf{Role} & \textbf{Message} \\
    \midrule
    Agent: & Hello, it's wonderful to meet you! I'm a health coaching chatbot and am excited that you're here to start this journey with me. How are you doing today? \\
    User: & Hi \\
    Agent: & Glad to see you here! I'm really looking forward to sharing this wellness journey with you. Could you please share your name and age with me? \\
    User: & \{NAME\} age \{AGE\} \\
    Agent: & It's wonderful to meet you, \{NAME\}! At your age, maintaining an active lifestyle is such a positive step towards ongoing health and wellness.
    \newline \newline
    I'd like to reassure you that while I'm here to assist, you'll be the one designing your own physical activity plan. Ultimately, the plan should be reflective of your interests, preferences, and available resources. This will ensure it's not only beneficial but also enjoyable and achievable. 
    \newline \newline
    I'll help you shape the specifics of your activity plan, based on the data from your wearable devices and your personal preferences. Are you understand, and do you have any specific questions or concerns before we proceed? \\
    \midrule
    User (simulated): & I’ve never really exercised before and I worry about what others might think of me. I feel ashamed. \\
    \midrule
    Vanilla GPT: & I'm really sorry to hear that you're feeling this way, \{NAME\}, but please know you're not alone. It's completely normal to feel a bit apprehensive when starting something new, especially when it comes to exercise. It's important to remember that everyone's fitness journey is unique and there's no need to compare yourself with others.
    \newline \newline
    Starting at your own pace and taking small steps are great ways to ease into an exercise routine. Things like walking in your neighborhood, light stretch exercises in your living room, or gardening can be wonderful starting points. 
    \newline \newline
    And remember, it's your journey. For professional help with these feelings of shame, you might want to consider reaching out to a mental health professional - they are trained to navigate these types of concerns. As a health coach, I'm here to support you in your physical activity journey, and together we can make this a positive and rewarding experience for you. \\
    GPTCoach: & I'm really sorry to hear that you're feeling this way, \{NAME\}. It's completely normal to feel a bit self-conscious when starting something new, especially when it involves exercising. But remember, everyone is there for their own health and well-being, just like you. 
    \newline \newline
    Could you share what type of physical activity, if any, you've tried in the past? How often and how long were these activities? Let's start there and see how we can make your exercise experience better and more comfortable. \\
    \bottomrule
\end{tabular}
\end{table}

\subsection{MISC Coding (LLM-Based)}
\label{appendix:misc-coding}
We draw from Chiu et al.'s~\cite{chiu2024computational} assessment of LLM psychotherapists and use an external GPT-4 evaluator to code the chatbot's utterances. Since Chiu et al.'s~\cite{chiu2024computational} coding scheme applies to psychotherapy, we used a different set of 19 motivational interviewing strategies from the Motivational Interviewing Skill Code (MISC)~\cite{moyers2003assessing}.\footnote{\url{https://casaa.unm.edu/assets/docs/misc3.pdf}} Each strategy is either MI-consistent, MI-inconsistent, or neutral. We construct three few-shot examples for each strategy, drawing from the MISC manual and adapting them to physical activity promotion. All 19 strategies along with their definitions are provided below in Table \ref{appendix:misc-table}.

\input{sections/appendix/misc-code-table}

We use an analogous prompt template to \citet{chiu2024computational} for strategy classification:

\begin{promptbox}
    What are all possible strategies of this coach utterance: \texttt{\{UTTERANCE\}}? \newline
    Strategy: \newline
    \texttt{\{STRATEGY\}}: \texttt{\{DEFINITION\}} Positive examples: \texttt{\{EXAMPLE 1\}} \texttt{\{EXAMPLE 2\}} \texttt{\{EXAMPLE 3\}} \newline
    ... \newline
    Only choose from this list [\texttt{\{STRATEGIES\}}] \newline
    Please say unknown only if you cannot find an answer from the list. Format:[strategies\_list] 
\end{promptbox}

To code an agent response, we first split the response into individual sentences using the spacy Sentencizer.\footnote{\url{https://spacy.io/api/sentencizer}} We then code each sentence using the prompt above and merge codes across sentences to produce a set of MI codes for each response.

\subsubsection{Results}
In Figures \ref{fig:external-misc-strategies}, \ref{fig:misc-consistency}, and \ref{fig:counterfactual-llm} below, we replicate Figures \ref{fig:external-strategies}, \ref{fig:mi-consistency}, and \ref{fig:counterfactual} from the main text with our LLM coding results. Overall, we find that our LLM-based coding results are highly analogous to our human coding results, revealing similar patterns in MI-consistency and MI strategy usage.

On average, each agent response contained 4.5 different MISC strategies. As shown in Figure \ref{fig:external-misc-strategies}, the most frequent strategies are \textsc{Affirm} (18.8\% of all MI codes) and \textsc{Open Question} (18.0\%). However, we also encounter undesirable strategies like \textit{Advise Without Permission} across 12.9\% of all codes.
Aggregating these strategies into MI consistent strategies (\textsc{Advise With Permission, Affirm, Emphasize Control, Open Question, Simple Reflection, Complex Reflection, Reframe, Support}), inconsistent strategies (\textsc{Advise Without Permission, Confront, Direct, Raise Concern without Permission, Warn}), and neutral strategies (all others) based on MISC~\cite{moyers2003assessing}, we see in Figure \ref{fig:misc-consistency} that only 15.9\% of total codes corresponding to our chatbot's responses are inconsistent with MI principles. 

In the counterfactual analysis, as shown in Figure \ref{fig:counterfactual-misc-consistency}, 62.6\% of \textsf{GPTCoach}'s codes are MI-consistent, compared to 55.0\% vanilla GPT-4's codes. Meanwhile, 14.5\% of \textsf{GPTCoach}'s codes are MI-inconsistent, compared to 22.9\% of vanilla GPT-4's codes. We found that the top three codes driving this difference in MI-consistency are \textsc{Advise with Permission}, \textsc{Advise without Permission}, and \textsc{Open Question}, matching our intuitions. In Figure \ref{fig:counterfactual-strategy-frequency-llm}, we plot the percentage of messages that contain each of these codes. We find that GPT-4 contains \textsc{Advise without Permission} in every response, compared to 52.5\% of \textsf{GPTCoach}'s responses. Meanwhile, \textsf{GPTCoach} contains \textsc{Open Question} in every response, compared to 39.4\% of vanilla GPT-4's. Lastly, we also find that vanilla GPT-4 contains \textsc{Advise with Permission} in 37.5\% of its replies, compared to 13.1\% of \textsf{GPTCoach}'s replies. Note that the same response can contain advice both with and without permission when the agent gives advice on two different topics, e.g., \textit{``Try incorporating workouts into your daily activities like brisk walking during lunch breaks, taking the stairs when you can or doing some body weight exercises at home.[...] If you feel you're struggling to manage stress or time due to these commitments, however, I recommend seeking advice from a professional counselor or psychologist who can help address these concerns better. Would that be helpful?''}. While both models are biased towards advice without permission, \textsf{GPTCoach} is far more inclined to ask open questions rather than jumping to unsolicited advice. 

\begin{figure}[H]
    \centering
    \includegraphics[width=\textwidth]{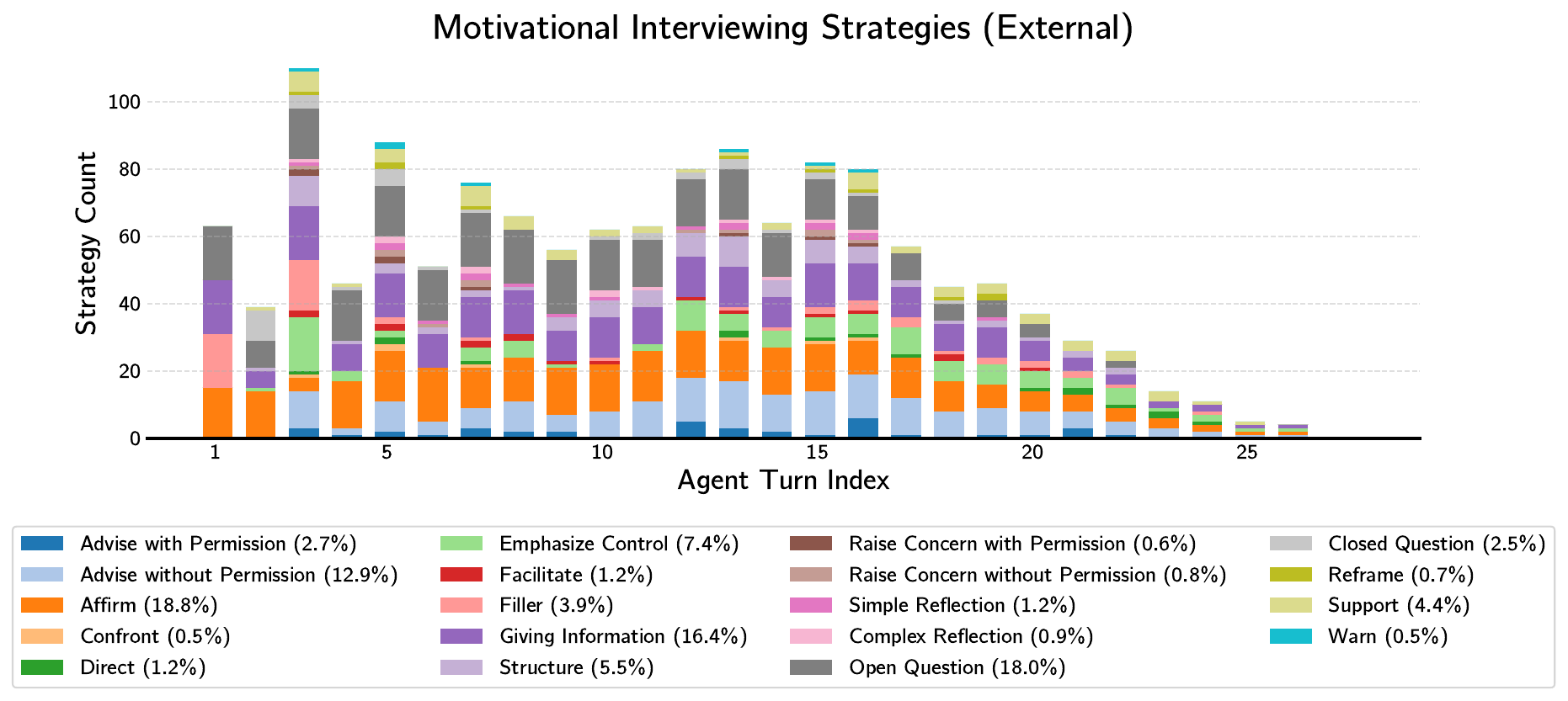} 
    \caption{
    \textbf{Distribution of \textsf{GPTCoach}'s External MI Strategies by Turn Index (LLM-Based MISC Coding).} We find that \textsf{GPTCoach} most frequently engages in \textsc{Affirm}, \textsc{Open Question}, and \textsc{Giving Information}, which is aligned with MI principles. However, \textsf{GPTCoach} still engages in \textsc{Advise without Permission}, which is not aligned with MI principles. External strategy defintions are provided in Table \ref{appendix:misc-table}.}
    \label{fig:external-misc-strategies}
    \Description{Distribution of GPTCoach’s External MI Strategies by Turn Index (LLM-Based MISC Coding). A stacked bar chart shows the distribution of external MI strategies employed by GPTCoach over the course of a conversation, with the Y-axis representing the strategy count (ranging from 0-100) and the X-axis representing the agent turn index (1-25). Each color in the stacked bars corresponds to a specific MI strategy. This figure highlights that GPTCoach frequently uses strategies such as Affirm, Open Question, and Giving Information, which align with MI principles. However, it also occasionally uses Advise without Permission, which is less aligned with these principles.
}
\end{figure}

\begin{figure}[H]
    \centering
    \includegraphics[width=\textwidth]{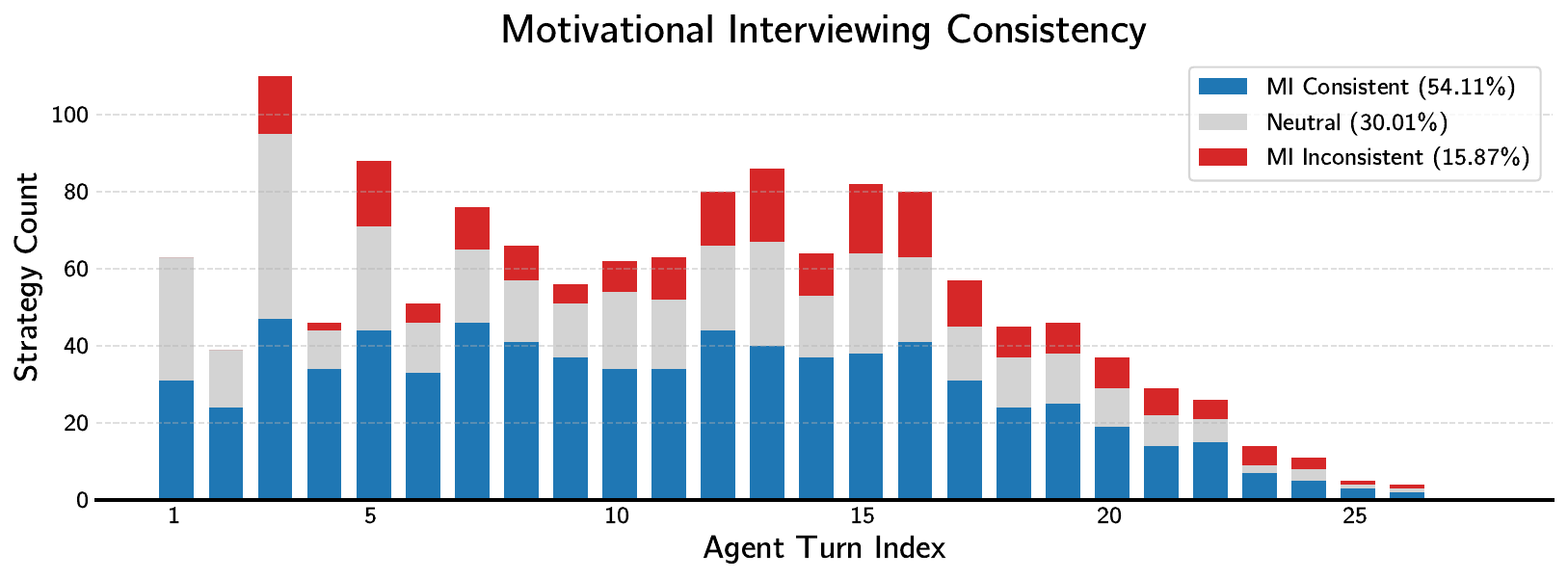}
    \caption{\textbf{\textsf{GPTCoach}'s MI Consistency by Turn Index (LLM-Based MISC Coding).} This figure aggregates external MI codes from Fig. \ref{fig:external-misc-strategies} into MI-consistent, MI-inconsistent, and neutral codes based on MISC~\cite{moyers2003assessing}. We find that only 15.9\% of \textsf{GPTCoach}'s total codes are inconsistent with MI principles.}
    \label{fig:misc-consistency}
    \Description{GPTCoach’s MI Consistency by Turn Index (LLM-Based MISC Coding). A stacked bar chart shows the proportion of MI-consistent, neutral, and MI-inconsistent strategies over the course of a conversation, with the Y-axis representing the strategy count (ranging from 0-100) and the X-axis representing the agent turn index (1-25). MI-consistent strategies are the most frequent, especially during the early and middle turns. MI-inconsistent strategies, although present, account for a smaller portion, peaking occasionally throughout the conversation.}
\end{figure}

\begin{figure}[H]
    \centering
    \begin{subfigure}{0.48\textwidth}
      \centering
        \includegraphics[width=\textwidth]{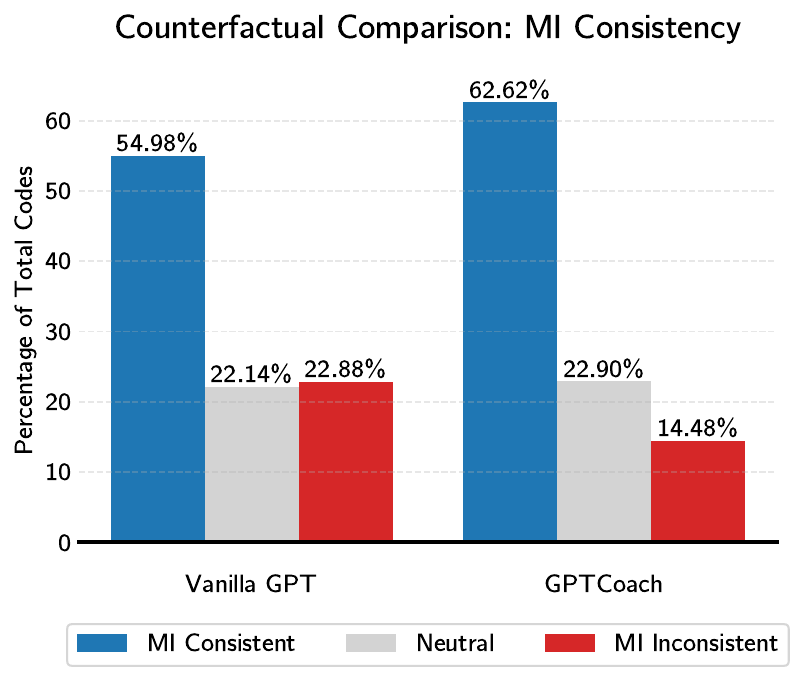}
        \caption{
        \textbf{Comparing the overall frequency of MI-constistent, MI-inconsistent, and neutral strategies in our counterfactual analysis.} Percentages represent the proportion of strategies that are MI-consistent/MI-inconsistent/neutral strategies among all MI strategy codes for a given agent. We find that \textsf{GPTCoach}'s responses contain more MI-consistent and fewer MI-inconsistent responses than vanilla GPT.\newline}
        \label{fig:counterfactual-misc-consistency}
    \end{subfigure}
    \hfill
    \begin{subfigure}{0.48\textwidth}
      \centering
        \includegraphics[width=\textwidth]{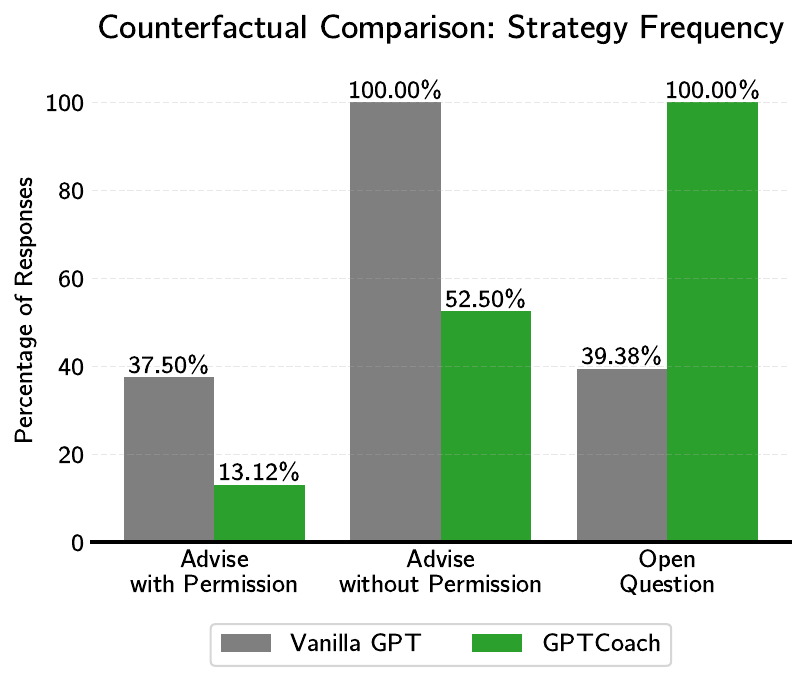}
        \caption{
        \textbf{Comparing the frequency of MI strategies by response in our counterfactual analysis.} These three strategies showed the largest differences in frequency across the two agents. Percentages represent the proportion messages that contain a given MI strategy across all of a given agent's counterfactual responses. We find that vanilla GPT gives advice without permission in all of its responses, while \textsf{GPTCoach} gives less advice and asks open questions in all of its responses.
        }
        \label{fig:counterfactual-strategy-frequency-llm}
    \end{subfigure}
    \caption{\textbf{Counterfactual comparison between \textsf{GPTCoach} and vanilla GPT-4 (LLM-Based MISC Coding).} The vanilla GPT condition includes the same system prompt, but with all prompt chains ablated. Illustrative examples are provided in Appendix \ref{appendix:illustrative-examples}.}
    \label{fig:counterfactual-llm}
    \Description{Counterfactual comparison between GPTCoach and vanilla GPT-4 (LLM-based MISC Coding). This figure consists of two bar charts. In panel A (left), the bar chart compares how the two models adhere to Motivational Interviewing (MI) principles. The Y-axis shows the percentage of total codes and the X-axis shows the two models being compared: Vanilla GPT and GPTCoach. The individual bars show the percentage of MI-consistent, neutral, and MI-inconsistent strategies for each agent. This comparison shows that GPTCoach has more MI-consistent responses and fewer MI-inconsistent responses compared to Vanilla GPT. In panel B (right), the bar chart compares the frequency of three MI strategies in each model’s responses. The Y-axis represents the percentage of responses that contain a particular strategy, while the X-axis shows three strategies: Advise with Permission, Advise without Permission, and Open Question. The individual bars show the percentage of responses containing the respective strategy for each model. This comparison demonstrates that Vanilla GPT frequently provides advice without permission, while GPTCoach uses Open Questions much more frequently, in line with MI principles.
}
\end{figure}

%% file: sections/appendix/miti-code-table.tex
\begin{table}[H]
\small
\begin{tabular*}{\linewidth}{@{\extracolsep{\fill}} l l p{10cm}}
\toprule
    \textbf{Behavior Code} & \textbf{Consistency} & \textbf{Definition} \\
    \midrule
    \textsc{Giving Information} & Neutral & Gives information, educates, provides feedback, or expresses a professional opinion without persuading, advising, or warning. \\
    \midrule
    \textsc{Questions} & MI-Consistent & Questions (open or closed). \\
    \midrule
    \textsc{Simple Reflections} & MI-Consistent & Reflects a client’s statement with little or no added meaning or emphasis. \\
    \midrule
    \textsc{Complex Reflections} & MI-Consistent & Reflects a client’s statement with added meaning or emphasis. \\
    \midrule
    \textsc{Affirm} & MI-Consistent & States something positive about the client’s strengths, efforts, intentions, or worth. \\
    \midrule
    \textsc{Emphasize Autonomy} & MI-Consistent & Highlights a client’s sense of control, freedom of choice, personal autonomy, ability, and obligation about change. \\
    \midrule
    \textsc{Seek Collaboration} & MI-Consistent & Attempts to share power or acknowledge the expertise of a client. \\
    \midrule
    \textsc{Persuade} & MI-Inconsistent & Overt attempts to change a client’s opinions, attitudes, or behaviors using tools such as logic, compelling arguments, self-disclosure, facts, biased information, advice, suggestions, tips, opinions, or solutions to problems. \\
    \midrule
    \textsc{Persuade with permission} & MI-Consistent & Emphasis on collaboration or autonomy support while using direct influence. \\
    \midrule
    \textsc{Confront} & MI-Consistent & Directly and unambiguously disagreeing, arguing, correcting, shaming, blaming, criticizing, labeling, warning, moralizing, ridiculing, or questioning a client’s honesty. \\
\bottomrule
\end{tabular*}
\vspace{1em}
\caption{MITI~\cite{moyers2016motivational} behavior codes used to code \textsf{GPTCoach}'s responses, along with their MI consistency and definition.}
\label{tab:miti-code-table}
\end{table}

%% file: sections/appendix/irr-table.tex
\begin{table}[H]
\small
\begin{tabular}{lcccc}
\toprule
    \textbf{Behavior Code} & \textbf{Total Count} & \textbf{ICC} & \textbf{Agreement} & \textbf{Max Diff. [Min, Max]}\\
    \midrule
    \textsc{Giving Information} & 645 & 0.89 & 50.0\% & 7 [7-21]\\
    \midrule
    \textsc{Questions} & 684 & 0.99 & 56.2\% & 4 [9-21] \\
    \midrule
    \textsc{Simple Reflections} & 44 & 0.63 & 77.1\% & 5 [0-5] \\
    \midrule
    \textsc{Complex Reflections} & 43 & 0.94 & 81.2\% & 2 [0-4] \\
    \midrule
    \textsc{Affirm} & 283 & 0.89 & 54.2\% & 5 [3-10] \\
    \midrule
    \textsc{Emphasize Autonomy} & 6 & 0.45 & 89.6\% & 1 [0-1]\\
    \midrule
    \textsc{Seek Collaboration} & 136 & 0.72 & 64.6\% & 7 [0-8] \\
    \midrule
    \textsc{Persuade} & 140 & 0.92 & 66.7\% & 4 [0-8] \\
    \midrule
    \textsc{Persuade with permission} & 95 & 0.94 & 62.5\% & 3 [0-4] \\
    \midrule
    \textsc{Confront} & 0 & - & - & - \\
\bottomrule
\end{tabular}
\vspace{1em}
\caption{Inter-rater reliability metrics for each behavior code, including ICC, absolute agreement, and maximum difference. \citet{cicchetti1981developing} provide the following benchmark for ICC values: 0.00–0.40 = poor, 0.40–0.59 = fair, 0.60–0.74 = good, and 0.75–1.00 = excellent.}
\label{tab:irr-table}
\end{table}

%% file: sections/appendix/misc-code-table.tex
\begin{table*}[b]
\tiny
\begin{tabular}{p{0.1\textwidth} p{0.1\textwidth} p{0.73\textwidth}}
\toprule
    \textbf{Strategy} & \textbf{Consistency} & \textbf{Definition} \\
    \midrule
    \textsc{Advise With Permission} & MI-consistent & The counselor gives advice, makes a suggestion, or offers a solution or possible action with client permission. These will usually contain language that indicates that advice is being given: should, why don't you, consider, try, suggest, advise, you could, etc. Prior permission can be in the form of a request from the client, or in the counselor asking the client's permission to offer it. Indirect forms of permission asking may also occur, such as a counselor statement that gives the client permission to disregard the advice ("This may or may not make sense to you"). \\
    \midrule
    \textsc{Advise Without Permission} & MI-inconsistent & The counselor gives advice, makes a suggestion, or offers a solution or possible action without client permission. \\
    \midrule
    \textsc{Affirm} & MI-consistent & The counselor says something positive or complimentary to the client. It may be in the form of expressed appreciation, confidence or reinforcement. \\
    \midrule
    \textsc{Confront} & MI-inconsistent & The counselor directly disagrees, argues, corrects, shames, blames, seeks to persuade, criticizes, judges, labels, moralizes, ridicules, or questions the client's honesty. These are the expert-like responses that have a particular negative-parent quality, an uneven power relationship accompanied by disapproval, disagreement, or negativity. There is a sense of “expert over-ride” of what the client says. \\
    \midrule
    \textsc{Direct} & MI-inconsistent & The counselor gives an order, command, or direction. The language is imperative. \\
    \midrule
    \textsc{Emphasize Control} & MI-consistent & The counselor directly acknowledges, honors, or emphasizes the client's freedom of choice, autonomy, personal responsibility, etc. There is no tone of blaming or faultfinding. \\
    \midrule
    \textsc{Facilitate} & Neutral & These are simple utterances that function as keep going acknowledgments. \\
    \midrule
    \textsc{Filler} & Neutral & This is a code for the few responses that are not codeable elsewhere: pleasantries, etc. It should not be used often. \\
    \midrule
    \textsc{Giving Information} & Neutral & The counselor gives information to the client, explains something, educates or provides feedback or discloses personal information. \\
    \midrule
    \textsc{Open Question} & MI-consistent & The counselor asks a question in order to gather information, understand, or elicit the client's story. Generally these begin with a question marker word: Who, What, Why, When, How, Where, etc. An open question is coded when the counselor asks a question that allows a wide range of possible answers. \\
    \midrule
    \textsc{Closed Quesiton} & Neutral & The counselor asks a question in order to gather information, understand, or elicit the client's story. Generally these begin with a question marker word: Who, What, Why, When, How, Where, etc. A closed question implies a short answer: Yes or no, a specific fact, a number, etc. \\
    \midrule
    \textsc{Raise Concern with Permission} & Neutral & The counselor points out a possible problem with a client's goal, plan, or intention with permission. Prior permission can be in the form of a request from the client or in the counselor asking the client's permission to offer it. Indirect forms of permission asking may also occur, such as a counselor’s statement that gives the client permission to disregard the counselor’s concern. \\
    \midrule
    \textsc{Raise Concern without Permission} & MI-inconsistent & The counselor points out a possible problem with a client's goal, plan, or intention without permission. \\
    \midrule
    \textsc{Simple Reflection} & MI-consistent & A reflection is a reflective listening statement made by the counselor in response to a client statement. Reflections capture and return to the client something that the client has said. Simple Reflections add little or no meaning or emphasis to what the client has said. \\
    \midrule
    \textsc{Complex Reflection} & MI-consistent & A reflection is a reflective listening statement made by the counselor in response to a client statement. Reflections capture and return to the client something that the client has said. Complex Reflections typically add substantial meaning or emphasis to what the client has said. \\
    \midrule
    \textsc{Reframe} & MI-consistent & The counselor suggests a different meaning for an experience expressed by the client, placing it in a new light. These generally have the quality of changing the emotional valence of meaning from negative to positive or from positive to negative. Reframes generally meet the criteria for Reflect but go further than adding meaning or emphasis by actually changing the valence of meaning and not just the depth. \\
    \midrule
    \textsc{Structure} & Neutral & To give information about what’s going to happen directly to the client throughout the course of treatment or within a study format, in this or subsequent sessions. To make a transition from one part of a session to another. \\
    \midrule
    \textsc{Support} & MI-consistent & These are generally sympathetic, compassionate, or understanding comments. They have the quality of agreeing or siding with the client. \\
    \midrule
    \textsc{Warn} & MI-inconsistent & The counselor provides a warning or threat, implying negative consequences unless the client takes a certain action. It may be a threat that the counselor has the perceived power to carry out or simply the prediction of a bad outcome if the client takes a certain course. \\
\bottomrule
\end{tabular}
\vspace{1em}
\caption{External MI strategies used to code agent responses for LLM-based coding. Strategies, definitions, and categories were sourced from MISC~\cite{moyers2003assessing}.}
\label{appendix:misc-table}
\end{table*}

%% file: sections/appendix/E_prompts.tex
\newpage
\section{Prompts}
\label{appendix:prompts}
In this section, we provide all of the prompts used by \textsf{GPTCoach}.

\input{sections/appendix/prompts/state_classifier}
\input{sections/appendix/prompts/dialogue_states}
\input{sections/appendix/prompts/strategy_prediction}
\input{sections/appendix/prompts/response_generation}
\input{sections/appendix/prompts/tool_call_prediction}
\input{sections/appendix/prompts/tool_call_generation}

%% file: sections/appendix/prompts/state_classifier.tex
\begin{figure}[ht]
\captionsetup{singlelinecheck=false}
\caption{Dialogue State Classification Prompt.}
\label{fig:dialogue-state-classification-prompt}
\begin{promptbox}
The following contains the dialogue history between a user and a health coach agent. Your task is to determine whether the agent has successfully completed the following task. Respond with only one word: `continue' or `completed'.
\newline
Task:
\newline
\texttt{\{DIALOGUE STATE PROMPT\}}
\end{promptbox}
\end{figure}

\begin{figure}[ht]
\captionsetup{singlelinecheck=false}
\caption{Dialogue State Classification Agent Prompt.}
\label{fig:dialogue-state-classification-agent-prompt}

\begin{promptbox}
Given this conversation history, respond only with `continue' or `completed' depending on whether the task has been successfully completed. 
\newline \newline
\texttt{\{DIALOGUE STATE PROMPT\}}
\end{promptbox}
\vspace{10pt}
\end{figure}

%% file: sections/appendix/prompts/dialogue_states.tex
\begin{figure}[ht]
\captionsetup{singlelinecheck=false}
\caption{Dialogue State Prompts. Instructions for each dialogue state was drawn from a validated health coaching program~\cite{king2002stanford}. 
\newline\newline
$^*$\textbf{Please note:} All dialogue state prompts below are copyrighted © 2025 by the Board of Trustees of the Leland Stanford Junior University. Any use or adaptation of the Stanford Active Choices materials requires prior written approval from the Stanford HEARTS Lab Faculty Director.
}
\label{fig:dialogue-state-prompts}
\end{figure}

\centering
\noindent \textbf{1. Onboarding$^*$}
\begin{promptbox}
    Your current task is to introduce yourself as a health coach if you have not already. After they have eased in, ask them for their name and age. At this point you should not be asking them to set goals or giving them advice.
\end{promptbox}

\noindent \textbf{2. Program$^*$} 
\begin{promptbox}
    Your current task is to welcome the client to the program and align expectations between them and you as the health coach. 
    \newline \newline
    First, inform the client that they will design their own physical activity plan, which should reflect their preferences, interests, and access to resources. With your assistance, they will determine the specifics of their activity plan. 
    \newline \newline
    Second, confirm their understanding and ask if they have any questions or concerns before getting started.
\end{promptbox}

\noindent \textbf{3. Past Experience$^*$} 
\begin{promptbox}
    Your current task is to acquire specific information about the client's past experiences with physical activity.
    \newline \newline
    First, you should ask the client what types of activities did they do and for how long?
    \newline \newline
    Second, you should ask them worked well about their previous exerices?
    \newline \newline
    Third, were there any difficulties they encountered?
    \newline \newline
    Why is this task important? \newline
    Understanding their history helps gauge their knowledge and tailor guidance, especially for beginners needing additional guidance on basics like endurance activities and warm-ups. 
    \newline \newline
    Handling certain situations \newline
    Some people may have had negative past experiences or faced several barriers with physical activity. This information can be used to their benefit now - their successful experiences can be used to address and overcome current barriers, such as discussing previous strategies for exercising during busy times.
\end{promptbox}

\noindent \textbf{4. Barriers$^*$}
\begin{promptbox}
    Your current task is to gather information regarding the barriers to physical activity that your client has faced in the past. 
    \newline \newline
    First, ask the client about their health or injury concerns. Follow up with specific questions if you require more information.
    \newline \newline
    Second, ask the client what their biggest obstacle is to doing physical activity. You should reference the conversation history to tailor this question to the client.
    \newline \newline
    Why is this task important? \newline
    Understanding their experiences and positive resources they have, such as knowledge, experience, equipment, or supportive friends, will aid their starting plan.
\end{promptbox}

\noindent \textbf{5. Motivation$^*$}
\begin{promptbox}
    Your current task is to determine what is motivating them to begin an exercise program now.
    \newline \newline
    First, ask the client what personal benefits do they hope to receive from regular exercise?
    \newline \newline
    Second, ask them what their main source of motivation is. Ask follow up questions if their response is vague.
    \newline \newline
    Third, ask them when they think in the long term, what kind of physical activity would they like to be able to do.
    \newline \newline
    Why is this task important? \newline
    This information will be referred to again and again during the course of the program, especially at times when the client may be struggling or losing sight of why they wanted to be more active.
\end{promptbox}

\noindent \textbf{6. Goal Setting$^*$} 
\begin{promptbox}
    Your current task is to help your client set a physical activity goal. 
    \newline \newline 
    First, help them set a short term goal, if they have not already identified one themselves.
    \newline \newline
    A good goal should adhere to the FITT (Frequency, Intensity, Time, Type) model to help them plan the specifics of an physical activity regimen. The goal the client identifies should adhere to the FITT model. \newline
    - Frequency: How many days of physical activity in the week? \newline
    - Intensity: Will it be light, moderate, or vigorous intensity? \newline
    - Time: How long will the physical activity session be? How many total minutes? What days of the week? What time of the day? \newline
    - Type: What kind of activities will the client do?
    \newline \newline
    You should assist the client in setting a FITT goal, asking one question at a time. 
    \newline \newline
    Let the client know that these goals can be changed as often as necessary. Encourage setting realistic goals and ask questions to probe if these goals are realistic, measurable, and specific, but don't tell the client what to do. Always provide justification for your suggestions.
    \newline \newline
    You have access to their health data using the `describe` and `visualize` functions. You should make use of this information to help them set realistic goals. 
    \newline \newline
    Why is this task important? \newline
    This will add to/build from the discussion of the resources or challenges they may have in store. Connecting their short term goal to larger motivations can help them stay motivated.
\end{promptbox}

\newpage
\noindent \textbf{7. Advice$^*$}
\begin{promptbox}
Your current task is to help the client overcome obstacles to their current goal.
\newline \newline
First, ask the client what resources they have available to reach their goals (e.g., available facilities, equipment, support). 
\newline \newline
Second, ask them if they anticipate any possible barriers or challenges.
\newline \newline
Third, ask them if they have any ideas for possible solutions. 
\newline \newline
As a facilitator, an important part of your job is tuning into the negative, self-destructive thoughts, helping the client become more aware of their negative influence on motivation. If the client expresses negative or self-defeating thoughts, suggest ways to replace negative thoughts with balanced, positive ones.
\newline \newline
Problem-solve with the client to make their activity more enjoyable baed on their circumstances, life-constraints and inferences from their health data.
\newline \newline
Problem: Discomfort \newline 
Reframing: Muscle soreness from inactivity is normal.\newline 
Solution: Walk lightly for 5 minutes before and after exercise. Consider light stretching.
\newline \newline 
Problem: Lack of Motivation \newline 
Reframing: It's common to have varying motivation levels.\newline 
Solution: Reflect on your goals and benefits of activity, reward progress, recall past motivations, and take incremental steps.
\newline \newline 
Problem: No Energy \newline 
Reframing: Exercise can boost energy levels.\newline 
Solution: Remember how revitalized you felt after previous walks.
\newline \newline 
Problem: No Time \newline 
Reframing: Inactive people have as much free time as those who exercise.\newline 
Solution: Schedule exercise, walk during breaks, and integrate walking into daily routines, like taking stairs or parking farther away.
\newline \newline 
Problem: Feeling Sick\newline 
Reframing: Illness can disrupt exercise routines.\newline 
Solution: Gradually increase activity in short sessions throughout the day.
\newline \newline 
Problem: Stress\newline 
Reframing: Exercise is an effective stress reliever.\newline 
Solution: Take brisk walks, reflecting on post-exercise relaxation.
\newline \newline 
Problem: Feeling Ashamed\newline 
Reframing: Starting to exercise can feel daunting.\newline 
Solution: Focus on health over others' opinions. Remind yourself each session will get easier.
\newline \newline 
Problem: Feeling Unsafe\newline 
Reframing: Concerns about safety can deter walking.\newline 
Solution: Follow safety tips like wearing visible clothing, walking in populated areas, and sharing your route with someone.
\newline \newline 
Problem: Feeling Unsupported\newline 
Reframing: Lack of social support can affect motivation.\newline 
Solution: Seek encouragement from friends or groups, join a walking club, and value personal exercise time.
\newline \newline 
Problem: Weather\newline 
Reframing: Don't let weather conditions stop you.\newline 
Solution: Walk indoors, dress appropriately for the weather, and stay hydrated.
\end{promptbox}

\noindent \textbf{8. Goodbye$^*$} 
\begin{promptbox}
Your current task is to answer any remaining questions, and wrap up the conversation after the client is done.
Before your client leaves, wish them good luck and that you have confidence in their ability to succeed. You can mention that you are always available to chat, but do not imply that there will be another scheduled session. This is the only session.
\end{promptbox}

%% file: sections/appendix/prompts/strategy_prediction.tex
\begin{figure}[ht]
\captionsetup{singlelinecheck=false}
\caption{System prompt used across all of our LLM prompts. Our prompt was inspired by the system prompt used by Chiu et al. \cite{chiu2024computational}, which we modified for physical activity coaching.}
\label{fig:system-prompt}

\begin{promptbox}
Act as if you're a professional health coach. You provide evidence-based support to clients seeking help with physical activity behavior change. You should maintain your health coach persona while responding. 
\newline \newline
You must maintain a friendly, warm, and empathetic tone. You must not give advice for medical or mental health concerns. Instead, you must respond empathetically and refer them to a professional. 
\newline \newline
Today's date is \texttt{\{DATE\_STRING\}}. Keep your responses brief and conversational.
\newline \newline
The following describes your instructions for the current stage of the conversation. Do not do anything that you are not asked to do. 
\end{promptbox}
\end{figure}

\begin{figure}[ht]
\captionsetup{singlelinecheck=false}
\caption{Prompt for strategy prediction instructions using strategies from motivational interviewing.}
\label{fig:strategy-prediction-instructions-prompt}

\begin{promptbox}
The following contains the dialogue history between a user and a health coach agent. Your task is to predict what strategy the agent should use to respond in the conversation.
\newline \newline
Please choose from one of 11 strategies described below (Advise with Permission, Affirm, Facilitate, Filler, Giving Information, Question, Raise Concern, Reflect, Reframe, Support, Structure) and output only one strategy from this list. 
\end{promptbox}
\end{figure}

\begin{figure}[ht]
\captionsetup{singlelinecheck=false}
\caption{Motivational Interviewing strategy codes and examples, selected and adapted from MISC~\cite{moyers2003assessing}.}
\label{fig:strategy-prediction-description-prompt}

\begin{promptbox}
Strategies
\newline \newline
Advise with Permission: Offering advice or suggestions after gaining permission, such as "Would it be alright if I suggested something?"
\newline \newline
Affirm: Positive reinforcement, appreciating client's efforts or strengths, such as "You're a very resourceful person."
\newline \newline
Facilitate: Simple responses to encourage further conversation, such as "Tell me more."
\newline \newline
Filler: General pleasantries or small talk, such as "Good morning, John."
\newline \newline
Giving Information: Provides explanations, feedback, or educational details, which can be personalized using health data, such as "Your heart rate was higher during today's workout."
\newline \newline
Question: Gathering information through open-ended questions, such as "How do you feel about that?"
\newline \newline
Raise Concern: Expressing concerns about the client's plans, such as "I'm worried about your plan to decrease your workout days."
\newline \newline
Reflect: Reflecting back the client's statements, simple or complex, such as  "You're looking for a relaxed gym environment." (simple) or "You see the benefits of exercise, yet find it unengaging." (complex)
\newline \newline
Reframe: Suggesting new perspectives on the client's experiences, such as reframing "nagging" as "concern."
\newline \newline
Support: Showing sympathy, compassion, or understanding, such as "That must have been difficult."
\newline \newline
Structure: Informing about session formats or transitions, such as "What we normally do is start by asking about your physical activity habits."
\end{promptbox}
\end{figure}

\begin{figure}[ht]
\captionsetup{singlelinecheck=false}
\caption{Motivational Interviewing strategy prediction agent prompt}
\label{fig:strategy-prediction-agent-prompt}

\begin{promptbox}
\texttt{\{DIALOGUE STATE PROMPT\}}
\newline \newline
Select one of the strategies from the list (\texttt{\{STRATEGIES\}}) to best achieve the given task while adhering to the natural flow of the dialogue. Output only one strategy from this list. 
\newline \newline
Strategy:
\end{promptbox}
\end{figure}

%% file: sections/appendix/prompts/response_generation.tex
\begin{figure}[ht]
\captionsetup{singlelinecheck=false}
\caption{Prompt for response generation instructions}
\label{fig:response-generation-instruction-prompts}

\begin{promptbox}
You will be given the dialogue history with the conversation between a user and a well being support agent acting as a health coach.  In order to complete this task, you should use the strategy specified. 
Given this strategy, please generate a response to the user. 
\end{promptbox}
\end{figure}

\begin{figure}[ht]
\captionsetup{singlelinecheck=false}
\caption{Prompt for few-shot tool call exampples}
\label{fig:response-generation-fewshot-tool-call-examples-prompts}

\begin{promptbox}
You are equipped to analyze and interpret sensor data from mobile phones and wearable devices. You have access to a function called describe and a function called visualize. When you call visualize, you will always see the output to describe. 
\newline \newline
Below are few examples of the input-output pairs for you to consider. Your job is to help people in interpreting this data. Always consider how this data relates to their broader life circumstances and physical activity goals. It is generally more insightful to look at long term trends than short term variations. You should keep in mind that the data may come from various sources and may not be fully accurate.
\newline \newline
\texttt{> describe(data\_source\_name="health.stepcount", start="2024-02-23 00:00:00", end="2024-02-23 23:59:59", granularity="day")
2024-02-23-00-00 to 2024-02-23-23-59: 10968.00 steps from Apple Watch (1 entries)}
\newline \newline 
\texttt{> describe(data\_source\_name="health.stepcount", start="2024-02-23 00:00:00", end="2024-02-23 23:59:59", granularity="hour")
2024-02-23-00-00 to 2024-02-23-00-59: 13.00 steps from iPhone (1 entries)
2024-02-23-01-00 to 2024-02-23-01-59: 34.00 steps from Apple Watch (1 entries)
2024-02-23-08-00 to 2024-02-23-08-59: 122.00 steps from Apple Watch (1 entries)
2024-02-23-09-00 to 2024-02-23-09-59: 988.00 steps from Apple Watch (19 entries)
... (output truncated)}
\newline \newline 
\texttt{> describe(data\_source\_name="health.workout", start="2024-03-01", end="2024-03-31", granularity="month")
 - cycling: 29 workouts, 21.14 mins/workout, 613.00 mins  (10h13m)  total
 - running: 7 workouts, 71.14 mins/workout, 497.96 mins  (8h17m)  total
 - walking: 50 workouts, 19.07 mins/workout, 953.44 mins  (15h53m)  total
 - traditionalStrengthTraining: 2 workouts, 64.31 mins/workout, 128.63 mins  (2h8m)  total
 - hiking: 2 workouts, 46.39 mins/workout, 92.79 mins  (1h32m)  total}
\newline \newline 
\texttt{visualize(data\_source\_name="health.stepcount", date="2024-03-01", granularity="month")
Returns the same output as \\ describe(data\_source\_name="health.stepcount", start="2024-03-01", end="2024-03-31", granularity="month") and shows a visualization to the user.}
\end{promptbox}
\end{figure}

\begin{figure}[ht]
\captionsetup{singlelinecheck=false}
\caption{Response Generation Agent Prompt.}
\label{fig:response-generation-agent-prompts}

\begin{promptbox}
\texttt{\{DIALOGUE STATE PROMPT\}} 
\newline \newline
The strategy you should use is:
\texttt{\{STRATEGY\_DESCRIPTION\}} 
\newline \newline
Output the response given this strategy. Keep your response brief. Only ask the client for one piece of information at a time. If your task includes asking multiple questions, break them up. If the user response is unrelated to the current task, acknowledge their response and nudge the conversation back to the current task.
\end{promptbox}
\end{figure}

%% file: sections/appendix/prompts/tool_call_prediction.tex
\begin{figure}[ht]
\captionsetup{singlelinecheck=false}
\caption{Prompts used for tool call prediction instructions.}
\label{fig:tool-call-prediction-instruction-prompts}

\begin{promptbox}
The following contains the dialogue history between a user and a health coach agent. The response is generated by an agent that did not use the describe or visualize functions. Your task is to decide whether this response should be augmented with the user’s health data. 
\newline \newline
You should respond 'yes' if the response from the agent is generic and could be more personalized to the client by looking at the health data. This is particularly useful when the client doesn't know when they should get physical activity, what kind of activities they could engage in, and their baseline levels of activity. It is not useful when the agent is explaining details about the program or is not giving specific advice. Respond 'no' otherwise.
\end{promptbox}
\end{figure}

\begin{figure}[ht]
\captionsetup{singlelinecheck=false}
\caption{Tool Call Prediction Agent Prompt.}
\label{fig:tool-call-prediction-agent-prompt}

\begin{promptbox}
\texttt{\{DIALOGUE STATE PROMPT\}} 
\newline \newline
The strategy being used in the conversation is:
\texttt{\{STRATEGY\_DESCRIPTION\}} 
\newline \newline
Given the current task, strategy, and the agent's response, your task is to detetmine whether the agent's response should be augmented with the user’s health data. 
\newline \newline
Respond with only one word: 'yes' or 'no'.
\end{promptbox}
\end{figure}

%% file: sections/appendix/prompts/tool_call_generation.tex
\begin{figure}[ht]
\captionsetup{singlelinecheck=false}
\caption{Prompts used for tool call generation instructions.}
\label{fig:tool-call-generation-instruction-prompts}

\begin{promptbox}
The following contains the dialogue history between a user and a health coach agent. The response is generated by an agent that did not use the describe or visualize functions. Another agent has decided that this response should be augmented with the user’s health data and use the function calls. Your task is to perform the appropriate tool call.
\end{promptbox}
\end{figure}

\begin{figure}[ht]
\captionsetup{singlelinecheck=false}
\caption{Tool Call Generation Agent Prompt.}
\label{fig:tool-call-generation-agent-prompt}

\begin{promptbox}
\texttt{\{DIALOGUE STATE PROMPT\}} 
\newline \newline
The strategy being used in the conversation is:
\texttt{\{STRATEGY\_DESCRIPTION\}} 
\newline \newline
Given this strategy and the agent's response, your task is to determine which tool call to use in the conversation to augment this agent's response.
\end{promptbox}
\end{figure}

%% file: main.bbl

\begin{thebibliography}{136}


\ifx \showCODEN    \undefined \def \showCODEN     #1{\unskip}     \fi
\ifx \showDOI      \undefined \def \showDOI       #1{#1}\fi
\ifx \showISBNx    \undefined \def \showISBNx     #1{\unskip}     \fi
\ifx \showISBNxiii \undefined \def \showISBNxiii  #1{\unskip}     \fi
\ifx \showISSN     \undefined \def \showISSN      #1{\unskip}     \fi
\ifx \showLCCN     \undefined \def \showLCCN      #1{\unskip}     \fi
\ifx \shownote     \undefined \def \shownote      #1{#1}          \fi
\ifx \showarticletitle \undefined \def \showarticletitle #1{#1}   \fi
\ifx \showURL      \undefined \def \showURL       {\relax}        \fi
\providecommand\bibfield[2]{#2}
\providecommand\bibinfo[2]{#2}
\providecommand\natexlab[1]{#1}
\providecommand\showeprint[2][]{arXiv:#2}

\bibitem[Abbasian et~al\mbox{.}(2023)]%
        {abbasian2023conversational}
\bibfield{author}{\bibinfo{person}{Mahyar Abbasian}, \bibinfo{person}{Iman
  Azimi}, \bibinfo{person}{Amir~M Rahmani}, {and} \bibinfo{person}{Ramesh
  Jain}.} \bibinfo{year}{2023}\natexlab{}.
\newblock \showarticletitle{Conversational Health Agents: A Personalized
  LLM-Powered Agent Framework}.
\newblock \bibinfo{journal}{\emph{arXiv preprint arXiv:2310.02374}}
  (\bibinfo{year}{2023}).
\newblock


\bibitem[Abbaspourazad et~al\mbox{.}(2024)]%
        {abbaspourazad2023large}
\bibfield{author}{\bibinfo{person}{Salar Abbaspourazad},
  \bibinfo{person}{Oussama Elachqar}, \bibinfo{person}{Andrew~C. Miller},
  \bibinfo{person}{Saba Emrani}, \bibinfo{person}{Udhyakumar Nallasamy}, {and}
  \bibinfo{person}{Ian Shapiro}.} \bibinfo{year}{2024}\natexlab{}.
\newblock \showarticletitle{Large-scale Training of Foundation Models for
  Wearable Biosignals}. In \bibinfo{booktitle}{\emph{The Twelfth International
  Conference on Learning Representations, {ICLR} 2024, Vienna, Austria, May
  7-11, 2024}}.
\newblock


\bibitem[Abdin et~al\mbox{.}(2024)]%
        {abdin2024phi}
\bibfield{author}{\bibinfo{person}{Marah Abdin}, \bibinfo{person}{Sam~Ade
  Jacobs}, \bibinfo{person}{Ammar~Ahmad Awan}, \bibinfo{person}{Jyoti Aneja},
  \bibinfo{person}{Ahmed Awadallah}, \bibinfo{person}{Hany Awadalla},
  \bibinfo{person}{Nguyen Bach}, \bibinfo{person}{Amit Bahree},
  \bibinfo{person}{Arash Bakhtiari}, \bibinfo{person}{Harkirat Behl},
  {et~al\mbox{.}}} \bibinfo{year}{2024}\natexlab{}.
\newblock \showarticletitle{Phi-3 Technical Report: A Highly Capable Language
  Model Locally on Your Phone}.
\newblock \bibinfo{journal}{\emph{arXiv preprint arXiv:2404.14219}}
  (\bibinfo{year}{2024}).
\newblock


\bibitem[Achiam et~al\mbox{.}(2023)]%
        {achiam2023gpt}
\bibfield{author}{\bibinfo{person}{Josh Achiam}, \bibinfo{person}{Steven
  Adler}, \bibinfo{person}{Sandhini Agarwal}, \bibinfo{person}{Lama Ahmad},
  \bibinfo{person}{Ilge Akkaya}, \bibinfo{person}{Florencia~Leoni Aleman},
  \bibinfo{person}{Diogo Almeida}, \bibinfo{person}{Janko Altenschmidt},
  \bibinfo{person}{Sam Altman}, \bibinfo{person}{Shyamal Anadkat},
  {et~al\mbox{.}}} \bibinfo{year}{2023}\natexlab{}.
\newblock \showarticletitle{GPT-4 technical report}.
\newblock \bibinfo{journal}{\emph{arXiv preprint arXiv:2303.08774}}
  (\bibinfo{year}{2023}).
\newblock


\bibitem[Aggarwal et~al\mbox{.}(2023)]%
        {aggarwal2023artificial}
\bibfield{author}{\bibinfo{person}{Abhishek Aggarwal},
  \bibinfo{person}{Cheuk~Chi Tam}, \bibinfo{person}{Dezhi Wu},
  \bibinfo{person}{Xiaoming Li}, {and} \bibinfo{person}{Shan Qiao}.}
  \bibinfo{year}{2023}\natexlab{}.
\newblock \showarticletitle{Artificial intelligence--based chatbots for
  promoting health behavioral changes: Systematic review}.
\newblock \bibinfo{journal}{\emph{Journal of Medical Internet Research}}
  \bibinfo{volume}{25} (\bibinfo{year}{2023}), \bibinfo{pages}{e40789}.
\newblock


\bibitem[Alizadeh et~al\mbox{.}(2024)]%
        {alizadeh2023llm}
\bibfield{author}{\bibinfo{person}{Keivan Alizadeh},
  \bibinfo{person}{Seyed~Iman Mirzadeh}, \bibinfo{person}{Dmitry Belenko},
  \bibinfo{person}{S. Khatamifard}, \bibinfo{person}{Minsik Cho},
  \bibinfo{person}{Carlo~C Del~Mundo}, \bibinfo{person}{Mohammad Rastegari},
  {and} \bibinfo{person}{Mehrdad Farajtabar}.} \bibinfo{year}{2024}\natexlab{}.
\newblock \showarticletitle{{LLM} in a flash: Efficient Large Language Model
  Inference with Limited Memory}. In \bibinfo{booktitle}{\emph{Proceedings of
  the 62nd Annual Meeting of the Association for Computational Linguistics
  (Volume 1: Long Papers)}}, \bibfield{editor}{\bibinfo{person}{Lun-Wei Ku},
  \bibinfo{person}{Andre Martins}, {and} \bibinfo{person}{Vivek Srikumar}}
  (Eds.). \bibinfo{publisher}{Association for Computational Linguistics},
  \bibinfo{address}{Bangkok, Thailand}, \bibinfo{pages}{12562--12584}.
\newblock
\urldef\tempurl%
\url{https://doi.org/10.18653/v1/2024.acl-long.678}
\showDOI{\tempurl}


\bibitem[Anderson et~al\mbox{.}(2007)]%
        {anderson2007shakra}
\bibfield{author}{\bibinfo{person}{Ian Anderson}, \bibinfo{person}{Julie
  Maitland}, \bibinfo{person}{Scott Sherwood}, \bibinfo{person}{Louise
  Barkhuus}, \bibinfo{person}{Matthew Chalmers}, \bibinfo{person}{Malcolm
  Hall}, \bibinfo{person}{Barry Brown}, {and} \bibinfo{person}{Henk Muller}.}
  \bibinfo{year}{2007}\natexlab{}.
\newblock \showarticletitle{Shakra: Tracking and sharing daily activity levels
  with unaugmented mobile phones}.
\newblock \bibinfo{journal}{\emph{Mobile Networks and Applications}}
  \bibinfo{volume}{12} (\bibinfo{year}{2007}), \bibinfo{pages}{185--199}.
\newblock


\bibitem[Andukuri et~al\mbox{.}(2024)]%
        {andukuri2024star}
\bibfield{author}{\bibinfo{person}{Chinmaya Andukuri},
  \bibinfo{person}{Jan-Philipp Fr{\"a}nken}, \bibinfo{person}{Tobias
  Gerstenberg}, {and} \bibinfo{person}{Noah~D Goodman}.}
  \bibinfo{year}{2024}\natexlab{}.
\newblock \showarticletitle{Star-gate: Teaching language models to ask
  clarifying questions}. In \bibinfo{booktitle}{\emph{Conference on Language
  Modeling (COLM)}}.
\newblock


\bibitem[Aseniero et~al\mbox{.}(2020)]%
        {aseniero2020activity}
\bibfield{author}{\bibinfo{person}{Bon~Adriel Aseniero},
  \bibinfo{person}{Charles Perin}, \bibinfo{person}{Wesley Willett},
  \bibinfo{person}{Anthony Tang}, {and} \bibinfo{person}{Sheelagh Carpendale}.}
  \bibinfo{year}{2020}\natexlab{}.
\newblock \showarticletitle{Activity River: Visualizing planned and logged
  personal activities for reflection}. In \bibinfo{booktitle}{\emph{Proceedings
  of the International Conference on Advanced Visual Interfaces}}.
  \bibinfo{pages}{1--9}.
\newblock


\bibitem[Baglivo et~al\mbox{.}(2023)]%
        {baglivo2023exploring}
\bibfield{author}{\bibinfo{person}{Francesco Baglivo}, \bibinfo{person}{Luigi
  De~Angelis}, \bibinfo{person}{Virginia Casigliani},
  \bibinfo{person}{Guglielmo Arzilli}, \bibinfo{person}{Gaetano~Pierpaolo
  Privitera}, {and} \bibinfo{person}{Caterina Rizzo}.}
  \bibinfo{year}{2023}\natexlab{}.
\newblock \showarticletitle{Exploring the Possible Use of AI Chatbots in Public
  Health Education: Feasibility Study}.
\newblock \bibinfo{journal}{\emph{JMIR Medical Education}}  \bibinfo{volume}{9}
  (\bibinfo{year}{2023}), \bibinfo{pages}{e51421}.
\newblock


\bibitem[Bai et~al\mbox{.}(2022)]%
        {bai2022training}
\bibfield{author}{\bibinfo{person}{Yuntao Bai}, \bibinfo{person}{Andy Jones},
  \bibinfo{person}{Kamal Ndousse}, \bibinfo{person}{Amanda Askell},
  \bibinfo{person}{Anna Chen}, \bibinfo{person}{Nova DasSarma},
  \bibinfo{person}{Dawn Drain}, \bibinfo{person}{Stanislav Fort},
  \bibinfo{person}{Deep Ganguli}, \bibinfo{person}{Tom Henighan},
  {et~al\mbox{.}}} \bibinfo{year}{2022}\natexlab{}.
\newblock \showarticletitle{Training a helpful and harmless assistant with
  reinforcement learning from human feedback}.
\newblock \bibinfo{journal}{\emph{arXiv preprint arXiv:2204.05862}}
  (\bibinfo{year}{2022}).
\newblock


\bibitem[Bandura(1999)]%
        {bandura1999social}
\bibfield{author}{\bibinfo{person}{Albert Bandura}.}
  \bibinfo{year}{1999}\natexlab{}.
\newblock \showarticletitle{Social cognitive theory: An agentic perspective}.
\newblock \bibinfo{journal}{\emph{Asian Journal of Social Psychology}}
  \bibinfo{volume}{2}, \bibinfo{number}{1} (\bibinfo{year}{1999}),
  \bibinfo{pages}{21--41}.
\newblock


\bibitem[Baumer et~al\mbox{.}(2014)]%
        {baumer2014reviewing}
\bibfield{author}{\bibinfo{person}{Eric~PS Baumer}, \bibinfo{person}{Vera
  Khovanskaya}, \bibinfo{person}{Mark Matthews}, \bibinfo{person}{Lindsay
  Reynolds}, \bibinfo{person}{Victoria Schwanda~Sosik}, {and}
  \bibinfo{person}{Geri Gay}.} \bibinfo{year}{2014}\natexlab{}.
\newblock \showarticletitle{Reviewing Reflection: On the use of reflection in
  interactive system design}. In \bibinfo{booktitle}{\emph{Proceedings of the
  2014 Conference on Designing Interactive Systems}}. \bibinfo{pages}{93--102}.
\newblock


\bibitem[Bentvelzen et~al\mbox{.}(2021)]%
        {bentvelzen2021development}
\bibfield{author}{\bibinfo{person}{Marit Bentvelzen}, \bibinfo{person}{Jasmin
  Niess}, \bibinfo{person}{Miko{\l}aj~P Wo{\'z}niak}, {and}
  \bibinfo{person}{Pawe{\l}~W Wo{\'z}niak}.} \bibinfo{year}{2021}\natexlab{}.
\newblock \showarticletitle{The Development and Validation of the
  Technology-Supported Reflection Inventory}. In
  \bibinfo{booktitle}{\emph{Proceedings of the 2021 CHI Conference on Human
  Factors in Computing Systems}}. \bibinfo{pages}{1--8}.
\newblock


\bibitem[Bentvelzen et~al\mbox{.}(2022)]%
        {bentvelzen2022revisiting}
\bibfield{author}{\bibinfo{person}{Marit Bentvelzen},
  \bibinfo{person}{Pawe{\l}~W Wo{\'z}niak}, \bibinfo{person}{Pia~SF Herbes},
  \bibinfo{person}{Evropi Stefanidi}, {and} \bibinfo{person}{Jasmin Niess}.}
  \bibinfo{year}{2022}\natexlab{}.
\newblock \showarticletitle{Revisiting Reflection in HCI: Four Design Resources
  for Technologies that Support Reflection}.
\newblock \bibinfo{journal}{\emph{Proceedings of the ACM on Interactive,
  Mobile, Wearable and Ubiquitous Technologies}} \bibinfo{volume}{6},
  \bibinfo{number}{1} (\bibinfo{year}{2022}), \bibinfo{pages}{1--27}.
\newblock


\bibitem[Bickmore and Giorgino(2006)]%
        {bickmore2006health}
\bibfield{author}{\bibinfo{person}{Timothy~W. Bickmore} {and}
  \bibinfo{person}{Toni Giorgino}.} \bibinfo{year}{2006}\natexlab{}.
\newblock \showarticletitle{Health dialog systems for patients and consumers}.
\newblock \bibinfo{journal}{\emph{Journal of Biomedical Informatics}}
  \bibinfo{volume}{39}, \bibinfo{number}{5} (\bibinfo{year}{2006}),
  \bibinfo{pages}{556--571}.
\newblock


\bibitem[Bickmore and Picard(2005)]%
        {bickmore2005establishing}
\bibfield{author}{\bibinfo{person}{Timothy~W. Bickmore} {and}
  \bibinfo{person}{Rosalind~W. Picard}.} \bibinfo{year}{2005}\natexlab{}.
\newblock \showarticletitle{Establishing and maintaining long-term
  human-computer relationships}.
\newblock \bibinfo{journal}{\emph{ACM Transactions on Computer-Human
  Interaction (TOCHI)}} \bibinfo{volume}{12}, \bibinfo{number}{2}
  (\bibinfo{date}{June} \bibinfo{year}{2005}), \bibinfo{pages}{293–327}.
\newblock
\showISSN{1073-0516}
\urldef\tempurl%
\url{https://doi.org/10.1145/1067860.1067867}
\showDOI{\tempurl}


\bibitem[Bickmore et~al\mbox{.}(2011)]%
        {bickmore2011reusable}
\bibfield{author}{\bibinfo{person}{Timothy~W. Bickmore},
  \bibinfo{person}{Daniel Schulman}, {and} \bibinfo{person}{Candace~L.
  Sidner}.} \bibinfo{year}{2011}\natexlab{}.
\newblock \showarticletitle{A reusable framework for health counseling dialogue
  systems based on a behavioral medicine ontology}.
\newblock \bibinfo{journal}{\emph{Journal of Biomedical Informatics}}
  \bibinfo{volume}{44}, \bibinfo{number}{2} (\bibinfo{year}{2011}),
  \bibinfo{pages}{183--197}.
\newblock


\bibitem[Blodgett et~al\mbox{.}(2020)]%
        {blodgett2020language}
\bibfield{author}{\bibinfo{person}{Su~Lin Blodgett}, \bibinfo{person}{Solon
  Barocas}, \bibinfo{person}{Hal Daum{\'e}~III}, {and} \bibinfo{person}{Hanna
  Wallach}.} \bibinfo{year}{2020}\natexlab{}.
\newblock \showarticletitle{Language (Technology) is Power: A Critical Survey
  of {\textquotedblleft}Bias{\textquotedblright} in {NLP}}. In
  \bibinfo{booktitle}{\emph{Proceedings of the 58th Annual Meeting of the
  Association for Computational Linguistics}},
  \bibfield{editor}{\bibinfo{person}{Dan Jurafsky}, \bibinfo{person}{Joyce
  Chai}, \bibinfo{person}{Natalie Schluter}, {and} \bibinfo{person}{Joel
  Tetreault}} (Eds.). \bibinfo{publisher}{Association for Computational
  Linguistics}, \bibinfo{address}{Online}, \bibinfo{pages}{5454--5476}.
\newblock
\urldef\tempurl%
\url{https://doi.org/10.18653/v1/2020.acl-main.485}
\showDOI{\tempurl}


\bibitem[Bommasani et~al\mbox{.}(2022)]%
        {bommasani2022opportunities}
\bibfield{author}{\bibinfo{person}{Rishi Bommasani}, \bibinfo{person}{Drew~A
  Hudson}, \bibinfo{person}{Ehsan Adeli}, \bibinfo{person}{Russ Altman},
  \bibinfo{person}{Simran Arora}, \bibinfo{person}{Sydney von Arx},
  \bibinfo{person}{Michael~S Bernstein}, \bibinfo{person}{Jeannette Bohg},
  \bibinfo{person}{Antoine Bosselut}, \bibinfo{person}{Emma Brunskill},
  {et~al\mbox{.}}} \bibinfo{year}{2022}\natexlab{}.
\newblock \bibinfo{title}{On the Opportunities and Risks of Foundation Models}.
\newblock
\newblock


\bibitem[Braun and Clarke(2012)]%
        {braun2012thematic}
\bibfield{author}{\bibinfo{person}{Virginia Braun} {and}
  \bibinfo{person}{Victoria Clarke}.} \bibinfo{year}{2012}\natexlab{}.
\newblock \bibinfo{booktitle}{\emph{Thematic analysis.}}
\newblock \bibinfo{publisher}{American Psychological Association}.
\newblock


\bibitem[Castro et~al\mbox{.}(2011)]%
        {castro2011physical}
\bibfield{author}{\bibinfo{person}{Cynthia~M Castro}, \bibinfo{person}{Leslie~A
  Pruitt}, \bibinfo{person}{Matthew~P Buman}, {and} \bibinfo{person}{Abby~C
  King}.} \bibinfo{year}{2011}\natexlab{}.
\newblock \showarticletitle{Physical activity program delivery by professionals
  versus volunteers: The TEAM randomized trial.}
\newblock \bibinfo{journal}{\emph{Health Psychology}} \bibinfo{volume}{30},
  \bibinfo{number}{3} (\bibinfo{year}{2011}), \bibinfo{pages}{285}.
\newblock


\bibitem[Chang et~al\mbox{.}(2024)]%
        {chang2024red}
\bibfield{author}{\bibinfo{person}{Crystal Tin-Tin Chang},
  \bibinfo{person}{Hodan Farah}, \bibinfo{person}{Haiwen Gui},
  \bibinfo{person}{Shawheen~Justin Rezaei}, \bibinfo{person}{Charbel
  Bou-Khalil}, \bibinfo{person}{Ye-Jean Park}, \bibinfo{person}{Akshay
  Swaminathan}, \bibinfo{person}{Jesutofunmi~A Omiye}, \bibinfo{person}{Akaash
  Kolluri}, \bibinfo{person}{Akash Chaurasia}, {et~al\mbox{.}}}
  \bibinfo{year}{2024}\natexlab{}.
\newblock \showarticletitle{Red Teaming Large Language Models in Medicine:
  Real-World Insights on Model Behavior}.
\newblock \bibinfo{journal}{\emph{medRxiv}} (\bibinfo{year}{2024}),
  \bibinfo{pages}{2024--04}.
\newblock


\bibitem[Chaszczewicz et~al\mbox{.}(2024)]%
        {chaszczewicz2024multi}
\bibfield{author}{\bibinfo{person}{Alicja Chaszczewicz}, \bibinfo{person}{Raj
  Shah}, \bibinfo{person}{Ryan Louie}, \bibinfo{person}{Bruce Arnow},
  \bibinfo{person}{Robert Kraut}, {and} \bibinfo{person}{Diyi Yang}.}
  \bibinfo{year}{2024}\natexlab{}.
\newblock \showarticletitle{Multi-Level Feedback Generation with Large Language
  Models for Empowering Novice Peer Counselors}. In
  \bibinfo{booktitle}{\emph{Proceedings of the 62nd Annual Meeting of the
  Association for Computational Linguistics (Volume 1: Long Papers)}},
  \bibfield{editor}{\bibinfo{person}{Lun-Wei Ku}, \bibinfo{person}{Andre
  Martins}, {and} \bibinfo{person}{Vivek Srikumar}} (Eds.).
  \bibinfo{publisher}{Association for Computational Linguistics},
  \bibinfo{address}{Bangkok, Thailand}, \bibinfo{pages}{4130--4161}.
\newblock
\urldef\tempurl%
\url{https://doi.org/10.18653/v1/2024.acl-long.227}
\showDOI{\tempurl}


\bibitem[Chiu et~al\mbox{.}(2024a)]%
        {chiu2024culturalteaming}
\bibfield{author}{\bibinfo{person}{Yu~Ying Chiu}, \bibinfo{person}{Liwei
  Jiang}, \bibinfo{person}{Maria Antoniak}, \bibinfo{person}{Chan~Young Park},
  \bibinfo{person}{Shuyue~Stella Li}, \bibinfo{person}{Mehar Bhatia},
  \bibinfo{person}{Sahithya Ravi}, \bibinfo{person}{Yulia Tsvetkov},
  \bibinfo{person}{Vered Shwartz}, {and} \bibinfo{person}{Yejin Choi}.}
  \bibinfo{year}{2024}\natexlab{a}.
\newblock \showarticletitle{CulturalTeaming: AI-Assisted Interactive
  Red-Teaming for Challenging LLMs' (Lack of) Multicultural Knowledge}.
\newblock \bibinfo{journal}{\emph{arXiv preprint arXiv:2404.06664}}
  (\bibinfo{year}{2024}).
\newblock


\bibitem[Chiu et~al\mbox{.}(2024b)]%
        {chiu2024computational}
\bibfield{author}{\bibinfo{person}{Yu~Ying Chiu}, \bibinfo{person}{Ashish
  Sharma}, \bibinfo{person}{Inna~Wanyin Lin}, {and} \bibinfo{person}{Tim
  Althoff}.} \bibinfo{year}{2024}\natexlab{b}.
\newblock \showarticletitle{A Computational Framework for Behavioral Assessment
  of LLM Therapists}.
\newblock \bibinfo{journal}{\emph{arXiv preprint arXiv:2401.00820}}
  (\bibinfo{year}{2024}).
\newblock


\bibitem[Cho et~al\mbox{.}(2022)]%
        {cho2022reflection}
\bibfield{author}{\bibinfo{person}{Janghee Cho}, \bibinfo{person}{Tian Xu},
  \bibinfo{person}{Abigail Zimmermann-Niefield}, {and} \bibinfo{person}{Stephen
  Voida}.} \bibinfo{year}{2022}\natexlab{}.
\newblock \showarticletitle{Reflection in theory and reflection in practice: An
  exploration of the gaps in reflection support among personal informatics
  apps}. In \bibinfo{booktitle}{\emph{Proceedings of the 2022 CHI Conference on
  Human Factors in Computing Systems}}. \bibinfo{pages}{1--23}.
\newblock


\bibitem[Choe et~al\mbox{.}(2015)]%
        {choe2015characterizing}
\bibfield{author}{\bibinfo{person}{Eun~Kyoung Choe}, \bibinfo{person}{Bongshin
  Lee}, {and} \bibinfo{person}{m.c. schraefel}.}
  \bibinfo{year}{2015}\natexlab{}.
\newblock \showarticletitle{Characterizing visualization insights from
  quantified selfers' personal data presentations}.
\newblock \bibinfo{journal}{\emph{IEEE Computer Graphics and Applications}}
  \bibinfo{volume}{35}, \bibinfo{number}{4} (\bibinfo{year}{2015}),
  \bibinfo{pages}{28--37}.
\newblock


\bibitem[Choe et~al\mbox{.}(2017)]%
        {choe2017understanding}
\bibfield{author}{\bibinfo{person}{Eun~Kyoung Choe}, \bibinfo{person}{Bongshin
  Lee}, \bibinfo{person}{Haining Zhu}, \bibinfo{person}{Nathalie~Henry Riche},
  {and} \bibinfo{person}{Dominikus Baur}.} \bibinfo{year}{2017}\natexlab{}.
\newblock \showarticletitle{Understanding Self-reflection: How people reflect
  on personal data through visual data exploration}. In
  \bibinfo{booktitle}{\emph{Proceedings of the 11th EAI International
  Conference on Pervasive Computing Technologies for Healthcare}}.
  \bibinfo{pages}{173--182}.
\newblock


\bibitem[Choe et~al\mbox{.}(2014)]%
        {choe2014understanding}
\bibfield{author}{\bibinfo{person}{Eun~Kyoung Choe}, \bibinfo{person}{Nicole~B
  Lee}, \bibinfo{person}{Bongshin Lee}, \bibinfo{person}{Wanda Pratt}, {and}
  \bibinfo{person}{Julie~A Kientz}.} \bibinfo{year}{2014}\natexlab{}.
\newblock \showarticletitle{Understanding quantified-selfers' practices in
  collecting and exploring personal data}. In
  \bibinfo{booktitle}{\emph{Proceedings of the SIGCHI Conference on Human
  Factors in Computing Systems}}. \bibinfo{pages}{1143--1152}.
\newblock


\bibitem[Chung et~al\mbox{.}(2015)]%
        {chung2015more}
\bibfield{author}{\bibinfo{person}{Chia-Fang Chung}, \bibinfo{person}{Jonathan
  Cook}, \bibinfo{person}{Elizabeth Bales}, \bibinfo{person}{Jasmine Zia},
  {and} \bibinfo{person}{Sean~A Munson}.} \bibinfo{year}{2015}\natexlab{}.
\newblock \showarticletitle{More than telemonitoring: Health provider use and
  nonuse of life-log data in irritable bowel syndrome and weight management}.
\newblock \bibinfo{journal}{\emph{Journal of Medical Internet Research}}
  \bibinfo{volume}{17}, \bibinfo{number}{8} (\bibinfo{year}{2015}),
  \bibinfo{pages}{e203}.
\newblock


\bibitem[Cicchetti and Sparrow(1981)]%
        {cicchetti1981developing}
\bibfield{author}{\bibinfo{person}{Domenic~V Cicchetti} {and}
  \bibinfo{person}{Sara~A Sparrow}.} \bibinfo{year}{1981}\natexlab{}.
\newblock \showarticletitle{Developing criteria for establishing interrater
  reliability of specific items: Applications to assessment of adaptive
  behavior.}
\newblock \bibinfo{journal}{\emph{American Journal of Mental Deficiency}}
  \bibinfo{volume}{86}, \bibinfo{number}{2} (\bibinfo{year}{1981}),
  \bibinfo{pages}{127--137}.
\newblock


\bibitem[Clavel et~al\mbox{.}(2018)]%
        {clavel2018wenner}
\bibfield{author}{\bibinfo{person}{C{\'e}line Clavel}, \bibinfo{person}{Steve
  Whittaker}, \bibinfo{person}{Ana{\"\i}s Blacodon}, {and}
  \bibinfo{person}{Jean-Claude Martin}.} \bibinfo{year}{2018}\natexlab{}.
\newblock \showarticletitle{WEnner: A theoretically motivated approach for
  tailored coaching about physical activity}. In
  \bibinfo{booktitle}{\emph{Proceedings of the 2018 ACM International Joint
  Conference and 2018 International Symposium on Pervasive and Ubiquitous
  Computing and Wearable Computers}}. \bibinfo{pages}{1669--1675}.
\newblock


\bibitem[Cochran and Cochran(2015)]%
        {cochran2015heart}
\bibfield{author}{\bibinfo{person}{Jeff~L Cochran} {and}
  \bibinfo{person}{Nancy~H Cochran}.} \bibinfo{year}{2015}\natexlab{}.
\newblock \bibinfo{booktitle}{\emph{The heart of counseling: Counseling skills
  through therapeutic relationships}}.
\newblock \bibinfo{publisher}{Routledge}.
\newblock


\bibitem[Cohen and Sherman(2014)]%
        {cohen2014psychology}
\bibfield{author}{\bibinfo{person}{Geoffrey~L Cohen} {and}
  \bibinfo{person}{David~K Sherman}.} \bibinfo{year}{2014}\natexlab{}.
\newblock \showarticletitle{The psychology of change: Self-affirmation and
  social psychological intervention}.
\newblock \bibinfo{journal}{\emph{Annual Review of Psychology}}
  \bibinfo{volume}{65} (\bibinfo{year}{2014}), \bibinfo{pages}{333--371}.
\newblock


\bibitem[Consolvo et~al\mbox{.}(2008)]%
        {consolvo2008activity}
\bibfield{author}{\bibinfo{person}{Sunny Consolvo}, \bibinfo{person}{David~W
  McDonald}, \bibinfo{person}{Tammy Toscos}, \bibinfo{person}{Mike~Y Chen},
  \bibinfo{person}{Jon Froehlich}, \bibinfo{person}{Beverly Harrison},
  \bibinfo{person}{Predrag Klasnja}, \bibinfo{person}{Anthony LaMarca},
  \bibinfo{person}{Louis LeGrand}, \bibinfo{person}{Ryan Libby},
  {et~al\mbox{.}}} \bibinfo{year}{2008}\natexlab{}.
\newblock \showarticletitle{Activity sensing in the wild: A field trial of
  ubifit garden}. In \bibinfo{booktitle}{\emph{Proceedings of the SIGCHI
  Conference on Human Factors in Computing Systems}}.
  \bibinfo{pages}{1797--1806}.
\newblock


\bibitem[Cosentino et~al\mbox{.}(2024)]%
        {cosentino2024towards}
\bibfield{author}{\bibinfo{person}{Justin Cosentino},
  \bibinfo{person}{Anastasiya Belyaeva}, \bibinfo{person}{Xin Liu},
  \bibinfo{person}{Nicholas~A Furlotte}, \bibinfo{person}{Zhun Yang},
  \bibinfo{person}{Chace Lee}, \bibinfo{person}{Erik Schenck},
  \bibinfo{person}{Yojan Patel}, \bibinfo{person}{Jian Cui},
  \bibinfo{person}{Logan~Douglas Schneider}, {et~al\mbox{.}}}
  \bibinfo{year}{2024}\natexlab{}.
\newblock \showarticletitle{Towards a Personal Health Large Language Model}.
\newblock \bibinfo{journal}{\emph{arXiv preprint arXiv:2406.06474}}
  (\bibinfo{year}{2024}).
\newblock


\bibitem[Craig et~al\mbox{.}(2003)]%
        {craig2003international}
\bibfield{author}{\bibinfo{person}{Cora~L Craig}, \bibinfo{person}{Alison~L
  Marshall}, \bibinfo{person}{Michael Sj{\"o}str{\"o}m},
  \bibinfo{person}{Adrian~E Bauman}, \bibinfo{person}{Michael~L Booth},
  \bibinfo{person}{Barbara~E Ainsworth}, \bibinfo{person}{Michael Pratt},
  \bibinfo{person}{ULF Ekelund}, \bibinfo{person}{Agneta Yngve},
  \bibinfo{person}{James~F Sallis}, {et~al\mbox{.}}}
  \bibinfo{year}{2003}\natexlab{}.
\newblock \showarticletitle{International physical activity questionnaire:
  12-country reliability and validity}.
\newblock \bibinfo{journal}{\emph{Medicine \& Science in Sports \& Exercise}}
  \bibinfo{volume}{35}, \bibinfo{number}{8} (\bibinfo{year}{2003}),
  \bibinfo{pages}{1381--1395}.
\newblock


\bibitem[Crum and Langer(2007)]%
        {crum2007mind}
\bibfield{author}{\bibinfo{person}{Alia~J Crum} {and} \bibinfo{person}{Ellen~J
  Langer}.} \bibinfo{year}{2007}\natexlab{}.
\newblock \showarticletitle{Mind-set matters: Exercise and the placebo effect}.
\newblock \bibinfo{journal}{\emph{Psychological Science}} \bibinfo{volume}{18},
  \bibinfo{number}{2} (\bibinfo{year}{2007}), \bibinfo{pages}{165--171}.
\newblock


\bibitem[Cuadra et~al\mbox{.}(2024)]%
        {cuadra2024illusion}
\bibfield{author}{\bibinfo{person}{Andrea Cuadra}, \bibinfo{person}{Maria
  Wang}, \bibinfo{person}{Lynn~Andrea Stein}, \bibinfo{person}{Malte~F Jung},
  \bibinfo{person}{Nicola Dell}, \bibinfo{person}{Deborah Estrin}, {and}
  \bibinfo{person}{James~A Landay}.} \bibinfo{year}{2024}\natexlab{}.
\newblock \showarticletitle{The Illusion of Empathy? Notes on Displays of
  Emotion in Human-Computer Interaction}. In \bibinfo{booktitle}{\emph{ACM
  Conference on Human Factors in Computing Systems (CHI)}}.
\newblock


\bibitem[De~Choudhury et~al\mbox{.}(2023)]%
        {de2023benefits}
\bibfield{author}{\bibinfo{person}{Munmun De~Choudhury},
  \bibinfo{person}{Sachin~R Pendse}, {and} \bibinfo{person}{Neha Kumar}.}
  \bibinfo{year}{2023}\natexlab{}.
\newblock \showarticletitle{Benefits and harms of large language models in
  digital mental health}.
\newblock \bibinfo{journal}{\emph{arXiv preprint arXiv:2311.14693}}
  (\bibinfo{year}{2023}).
\newblock


\bibitem[Englhardt et~al\mbox{.}(2024)]%
        {englhardt2023classification}
\bibfield{author}{\bibinfo{person}{Zachary Englhardt},
  \bibinfo{person}{Chengqian Ma}, \bibinfo{person}{Margaret~E. Morris},
  \bibinfo{person}{Chun-Cheng Chang}, \bibinfo{person}{Xuhai~"Orson" Xu},
  \bibinfo{person}{Lianhui Qin}, \bibinfo{person}{Daniel McDuff},
  \bibinfo{person}{Xin Liu}, \bibinfo{person}{Shwetak Patel}, {and}
  \bibinfo{person}{Vikram Iyer}.} \bibinfo{year}{2024}\natexlab{}.
\newblock \showarticletitle{From Classification to Clinical Insights: Towards
  Analyzing and Reasoning About Mobile and Behavioral Health Data With Large
  Language Models}.
\newblock \bibinfo{journal}{\emph{Proceedings of the ACM on Interactive,
  Mobile, Wearable and Ubiquitous Technologies}} \bibinfo{volume}{8},
  \bibinfo{number}{2}, Article \bibinfo{articleno}{56} (\bibinfo{date}{May}
  \bibinfo{year}{2024}), \bibinfo{numpages}{25}~pages.
\newblock
\urldef\tempurl%
\url{https://doi.org/10.1145/3659604}
\showDOI{\tempurl}


\bibitem[Epstein et~al\mbox{.}(2020)]%
        {epstein2020mapping}
\bibfield{author}{\bibinfo{person}{Daniel~A Epstein}, \bibinfo{person}{Clara
  Caldeira}, \bibinfo{person}{Mayara~Costa Figueiredo}, \bibinfo{person}{Xi
  Lu}, \bibinfo{person}{Lucas~M Silva}, \bibinfo{person}{Lucretia Williams},
  \bibinfo{person}{Jong~Ho Lee}, \bibinfo{person}{Qingyang Li},
  \bibinfo{person}{Simran Ahuja}, \bibinfo{person}{Qiuer Chen},
  {et~al\mbox{.}}} \bibinfo{year}{2020}\natexlab{}.
\newblock \showarticletitle{Mapping and taking stock of the personal
  informatics literature}.
\newblock \bibinfo{journal}{\emph{Proceedings of the ACM on Interactive,
  Mobile, Wearable and Ubiquitous Technologies}} \bibinfo{volume}{4},
  \bibinfo{number}{4} (\bibinfo{year}{2020}), \bibinfo{pages}{1--38}.
\newblock


\bibitem[Fang et~al\mbox{.}({[n.\,d.]})]%
        {fang2024physiollm}
\bibfield{author}{\bibinfo{person}{Cathy~Mengying Fang},
  \bibinfo{person}{Valdemar Danry}, \bibinfo{person}{Nathan Whitmore},
  \bibinfo{person}{Andria Bao}, \bibinfo{person}{Andrew Hutchison},
  \bibinfo{person}{Cayden Pierce}, {and} \bibinfo{person}{Patricia Maes}.}
  \bibinfo{year}{[n.\,d.]}\natexlab{}.
\newblock \showarticletitle{PhysioLLM: Supporting Personalized Health Insights
  with Wearables and Large Language Models}. In
  \bibinfo{booktitle}{\emph{IEEE-EMBS International Conference on Biomedical
  and Health Informatics}}.
\newblock


\bibitem[for Disease~Control and Prevention(2022)]%
        {cdc2022}
\bibfield{author}{\bibinfo{person}{Centers for Disease~Control} {and}
  \bibinfo{person}{Prevention}.} \bibinfo{year}{2022}\natexlab{}.
\newblock \bibinfo{title}{Physical Activity}.
\newblock
\newblock
\urldef\tempurl%
\url{https://www.cdc.gov/physicalactivity/index.html}
\showURL{%
\tempurl}


\bibitem[Ganguli et~al\mbox{.}(2022)]%
        {ganguli2022red}
\bibfield{author}{\bibinfo{person}{Deep Ganguli}, \bibinfo{person}{Liane
  Lovitt}, \bibinfo{person}{Jackson Kernion}, \bibinfo{person}{Amanda Askell},
  \bibinfo{person}{Yuntao Bai}, \bibinfo{person}{Saurav Kadavath},
  \bibinfo{person}{Ben Mann}, \bibinfo{person}{Ethan Perez},
  \bibinfo{person}{Nicholas Schiefer}, \bibinfo{person}{Kamal Ndousse},
  {et~al\mbox{.}}} \bibinfo{year}{2022}\natexlab{}.
\newblock \showarticletitle{Red teaming language models to reduce harms:
  Methods, scaling behaviors, and lessons learned}.
\newblock \bibinfo{journal}{\emph{arXiv preprint arXiv:2209.07858}}
  (\bibinfo{year}{2022}).
\newblock


\bibitem[Google(2024)]%
        {gemma2024}
\bibfield{author}{\bibinfo{person}{Google}.} \bibinfo{year}{2024}\natexlab{}.
\newblock \bibinfo{title}{Gemma - A Family of Lightweight, State-of-the Art
  Open Models from Google}.
\newblock
\newblock
\urldef\tempurl%
\url{https://ai.google.dev/gemma}
\showURL{%
\tempurl}


\bibitem[Grammel et~al\mbox{.}(2010)]%
        {grammel2010information}
\bibfield{author}{\bibinfo{person}{Lars Grammel}, \bibinfo{person}{Melanie
  Tory}, {and} \bibinfo{person}{Margaret-Anne Storey}.}
  \bibinfo{year}{2010}\natexlab{}.
\newblock \showarticletitle{How information visualization novices construct
  visualizations}.
\newblock \bibinfo{journal}{\emph{IEEE Transactions on Visualization and
  Computer Graphics}} \bibinfo{volume}{16}, \bibinfo{number}{6}
  (\bibinfo{year}{2010}), \bibinfo{pages}{943--952}.
\newblock


\bibitem[Hallgren(2012)]%
        {hallgren2012computing}
\bibfield{author}{\bibinfo{person}{Kevin~A Hallgren}.}
  \bibinfo{year}{2012}\natexlab{}.
\newblock \showarticletitle{Computing inter-rater reliability for observational
  data: An overview and tutorial}.
\newblock \bibinfo{journal}{\emph{Tutorials in Quantitative Methods for
  Psychology}} \bibinfo{volume}{8}, \bibinfo{number}{1} (\bibinfo{year}{2012}),
  \bibinfo{pages}{23}.
\newblock


\bibitem[{Health Level Seven International}(2023)]%
        {HL7FHIR}
\bibfield{author}{\bibinfo{person}{{Health Level Seven International}}.}
  \bibinfo{year}{2023}\natexlab{}.
\newblock \bibinfo{title}{{HL7 FHIR Release 4}}.
\newblock \bibinfo{howpublished}{\url{https://www.hl7.org/fhir/}}.
\newblock


\bibitem[Hegde et~al\mbox{.}(2024)]%
        {hegde2024infusing}
\bibfield{author}{\bibinfo{person}{Narayan Hegde}, \bibinfo{person}{Madhurima
  Vardhan}, \bibinfo{person}{Deepak Nathani}, \bibinfo{person}{Emily
  Rosenzweig}, \bibinfo{person}{Cathy Speed}, \bibinfo{person}{Alan
  Karthikesalingam}, {and} \bibinfo{person}{Martin Seneviratne}.}
  \bibinfo{year}{2024}\natexlab{}.
\newblock \showarticletitle{Infusing behavior science into large language
  models for activity coaching}.
\newblock \bibinfo{journal}{\emph{PLOS Digital Health}} \bibinfo{volume}{3},
  \bibinfo{number}{4} (\bibinfo{year}{2024}), \bibinfo{pages}{e0000431}.
\newblock


\bibitem[Hicks et~al\mbox{.}(2023)]%
        {hicks2023leveraging}
\bibfield{author}{\bibinfo{person}{Jennifer~L Hicks},
  \bibinfo{person}{Melissa~A Boswell}, \bibinfo{person}{Tim Althoff},
  \bibinfo{person}{Alia~J Crum}, \bibinfo{person}{Joy~P Ku},
  \bibinfo{person}{James~A Landay}, \bibinfo{person}{Paula~ML Moya},
  \bibinfo{person}{Elizabeth~L Murnane}, \bibinfo{person}{Michael~P Snyder},
  \bibinfo{person}{Abby~C King}, {et~al\mbox{.}}}
  \bibinfo{year}{2023}\natexlab{}.
\newblock \showarticletitle{Leveraging mobile technology for public health
  promotion: A multidisciplinary perspective}.
\newblock \bibinfo{journal}{\emph{Annual Review of Public Health}}
  \bibinfo{volume}{44} (\bibinfo{year}{2023}), \bibinfo{pages}{131--150}.
\newblock


\bibitem[Hone and Graham(2000)]%
        {hone2000towards}
\bibfield{author}{\bibinfo{person}{Kate~S Hone} {and} \bibinfo{person}{Robert
  Graham}.} \bibinfo{year}{2000}\natexlab{}.
\newblock \showarticletitle{Towards a tool for the subjective assessment of
  speech system interfaces (SASSI)}.
\newblock \bibinfo{journal}{\emph{Natural Language Engineering}}
  \bibinfo{volume}{6}, \bibinfo{number}{3-4} (\bibinfo{year}{2000}),
  \bibinfo{pages}{287--303}.
\newblock


\bibitem[Horvitz(1999)]%
        {horvitz1999principles}
\bibfield{author}{\bibinfo{person}{Eric Horvitz}.}
  \bibinfo{year}{1999}\natexlab{}.
\newblock \showarticletitle{Principles of mixed-initiative user interfaces}. In
  \bibinfo{booktitle}{\emph{Proceedings of the SIGCHI conference on Human
  Factors in Computing Systems}}. \bibinfo{pages}{159--166}.
\newblock


\bibitem[Hsu et~al\mbox{.}(2023)]%
        {hsu2023helping}
\bibfield{author}{\bibinfo{person}{Shang-Ling Hsu}, \bibinfo{person}{Raj~Sanjay
  Shah}, \bibinfo{person}{Prathik Senthil}, \bibinfo{person}{Zahra Ashktorab},
  \bibinfo{person}{Casey Dugan}, \bibinfo{person}{Werner Geyer}, {and}
  \bibinfo{person}{Diyi Yang}.} \bibinfo{year}{2023}\natexlab{}.
\newblock \showarticletitle{Helping the Helper: Supporting Peer Counselors via
  AI-empowered Practice and Feedback}.
\newblock \bibinfo{journal}{\emph{arXiv preprint arXiv:2305.08982}}
  (\bibinfo{year}{2023}).
\newblock


\bibitem[Huang et~al\mbox{.}(2016)]%
        {huang2016field}
\bibfield{author}{\bibinfo{person}{Dandan Huang}, \bibinfo{person}{Melanie
  Tory}, {and} \bibinfo{person}{Lyn Bartram}.} \bibinfo{year}{2016}\natexlab{}.
\newblock \showarticletitle{A Field Study of On-Calendar Visualizations}. In
  \bibinfo{booktitle}{\emph{Proceedings of the 42nd Graphics Interface
  Conference}}. \bibinfo{pages}{13--20}.
\newblock


\bibitem[Hutchinson et~al\mbox{.}(2003)]%
        {hutchinson2003technology}
\bibfield{author}{\bibinfo{person}{Hilary Hutchinson}, \bibinfo{person}{Wendy
  Mackay}, \bibinfo{person}{Bo Westerlund}, \bibinfo{person}{Benjamin~B
  Bederson}, \bibinfo{person}{Allison Druin}, \bibinfo{person}{Catherine
  Plaisant}, \bibinfo{person}{Michel Beaudouin-Lafon},
  \bibinfo{person}{St{\'e}phane Conversy}, \bibinfo{person}{Helen Evans},
  \bibinfo{person}{Heiko Hansen}, {et~al\mbox{.}}}
  \bibinfo{year}{2003}\natexlab{}.
\newblock \showarticletitle{Technology probes: Inspiring design for and with
  families}. In \bibinfo{booktitle}{\emph{Proceedings of the SIGCHI Conference
  on Human Factors in Computing Systems}}. \bibinfo{pages}{17--24}.
\newblock


\bibitem[Inan et~al\mbox{.}(2023)]%
        {inan2023llama}
\bibfield{author}{\bibinfo{person}{Hakan Inan}, \bibinfo{person}{Kartikeya
  Upasani}, \bibinfo{person}{Jianfeng Chi}, \bibinfo{person}{Rashi Rungta},
  \bibinfo{person}{Krithika Iyer}, \bibinfo{person}{Yuning Mao},
  \bibinfo{person}{Michael Tontchev}, \bibinfo{person}{Qing Hu},
  \bibinfo{person}{Brian Fuller}, \bibinfo{person}{Davide Testuggine},
  {et~al\mbox{.}}} \bibinfo{year}{2023}\natexlab{}.
\newblock \showarticletitle{Llama Guard: LLM-based Input-Output Safeguard for
  Human-AI Conversations}.
\newblock \bibinfo{journal}{\emph{arXiv preprint arXiv:2312.06674}}
  (\bibinfo{year}{2023}).
\newblock


\bibitem[Jamalova(2019)]%
        {jamalova2019comparative}
\bibfield{author}{\bibinfo{person}{Maral Jamalova}.}
  \bibinfo{year}{2019}\natexlab{}.
\newblock \showarticletitle{The comparative study of the relationship between
  smartphone choice and socio-economic indicators}.
\newblock \bibinfo{journal}{\emph{International Journal of Marketing Studies}}
  \bibinfo{volume}{11}, \bibinfo{number}{11} (\bibinfo{year}{2019}),
  \bibinfo{pages}{10--5539}.
\newblock


\bibitem[Jin et~al\mbox{.}(2022)]%
        {jin2022self}
\bibfield{author}{\bibinfo{person}{Daoyan Jin}, \bibinfo{person}{Hallgeir
  Halvari}, \bibinfo{person}{Natalia Maehle}, {and} \bibinfo{person}{Anja~H
  Olafsen}.} \bibinfo{year}{2022}\natexlab{}.
\newblock \showarticletitle{Self-tracking behaviour in physical activity: A
  systematic review of drivers and outcomes of fitness tracking}.
\newblock \bibinfo{journal}{\emph{Behaviour \& Information Technology}}
  \bibinfo{volume}{41}, \bibinfo{number}{2} (\bibinfo{year}{2022}),
  \bibinfo{pages}{242--261}.
\newblock


\bibitem[Kamphorst(2017)]%
        {kamphorst2017coaching}
\bibfield{author}{\bibinfo{person}{Bart~A Kamphorst}.}
  \bibinfo{year}{2017}\natexlab{}.
\newblock \showarticletitle{E-coaching systems: What they are, and what they
  aren’t}.
\newblock \bibinfo{journal}{\emph{Personal and Ubiquitous Computing}}
  \bibinfo{volume}{21}, \bibinfo{number}{4} (\bibinfo{year}{2017}),
  \bibinfo{pages}{625--632}.
\newblock


\bibitem[Kersten-van Dijk et~al\mbox{.}(2017)]%
        {kersten2017personal}
\bibfield{author}{\bibinfo{person}{Elisabeth~T Kersten-van Dijk},
  \bibinfo{person}{Joyce~HDM Westerink}, \bibinfo{person}{Femke Beute}, {and}
  \bibinfo{person}{Wijnand~A IJsselsteijn}.} \bibinfo{year}{2017}\natexlab{}.
\newblock \showarticletitle{Personal informatics, self-insight, and behavior
  change: A critical review of current literature}.
\newblock \bibinfo{journal}{\emph{Human--Computer Interaction}}
  \bibinfo{volume}{32}, \bibinfo{number}{5-6} (\bibinfo{year}{2017}),
  \bibinfo{pages}{268--296}.
\newblock


\bibitem[Kim et~al\mbox{.}(2024)]%
        {kim2024health}
\bibfield{author}{\bibinfo{person}{Yubin Kim}, \bibinfo{person}{Xuhai Xu},
  \bibinfo{person}{Daniel McDuff}, \bibinfo{person}{Cynthia Breazeal}, {and}
  \bibinfo{person}{Hae~Won Park}.} \bibinfo{year}{2024}\natexlab{}.
\newblock \showarticletitle{Health-LLM: Large Language Models for Health
  Prediction via Wearable Sensor Data}. In
  \bibinfo{booktitle}{\emph{Proceedings of the Fifth Conference on Health,
  Inference, and Learning}} \emph{(\bibinfo{series}{Proceedings of Machine
  Learning Research}, Vol.~\bibinfo{volume}{248})},
  \bibfield{editor}{\bibinfo{person}{Tom Pollard}, \bibinfo{person}{Edward
  Choi}, \bibinfo{person}{Pankhuri Singhal}, \bibinfo{person}{Michael Hughes},
  \bibinfo{person}{Elena Sizikova}, \bibinfo{person}{Bobak Mortazavi},
  \bibinfo{person}{Irene Chen}, \bibinfo{person}{Fei Wang},
  \bibinfo{person}{Tasmie Sarker}, \bibinfo{person}{Matthew McDermott}, {and}
  \bibinfo{person}{Marzyeh Ghassemi}} (Eds.). \bibinfo{publisher}{PMLR},
  \bibinfo{pages}{522--539}.
\newblock
\urldef\tempurl%
\url{https://proceedings.mlr.press/v248/kim24b.html}
\showURL{%
\tempurl}


\bibitem[King et~al\mbox{.}(2020)]%
        {king2020effects}
\bibfield{author}{\bibinfo{person}{Abby~C King}, \bibinfo{person}{Maria~Ines
  Campero}, \bibinfo{person}{Jylana~L Sheats}, \bibinfo{person}{Cynthia
  M~Castro Sweet}, \bibinfo{person}{Michelle~E Hauser}, \bibinfo{person}{Dulce
  Garcia}, \bibinfo{person}{Aldo Chazaro}, \bibinfo{person}{German Blanco},
  \bibinfo{person}{Jorge Banda}, \bibinfo{person}{David~K Ahn},
  {et~al\mbox{.}}} \bibinfo{year}{2020}\natexlab{}.
\newblock \showarticletitle{Effects of counseling by peer human advisors vs
  computers to increase walking in underserved populations: The COMPASS
  randomized clinical trial}.
\newblock \bibinfo{journal}{\emph{JAMA Internal Medicine}}
  \bibinfo{volume}{180}, \bibinfo{number}{11} (\bibinfo{year}{2020}),
  \bibinfo{pages}{1481--1490}.
\newblock


\bibitem[King et~al\mbox{.}(2007)]%
        {king2007ongoing}
\bibfield{author}{\bibinfo{person}{Abby~C King}, \bibinfo{person}{Robert
  Friedman}, \bibinfo{person}{Bess Marcus}, \bibinfo{person}{Cynthia Castro},
  \bibinfo{person}{Melissa Napolitano}, \bibinfo{person}{David Ahn}, {and}
  \bibinfo{person}{Lawrence Baker}.} \bibinfo{year}{2007}\natexlab{}.
\newblock \showarticletitle{Ongoing physical activity advice by humans versus
  computers: the Community Health Advice by Telephone (CHAT) trial}.
\newblock \bibinfo{journal}{\emph{Health Psychology}} \bibinfo{volume}{26},
  \bibinfo{number}{6} (\bibinfo{year}{2007}), \bibinfo{pages}{718}.
\newblock


\bibitem[King et~al\mbox{.}(2002)]%
        {king2002stanford}
\bibfield{author}{\bibinfo{person}{Abby~C. King}, \bibinfo{person}{William~L.
  Haskell}, \bibinfo{person}{C.~Barr Taylor}, \bibinfo{person}{Robert DeBusk},
  \bibinfo{person}{Cynthia~M. Castro}, \bibinfo{person}{Leslie~A. Pruitt},
  {and} \bibinfo{person}{Stanford Prevention Research~Center staff}.}
  \bibinfo{year}{2002}\natexlab{}.
\newblock \bibinfo{title}{The Stanford Active Choices Program:
  Telephone-Assisted Counseling for Physical Activity}.
\newblock \bibinfo{howpublished}{Stanford Health Promotion Resource Center,
  Stanford Prevention Research Center, Stanford University}.
\newblock


\bibitem[King et~al\mbox{.}(2014)]%
        {king2014exercise}
\bibfield{author}{\bibinfo{person}{Abby~C King}, \bibinfo{person}{Eric~B
  Hekler}, \bibinfo{person}{Cynthia~M Castro}, \bibinfo{person}{Matthew~P
  Buman}, \bibinfo{person}{Bess~H Marcus}, \bibinfo{person}{Robert~H Friedman},
  {and} \bibinfo{person}{Melissa~A Napolitano}.}
  \bibinfo{year}{2014}\natexlab{}.
\newblock \showarticletitle{Exercise advice by humans versus computers:
  Maintenance effects at 18 months}.
\newblock \bibinfo{journal}{\emph{Health Psychology}} \bibinfo{volume}{33},
  \bibinfo{number}{2} (\bibinfo{year}{2014}), \bibinfo{pages}{192}.
\newblock


\bibitem[Klasnja et~al\mbox{.}(2011)]%
        {klasnja2011evaluate}
\bibfield{author}{\bibinfo{person}{Predrag Klasnja}, \bibinfo{person}{Sunny
  Consolvo}, {and} \bibinfo{person}{Wanda Pratt}.}
  \bibinfo{year}{2011}\natexlab{}.
\newblock \showarticletitle{How to evaluate technologies for health behavior
  change in HCI research}. In \bibinfo{booktitle}{\emph{Proceedings of the
  SIGCHI Conference on Human Factors in Computing Systems}}
  \emph{(\bibinfo{series}{CHI '11})}. \bibinfo{publisher}{Association for
  Computing Machinery}, \bibinfo{address}{New York, NY, USA},
  \bibinfo{pages}{3063–3072}.
\newblock
\showISBNx{9781450302289}
\urldef\tempurl%
\url{https://doi.org/10.1145/1978942.1979396}
\showDOI{\tempurl}


\bibitem[Kocielnik et~al\mbox{.}(2018)]%
        {kocielnik2018reflection}
\bibfield{author}{\bibinfo{person}{Rafal Kocielnik}, \bibinfo{person}{Lillian
  Xiao}, \bibinfo{person}{Daniel Avrahami}, {and} \bibinfo{person}{Gary
  Hsieh}.} \bibinfo{year}{2018}\natexlab{}.
\newblock \showarticletitle{Reflection Companion: A conversational system for
  engaging users in reflection on physical activity}.
\newblock \bibinfo{journal}{\emph{Proceedings of the ACM on Interactive,
  Mobile, Wearable and Ubiquitous Technologies}} \bibinfo{volume}{2},
  \bibinfo{number}{2} (\bibinfo{year}{2018}), \bibinfo{pages}{1--26}.
\newblock


\bibitem[Kouaho and Epstein(2024)]%
        {kouaho2024socioeconomic}
\bibfield{author}{\bibinfo{person}{Whitney-Jocelyn Kouaho} {and}
  \bibinfo{person}{Daniel~A Epstein}.} \bibinfo{year}{2024}\natexlab{}.
\newblock \showarticletitle{Socioeconomic Class in Physical Activity Wearables
  Research and Design}. In \bibinfo{booktitle}{\emph{Proceedings of the CHI
  Conference on Human Factors in Computing Systems}}. \bibinfo{pages}{1--15}.
\newblock


\bibitem[Kumar et~al\mbox{.}(2023)]%
        {kumar2022language}
\bibfield{author}{\bibinfo{person}{Sachin Kumar}, \bibinfo{person}{Vidhisha
  Balachandran}, \bibinfo{person}{Lucille Njoo}, \bibinfo{person}{Antonios
  Anastasopoulos}, {and} \bibinfo{person}{Yulia Tsvetkov}.}
  \bibinfo{year}{2023}\natexlab{}.
\newblock \showarticletitle{Language Generation Models Can Cause Harm: So What
  Can We Do About It? An Actionable Survey}. In
  \bibinfo{booktitle}{\emph{Proceedings of the 17th Conference of the European
  Chapter of the Association for Computational Linguistics}},
  \bibfield{editor}{\bibinfo{person}{Andreas Vlachos} {and}
  \bibinfo{person}{Isabelle Augenstein}} (Eds.).
  \bibinfo{publisher}{Association for Computational Linguistics},
  \bibinfo{address}{Dubrovnik, Croatia}, \bibinfo{pages}{3299--3321}.
\newblock
\urldef\tempurl%
\url{https://doi.org/10.18653/v1/2023.eacl-main.241}
\showDOI{\tempurl}


\bibitem[Lane et~al\mbox{.}(2012)]%
        {lane2012bewell}
\bibfield{author}{\bibinfo{person}{Nicholas Lane}, \bibinfo{person}{Mashfiqui
  Mohammod}, \bibinfo{person}{Mu Lin}, \bibinfo{person}{Xiaochao Yang},
  \bibinfo{person}{Hong Lu}, \bibinfo{person}{Shahid Ali},
  \bibinfo{person}{Afsaneh Doryab}, \bibinfo{person}{Ethan Berke},
  \bibinfo{person}{Tanzeem Choudhury}, {and} \bibinfo{person}{Andrew
  Campbell}.} \bibinfo{year}{2012}\natexlab{}.
\newblock \showarticletitle{Bewell: A smartphone application to monitor, model
  and promote wellbeing}. In \bibinfo{booktitle}{\emph{5th International ICST
  Conference on Pervasive Computing Technologies for Healthcare}}.
\newblock


\bibitem[Lewis et~al\mbox{.}(2020)]%
        {lewis2020rag}
\bibfield{author}{\bibinfo{person}{Patrick Lewis}, \bibinfo{person}{Ethan
  Perez}, \bibinfo{person}{Aleksandra Piktus}, \bibinfo{person}{Fabio Petroni},
  \bibinfo{person}{Vladimir Karpukhin}, \bibinfo{person}{Naman Goyal},
  \bibinfo{person}{Heinrich K{\"u}ttler}, \bibinfo{person}{Mike Lewis},
  \bibinfo{person}{Wen-tau Yih}, \bibinfo{person}{Tim Rockt{\"a}schel},
  {et~al\mbox{.}}} \bibinfo{year}{2020}\natexlab{}.
\newblock \showarticletitle{Retrieval-augmented generation for
  knowledge-intensive nlp tasks}.
\newblock \bibinfo{journal}{\emph{Advances in Neural Information Processing
  Systems}}  \bibinfo{volume}{33} (\bibinfo{year}{2020}),
  \bibinfo{pages}{9459--9474}.
\newblock


\bibitem[Li et~al\mbox{.}(2010)]%
        {li2010stage}
\bibfield{author}{\bibinfo{person}{Ian Li}, \bibinfo{person}{Anind Dey}, {and}
  \bibinfo{person}{Jodi Forlizzi}.} \bibinfo{year}{2010}\natexlab{}.
\newblock \showarticletitle{A stage-based model of personal informatics
  systems}. In \bibinfo{booktitle}{\emph{Proceedings of the SIGCHI Conference
  on Human Factors in Computing Systems}}. \bibinfo{pages}{557--566}.
\newblock


\bibitem[Lieberman(2021)]%
        {lieberman2021exercised}
\bibfield{author}{\bibinfo{person}{Daniel Lieberman}.}
  \bibinfo{year}{2021}\natexlab{}.
\newblock \bibinfo{booktitle}{\emph{Exercised: Why something we never evolved
  to do is healthy and rewarding}}.
\newblock \bibinfo{publisher}{Vintage}.
\newblock


\bibitem[Lin et~al\mbox{.}(2006)]%
        {lin2006fish}
\bibfield{author}{\bibinfo{person}{James~J Lin}, \bibinfo{person}{Lena
  Mamykina}, \bibinfo{person}{Silvia Lindtner}, \bibinfo{person}{Gregory
  Delajoux}, {and} \bibinfo{person}{Henry~B Strub}.}
  \bibinfo{year}{2006}\natexlab{}.
\newblock \showarticletitle{Fish’n’Steps: Encouraging physical activity
  with an interactive computer game}. In \bibinfo{booktitle}{\emph{UbiComp
  2006: Ubiquitous Computing: 8th International Conference, Orange County, CA,
  USA, September 17-21, 2006 Proceedings 8}}. Springer,
  \bibinfo{pages}{261--278}.
\newblock


\bibitem[Liu et~al\mbox{.}(2024)]%
        {liu2024lost}
\bibfield{author}{\bibinfo{person}{Nelson~F Liu}, \bibinfo{person}{Kevin Lin},
  \bibinfo{person}{John Hewitt}, \bibinfo{person}{Ashwin Paranjape},
  \bibinfo{person}{Michele Bevilacqua}, \bibinfo{person}{Fabio Petroni}, {and}
  \bibinfo{person}{Percy Liang}.} \bibinfo{year}{2024}\natexlab{}.
\newblock \showarticletitle{Lost in the middle: How language models use long
  contexts}.
\newblock \bibinfo{journal}{\emph{Transactions of the Association for
  Computational Linguistics}}  \bibinfo{volume}{12} (\bibinfo{year}{2024}),
  \bibinfo{pages}{157--173}.
\newblock


\bibitem[Liu et~al\mbox{.}(2023)]%
        {liu2023large}
\bibfield{author}{\bibinfo{person}{Xin Liu}, \bibinfo{person}{Daniel McDuff},
  \bibinfo{person}{Geza Kovacs}, \bibinfo{person}{Isaac Galatzer-Levy},
  \bibinfo{person}{Jacob Sunshine}, \bibinfo{person}{Jiening Zhan},
  \bibinfo{person}{Ming-Zher Poh}, \bibinfo{person}{Shun Liao},
  \bibinfo{person}{Paolo~Di Achille}, {and} \bibinfo{person}{Shwetak Patel}.}
  \bibinfo{year}{2023}\natexlab{}.
\newblock \showarticletitle{Large Language Models are Few-Shot Health
  Learners}.
\newblock \bibinfo{journal}{\emph{arXiv preprint arXiv:2305.15525}}
  (\bibinfo{year}{2023}).
\newblock


\bibitem[Luo et~al\mbox{.}(2021)]%
        {luo2021promoting}
\bibfield{author}{\bibinfo{person}{Tiffany~Christina Luo},
  \bibinfo{person}{Adrian Aguilera}, \bibinfo{person}{Courtney~Rees Lyles},
  {and} \bibinfo{person}{Caroline~Astrid Figueroa}.}
  \bibinfo{year}{2021}\natexlab{}.
\newblock \showarticletitle{Promoting physical activity through conversational
  agents: Mixed methods systematic review}.
\newblock \bibinfo{journal}{\emph{Journal of Medical Internet Research}}
  \bibinfo{volume}{23}, \bibinfo{number}{9} (\bibinfo{year}{2021}),
  \bibinfo{pages}{e25486}.
\newblock


\bibitem[Ma et~al\mbox{.}(2024)]%
        {ma2024evaluating}
\bibfield{author}{\bibinfo{person}{Zilin Ma}, \bibinfo{person}{Yiyang Mei},
  \bibinfo{person}{Yinru Long}, \bibinfo{person}{Zhaoyuan Su}, {and}
  \bibinfo{person}{Krzysztof~Z. Gajos}.} \bibinfo{year}{2024}\natexlab{}.
\newblock \showarticletitle{Evaluating the Experience of LGBTQ+ People Using
  Large Language Model Based Chatbots for Mental Health Support}. In
  \bibinfo{booktitle}{\emph{Proceedings of the 2024 CHI Conference on Human
  Factors in Computing Systems}} (Honolulu, HI, USA).
  \bibinfo{publisher}{Association for Computing Machinery},
  \bibinfo{address}{New York, NY, USA}, Article \bibinfo{articleno}{872},
  \bibinfo{numpages}{15}~pages.
\newblock
\showISBNx{9798400703300}
\urldef\tempurl%
\url{https://doi.org/10.1145/3613904.3642482}
\showDOI{\tempurl}


\bibitem[Ma et~al\mbox{.}(2023)]%
        {ma2023understanding}
\bibfield{author}{\bibinfo{person}{Zilin Ma}, \bibinfo{person}{Yiyang Mei},
  {and} \bibinfo{person}{Zhaoyuan Su}.} \bibinfo{year}{2023}\natexlab{}.
\newblock \showarticletitle{Understanding the benefits and challenges of using
  large language model-based conversational agents for mental well-being
  support}. In \bibinfo{booktitle}{\emph{AMIA Annual Symposium Proceedings}},
  Vol.~\bibinfo{volume}{2023}. American Medical Informatics Association,
  \bibinfo{pages}{1105}.
\newblock


\bibitem[Mandyam et~al\mbox{.}(2024)]%
        {mandyam2023adaptive}
\bibfield{author}{\bibinfo{person}{Aishwarya Mandyam}, \bibinfo{person}{Matthew
  J\"orke}, \bibinfo{person}{William Denton}, \bibinfo{person}{Barbara~E.
  Engelhardt}, {and} \bibinfo{person}{Emma Brunskill}.}
  \bibinfo{year}{2024}\natexlab{}.
\newblock \showarticletitle{Adaptive Interventions with User-Defined Goals for
  Health Behavior Change}. In \bibinfo{booktitle}{\emph{Proceedings of the
  Fifth Conference on Health, Inference, and Learning}}
  \emph{(\bibinfo{series}{Proceedings of Machine Learning Research},
  Vol.~\bibinfo{volume}{248})}, \bibfield{editor}{\bibinfo{person}{Tom
  Pollard}, \bibinfo{person}{Edward Choi}, \bibinfo{person}{Pankhuri Singhal},
  \bibinfo{person}{Michael Hughes}, \bibinfo{person}{Elena Sizikova},
  \bibinfo{person}{Bobak Mortazavi}, \bibinfo{person}{Irene Chen},
  \bibinfo{person}{Fei Wang}, \bibinfo{person}{Tasmie Sarker},
  \bibinfo{person}{Matthew McDermott}, {and} \bibinfo{person}{Marzyeh
  Ghassemi}} (Eds.). \bibinfo{publisher}{PMLR}, \bibinfo{pages}{597--618}.
\newblock
\urldef\tempurl%
\url{https://proceedings.mlr.press/v248/mandyam24a.html}
\showURL{%
\tempurl}


\bibitem[Marcus et~al\mbox{.}(1992)]%
        {marcus1992self}
\bibfield{author}{\bibinfo{person}{Bess~H Marcus}, \bibinfo{person}{Vanessa~C
  Selby}, \bibinfo{person}{Raymond~S Niaura}, {and} \bibinfo{person}{Joseph~S
  Rossi}.} \bibinfo{year}{1992}\natexlab{}.
\newblock \showarticletitle{Self-efficacy and the stages of exercise behavior
  change}.
\newblock \bibinfo{journal}{\emph{Research Quarterly for Exercise and Sport}}
  \bibinfo{volume}{63}, \bibinfo{number}{1} (\bibinfo{year}{1992}),
  \bibinfo{pages}{60--66}.
\newblock


\bibitem[Maynez et~al\mbox{.}(2020)]%
        {maynez2020faithfulness}
\bibfield{author}{\bibinfo{person}{Joshua Maynez}, \bibinfo{person}{Shashi
  Narayan}, \bibinfo{person}{Bernd Bohnet}, {and} \bibinfo{person}{Ryan
  McDonald}.} \bibinfo{year}{2020}\natexlab{}.
\newblock \showarticletitle{On Faithfulness and Factuality in Abstractive
  Summarization}. In \bibinfo{booktitle}{\emph{Proceedings of the 58th Annual
  Meeting of the Association for Computational Linguistics}},
  \bibfield{editor}{\bibinfo{person}{Dan Jurafsky}, \bibinfo{person}{Joyce
  Chai}, \bibinfo{person}{Natalie Schluter}, {and} \bibinfo{person}{Joel
  Tetreault}} (Eds.). \bibinfo{publisher}{Association for Computational
  Linguistics}, \bibinfo{address}{Online}, \bibinfo{pages}{1906--1919}.
\newblock
\urldef\tempurl%
\url{https://doi.org/10.18653/v1/2020.acl-main.173}
\showDOI{\tempurl}


\bibitem[Mazeika et~al\mbox{.}({[n.\,d.]})]%
        {mazeika2024harmbench}
\bibfield{author}{\bibinfo{person}{Mantas Mazeika}, \bibinfo{person}{Long
  Phan}, \bibinfo{person}{Xuwang Yin}, \bibinfo{person}{Andy Zou},
  \bibinfo{person}{Zifan Wang}, \bibinfo{person}{Norman Mu},
  \bibinfo{person}{Elham Sakhaee}, \bibinfo{person}{Nathaniel Li},
  \bibinfo{person}{Steven Basart}, \bibinfo{person}{Bo Li}, {et~al\mbox{.}}}
  \bibinfo{year}{[n.\,d.]}\natexlab{}.
\newblock \showarticletitle{HarmBench: A Standardized Evaluation Framework for
  Automated Red Teaming and Robust Refusal}. In
  \bibinfo{booktitle}{\emph{Forty-first International Conference on Machine
  Learning}}.
\newblock


\bibitem[McDuff et~al\mbox{.}(2012)]%
        {mcduff2012affectaura}
\bibfield{author}{\bibinfo{person}{Daniel McDuff}, \bibinfo{person}{Amy
  Karlson}, \bibinfo{person}{Ashish Kapoor}, \bibinfo{person}{Asta Roseway},
  {and} \bibinfo{person}{Mary Czerwinski}.} \bibinfo{year}{2012}\natexlab{}.
\newblock \showarticletitle{AffectAura: An intelligent system for emotional
  memory}. In \bibinfo{booktitle}{\emph{Proceedings of the SIGCHI Conference on
  Human Factors in Computing Systems}}. \bibinfo{pages}{849--858}.
\newblock


\bibitem[Merrill et~al\mbox{.}(2024a)]%
        {merrill2024transforming}
\bibfield{author}{\bibinfo{person}{Mike~A Merrill}, \bibinfo{person}{Akshay
  Paruchuri}, \bibinfo{person}{Naghmeh Rezaei}, \bibinfo{person}{Geza Kovacs},
  \bibinfo{person}{Javier Perez}, \bibinfo{person}{Yun Liu},
  \bibinfo{person}{Erik Schenck}, \bibinfo{person}{Nova Hammerquist},
  \bibinfo{person}{Jake Sunshine}, \bibinfo{person}{Shyam Tailor},
  {et~al\mbox{.}}} \bibinfo{year}{2024}\natexlab{a}.
\newblock \showarticletitle{Transforming wearable data into health insights
  using large language model agents}.
\newblock \bibinfo{journal}{\emph{arXiv preprint arXiv:2406.06464}}
  (\bibinfo{year}{2024}).
\newblock


\bibitem[Merrill et~al\mbox{.}(2024b)]%
        {merrill2024language}
\bibfield{author}{\bibinfo{person}{Mike~A Merrill}, \bibinfo{person}{Mingtian
  Tan}, \bibinfo{person}{Vinayak Gupta}, \bibinfo{person}{Thomas Hartvigsen},
  {and} \bibinfo{person}{Tim Althoff}.} \bibinfo{year}{2024}\natexlab{b}.
\newblock \showarticletitle{Language Models Still Struggle to Zero-shot Reason
  about Time Series}. In \bibinfo{booktitle}{\emph{Findings of the Association
  for Computational Linguistics: EMNLP 2024}},
  \bibfield{editor}{\bibinfo{person}{Yaser Al-Onaizan}, \bibinfo{person}{Mohit
  Bansal}, {and} \bibinfo{person}{Yun-Nung Chen}} (Eds.).
  \bibinfo{publisher}{Association for Computational Linguistics},
  \bibinfo{address}{Miami, Florida, USA}, \bibinfo{pages}{3512--3533}.
\newblock
\urldef\tempurl%
\url{https://doi.org/10.18653/v1/2024.findings-emnlp.201}
\showDOI{\tempurl}


\bibitem[Miller and Rollnick(2023)]%
        {miller2023motivational}
\bibfield{author}{\bibinfo{person}{W.R. Miller} {and} \bibinfo{person}{S.
  Rollnick}.} \bibinfo{year}{2023}\natexlab{}.
\newblock \bibinfo{booktitle}{\emph{Motivational Interviewing: Helping People
  Change and Grow} (\bibinfo{edition}{4th} ed.)}.
\newblock \bibinfo{publisher}{Guilford Publications}.
\newblock
\showISBNx{9781462552818}


\bibitem[Min et~al\mbox{.}(2022)]%
        {min-etal-2022-pair}
\bibfield{author}{\bibinfo{person}{Do~June Min}, \bibinfo{person}{Ver{\'o}nica
  P{\'e}rez-Rosas}, \bibinfo{person}{Kenneth Resnicow}, {and}
  \bibinfo{person}{Rada Mihalcea}.} \bibinfo{year}{2022}\natexlab{}.
\newblock \showarticletitle{{PAIR}: Prompt-Aware marg{I}n Ranking for Counselor
  Reflection Scoring in Motivational Interviewing}. In
  \bibinfo{booktitle}{\emph{Proceedings of the 2022 Conference on Empirical
  Methods in Natural Language Processing}},
  \bibfield{editor}{\bibinfo{person}{Yoav Goldberg}, \bibinfo{person}{Zornitsa
  Kozareva}, {and} \bibinfo{person}{Yue Zhang}} (Eds.).
  \bibinfo{publisher}{Association for Computational Linguistics},
  \bibinfo{address}{Abu Dhabi, United Arab Emirates},
  \bibinfo{pages}{148--158}.
\newblock
\urldef\tempurl%
\url{https://doi.org/10.18653/v1/2022.emnlp-main.11}
\showDOI{\tempurl}


\bibitem[Mitchell et~al\mbox{.}(2021)]%
        {mitchell2021automated}
\bibfield{author}{\bibinfo{person}{Elliot~G Mitchell}, \bibinfo{person}{Rosa
  Maimone}, \bibinfo{person}{Andrea Cassells}, \bibinfo{person}{Jonathan~N
  Tobin}, \bibinfo{person}{Patricia Davidson}, \bibinfo{person}{Arlene~M
  Smaldone}, {and} \bibinfo{person}{Lena Mamykina}.}
  \bibinfo{year}{2021}\natexlab{}.
\newblock \showarticletitle{Automated vs. Human Health Coaching: Exploring
  Participant and Practitioner Experiences}.
\newblock \bibinfo{journal}{\emph{Proceedings of the ACM on Human-Computer
  Interaction}} \bibinfo{volume}{5}, \bibinfo{number}{CSCW1}
  (\bibinfo{year}{2021}), \bibinfo{pages}{1--37}.
\newblock


\bibitem[Mollee et~al\mbox{.}(2017)]%
        {mollee2017evaluation}
\bibfield{author}{\bibinfo{person}{Julia~S. Mollee}, \bibinfo{person}{Anouk
  Middelweerd}, \bibinfo{person}{Saskia J.~te Velde}, {and}
  \bibinfo{person}{Michel C.~A. Klein}.} \bibinfo{year}{2017}\natexlab{}.
\newblock \showarticletitle{Evaluation of a personalized coaching system for
  physical activity: User appreciation and adherence}. In
  \bibinfo{booktitle}{\emph{Proceedings of the 11th EAI International
  Conference on Pervasive Computing Technologies for Healthcare}} (Barcelona,
  Spain) \emph{(\bibinfo{series}{PervasiveHealth '17})}.
  \bibinfo{publisher}{Association for Computing Machinery},
  \bibinfo{address}{New York, NY, USA}, \bibinfo{pages}{315–324}.
\newblock
\showISBNx{9781450363631}
\urldef\tempurl%
\url{https://doi.org/10.1145/3154862.3154933}
\showDOI{\tempurl}


\bibitem[Moyers et~al\mbox{.}(2003)]%
        {moyers2003assessing}
\bibfield{author}{\bibinfo{person}{Theresa Moyers}, \bibinfo{person}{Tim
  Martin}, \bibinfo{person}{Delwyn Catley}, \bibinfo{person}{Kari~Jo Harris},
  {and} \bibinfo{person}{Jasjit~S Ahluwalia}.} \bibinfo{year}{2003}\natexlab{}.
\newblock \showarticletitle{Assessing the integrity of motivational
  interviewing interventions: Reliability of the motivational interviewing
  skills code}.
\newblock \bibinfo{journal}{\emph{Behavioural and Cognitive Psychotherapy}}
  \bibinfo{volume}{31}, \bibinfo{number}{2} (\bibinfo{year}{2003}),
  \bibinfo{pages}{177--184}.
\newblock


\bibitem[Moyers et~al\mbox{.}(2016)]%
        {moyers2016motivational}
\bibfield{author}{\bibinfo{person}{Theresa~B Moyers}, \bibinfo{person}{Lauren~N
  Rowell}, \bibinfo{person}{Jennifer~K Manuel}, \bibinfo{person}{Denise Ernst},
  {and} \bibinfo{person}{Jon~M Houck}.} \bibinfo{year}{2016}\natexlab{}.
\newblock \showarticletitle{The motivational interviewing treatment integrity
  code (MITI 4): Rationale, preliminary reliability and validity}.
\newblock \bibinfo{journal}{\emph{Journal of Substance Abuse Treatment}}
  \bibinfo{volume}{65} (\bibinfo{year}{2016}), \bibinfo{pages}{36--42}.
\newblock


\bibitem[Munson and Consolvo(2012)]%
        {munson2012exploring}
\bibfield{author}{\bibinfo{person}{Sean~A Munson} {and} \bibinfo{person}{Sunny
  Consolvo}.} \bibinfo{year}{2012}\natexlab{}.
\newblock \showarticletitle{Exploring goal-setting, rewards, self-monitoring,
  and sharing to motivate physical activity}. In \bibinfo{booktitle}{\emph{2012
  6th International Conference on Pervasive Computing Technologies for
  Healthcare (PervasiveHealth) and Workshops}}. IEEE, \bibinfo{pages}{25--32}.
\newblock


\bibitem[Murnane et~al\mbox{.}(2020)]%
        {murnane2020designing}
\bibfield{author}{\bibinfo{person}{Elizabeth~L Murnane}, \bibinfo{person}{Xin
  Jiang}, \bibinfo{person}{Anna Kong}, \bibinfo{person}{Michelle Park},
  \bibinfo{person}{Weili Shi}, \bibinfo{person}{Connor Soohoo},
  \bibinfo{person}{Luke Vink}, \bibinfo{person}{Iris Xia}, \bibinfo{person}{Xin
  Yu}, \bibinfo{person}{John Yang-Sammataro}, {et~al\mbox{.}}}
  \bibinfo{year}{2020}\natexlab{}.
\newblock \showarticletitle{Designing ambient narrative-based interfaces to
  reflect and motivate physical activity}. In
  \bibinfo{booktitle}{\emph{Proceedings of the 2020 CHI Conference on Human
  Factors in Computing Systems}}. \bibinfo{pages}{1--14}.
\newblock


\bibitem[Nahum-Shani et~al\mbox{.}(2018)]%
        {nahum2018just}
\bibfield{author}{\bibinfo{person}{Inbal Nahum-Shani},
  \bibinfo{person}{Shawna~N Smith}, \bibinfo{person}{Bonnie~J Spring},
  \bibinfo{person}{Linda~M Collins}, \bibinfo{person}{Katie Witkiewitz},
  \bibinfo{person}{Ambuj Tewari}, {and} \bibinfo{person}{Susan~A Murphy}.}
  \bibinfo{year}{2018}\natexlab{}.
\newblock \showarticletitle{Just-in-time adaptive interventions (JITAIs) in
  mobile health: Key components and design principles for ongoing health
  behavior support}.
\newblock \bibinfo{journal}{\emph{Annals of Behavioral Medicine}}
  (\bibinfo{year}{2018}), \bibinfo{pages}{1--17}.
\newblock


\bibitem[Oh et~al\mbox{.}(2021)]%
        {oh2021systematic}
\bibfield{author}{\bibinfo{person}{Yoo~Jung Oh}, \bibinfo{person}{Jingwen
  Zhang}, \bibinfo{person}{Min-Lin Fang}, {and} \bibinfo{person}{Yoshimi
  Fukuoka}.} \bibinfo{year}{2021}\natexlab{}.
\newblock \showarticletitle{A systematic review of artificial intelligence
  chatbots for promoting physical activity, healthy diet, and weight loss}.
\newblock \bibinfo{journal}{\emph{International Journal of Behavioral Nutrition
  and Physical Activity}}  \bibinfo{volume}{18} (\bibinfo{year}{2021}),
  \bibinfo{pages}{1--25}.
\newblock


\bibitem[Olsen(2014)]%
        {olsen2014health}
\bibfield{author}{\bibinfo{person}{Jeanette~M Olsen}.}
  \bibinfo{year}{2014}\natexlab{}.
\newblock \showarticletitle{Health coaching: A concept analysis}. In
  \bibinfo{booktitle}{\emph{Nursing Forum}}, Vol.~\bibinfo{volume}{49}. Wiley
  Online Library, \bibinfo{pages}{18--29}.
\newblock


\bibitem[Olsen and Nesbitt(2010)]%
        {olsen2010health}
\bibfield{author}{\bibinfo{person}{Jeanette~M Olsen} {and}
  \bibinfo{person}{Bonnie~J Nesbitt}.} \bibinfo{year}{2010}\natexlab{}.
\newblock \showarticletitle{Health Coaching to Improve Healthy Lifestyle
  Behaviors: An Integrative Review}.
\newblock \bibinfo{journal}{\emph{American Journal of Health Promotion}}
  \bibinfo{volume}{25}, \bibinfo{number}{1} (\bibinfo{year}{2010}),
  \bibinfo{pages}{e1--e12}.
\newblock


\bibitem[Ong et~al\mbox{.}(2024)]%
        {ong2024advancing}
\bibfield{author}{\bibinfo{person}{Qi~Chwen Ong}, \bibinfo{person}{Chin-Siang
  Ang}, \bibinfo{person}{Davidson Zun~Yin Chee}, \bibinfo{person}{Ashwini
  Lawate}, \bibinfo{person}{Frederick Sundram}, \bibinfo{person}{Mayank
  Dalakoti}, \bibinfo{person}{Leonardo Pasalic}, \bibinfo{person}{Daniel To},
  \bibinfo{person}{Tatiana~Erlikh Fox}, \bibinfo{person}{Iva Bojic},
  {et~al\mbox{.}}} \bibinfo{year}{2024}\natexlab{}.
\newblock \showarticletitle{Advancing health coaching: A comparative study of
  large language model and health coaches}.
\newblock \bibinfo{journal}{\emph{Artificial Intelligence in Medicine}}
  \bibinfo{volume}{157} (\bibinfo{year}{2024}), \bibinfo{pages}{103004}.
\newblock


\bibitem[ONVY(2024)]%
        {onvy}
\bibfield{author}{\bibinfo{person}{ONVY}.} \bibinfo{year}{2024}\natexlab{}.
\newblock \bibinfo{howpublished}{\url{https://www.onvy.health}}.
\newblock


\bibitem[Organization(2022)]%
        {who2022}
\bibfield{author}{\bibinfo{person}{World~Health Organization}.}
  \bibinfo{year}{2022}\natexlab{}.
\newblock \bibinfo{title}{Physical Activity Fact Sheet}.
\newblock
\newblock
\urldef\tempurl%
\url{https://www.who.int/news-room/fact-sheets/detail/physical-activity}
\showURL{%
\tempurl}


\bibitem[Organization(2024)]%
        {who2024sarah}
\bibfield{author}{\bibinfo{person}{World~Health Organization}.}
  \bibinfo{year}{2024}\natexlab{}.
\newblock \bibinfo{title}{S.A.R.A.H, a Smart AI Resource Assistant for Health}.
\newblock
  \bibinfo{howpublished}{\url{https://www.who.int/campaigns/s-a-r-a-h}}.
\newblock


\bibitem[Ouyang et~al\mbox{.}(2022)]%
        {ouyang2022training}
\bibfield{author}{\bibinfo{person}{Long Ouyang}, \bibinfo{person}{Jeffrey Wu},
  \bibinfo{person}{Xu Jiang}, \bibinfo{person}{Diogo Almeida},
  \bibinfo{person}{Carroll Wainwright}, \bibinfo{person}{Pamela Mishkin},
  \bibinfo{person}{Chong Zhang}, \bibinfo{person}{Sandhini Agarwal},
  \bibinfo{person}{Katarina Slama}, \bibinfo{person}{Alex Ray},
  {et~al\mbox{.}}} \bibinfo{year}{2022}\natexlab{}.
\newblock \showarticletitle{Training language models to follow instructions
  with human feedback}.
\newblock \bibinfo{journal}{\emph{Advances in Neural Information Processing
  Systems}}  \bibinfo{volume}{35} (\bibinfo{year}{2022}),
  \bibinfo{pages}{27730--27744}.
\newblock


\bibitem[Park and Kulkarni(2023)]%
        {park2023thinking}
\bibfield{author}{\bibinfo{person}{Soya Park} {and} \bibinfo{person}{Chinmay
  Kulkarni}.} \bibinfo{year}{2023}\natexlab{}.
\newblock \showarticletitle{Thinking Assistants: LLM-Based Conversational
  Assistants that Help Users Think By Asking rather than Answering}.
\newblock \bibinfo{journal}{\emph{arXiv preprint arXiv:2312.06024}}
  (\bibinfo{year}{2023}).
\newblock


\bibitem[P{\'e}rez-Rosas et~al\mbox{.}(2016)]%
        {perez2016building}
\bibfield{author}{\bibinfo{person}{Ver{\'o}nica P{\'e}rez-Rosas},
  \bibinfo{person}{Rada Mihalcea}, \bibinfo{person}{Kenneth Resnicow},
  \bibinfo{person}{Satinder Singh}, {and} \bibinfo{person}{Lawrence An}.}
  \bibinfo{year}{2016}\natexlab{}.
\newblock \showarticletitle{Building a motivational interviewing dataset}. In
  \bibinfo{booktitle}{\emph{Proceedings of the Third Workshop on Computational
  Linguistics and Clinical Psychology}}. \bibinfo{pages}{42--51}.
\newblock


\bibitem[P{\'e}rez-Rosas et~al\mbox{.}(2017)]%
        {perez2017understanding}
\bibfield{author}{\bibinfo{person}{Ver{\'o}nica P{\'e}rez-Rosas},
  \bibinfo{person}{Rada Mihalcea}, \bibinfo{person}{Kenneth Resnicow},
  \bibinfo{person}{Satinder Singh}, {and} \bibinfo{person}{Lawrence An}.}
  \bibinfo{year}{2017}\natexlab{}.
\newblock \showarticletitle{Understanding and predicting empathic behavior in
  counseling therapy}. In \bibinfo{booktitle}{\emph{Proceedings of the 55th
  Annual Meeting of the Association for Computational Linguistics (Volume 1:
  Long Papers)}}. \bibinfo{pages}{1426--1435}.
\newblock


\bibitem[P{\'e}rez-Rosas et~al\mbox{.}(2019)]%
        {perez2019makes}
\bibfield{author}{\bibinfo{person}{Ver{\'o}nica P{\'e}rez-Rosas},
  \bibinfo{person}{Xinyi Wu}, \bibinfo{person}{Kenneth Resnicow}, {and}
  \bibinfo{person}{Rada Mihalcea}.} \bibinfo{year}{2019}\natexlab{}.
\newblock \showarticletitle{What makes a good counselor? Learning to
  distinguish between high-quality and low-quality counseling conversations}.
  In \bibinfo{booktitle}{\emph{Proceedings of the 57th Annual Meeting of the
  Association for Computational Linguistics}}. \bibinfo{pages}{926--935}.
\newblock


\bibitem[Prochaska and Velicer(1997)]%
        {prochaska1997transtheoretical}
\bibfield{author}{\bibinfo{person}{James~O Prochaska} {and}
  \bibinfo{person}{Wayne~F Velicer}.} \bibinfo{year}{1997}\natexlab{}.
\newblock \showarticletitle{The transtheoretical model of health behavior
  change}.
\newblock \bibinfo{journal}{\emph{American Journal of Health Promotion}}
  \bibinfo{volume}{12}, \bibinfo{number}{1} (\bibinfo{year}{1997}),
  \bibinfo{pages}{38--48}.
\newblock


\bibitem[Rapp and Cena(2016)]%
        {rapp2016personal}
\bibfield{author}{\bibinfo{person}{Amon Rapp} {and} \bibinfo{person}{Federica
  Cena}.} \bibinfo{year}{2016}\natexlab{}.
\newblock \showarticletitle{Personal informatics for everyday life: How users
  without prior self-tracking experience engage with personal data}.
\newblock \bibinfo{journal}{\emph{International Journal of Human-Computer
  Studies}}  \bibinfo{volume}{94} (\bibinfo{year}{2016}),
  \bibinfo{pages}{1--17}.
\newblock


\bibitem[Reeves and Nass(1996)]%
        {reeves1996media}
\bibfield{author}{\bibinfo{person}{Byron Reeves} {and}
  \bibinfo{person}{Clifford Nass}.} \bibinfo{year}{1996}\natexlab{}.
\newblock \showarticletitle{The media equation: How people treat computers,
  television, and new media like real people}.
\newblock \bibinfo{journal}{\emph{Cambridge, UK}} \bibinfo{volume}{10},
  \bibinfo{number}{10} (\bibinfo{year}{1996}).
\newblock


\bibitem[Rutjes et~al\mbox{.}(2022)]%
        {rutjes2022influence}
\bibfield{author}{\bibinfo{person}{Heleen Rutjes}, \bibinfo{person}{Martijn~C
  Willemsen}, \bibinfo{person}{Milou~A Feijt}, {and} \bibinfo{person}{Wijnand~A
  IJsselsteijn}.} \bibinfo{year}{2022}\natexlab{}.
\newblock \showarticletitle{The Influence of Personal Health Data on the Health
  Coaching Process}.
\newblock \bibinfo{journal}{\emph{Frontiers in Big Data}}  \bibinfo{volume}{5}
  (\bibinfo{year}{2022}), \bibinfo{pages}{678061}.
\newblock


\bibitem[Rutjes et~al\mbox{.}(2019)]%
        {rutjes2019beyond}
\bibfield{author}{\bibinfo{person}{Heleen Rutjes}, \bibinfo{person}{Martijn~C
  Willemsen}, {and} \bibinfo{person}{Wijnand~A IJsselsteijn}.}
  \bibinfo{year}{2019}\natexlab{}.
\newblock \showarticletitle{Beyond behavior: the coach's perspective on
  technology in health coaching}. In \bibinfo{booktitle}{\emph{Proceedings of
  the 2019 CHI Conference on Human Factors in Computing Systems}}.
  \bibinfo{pages}{1--14}.
\newblock


\bibitem[Schick et~al\mbox{.}(2023)]%
        {schick2023toolformer}
\bibfield{author}{\bibinfo{person}{Timo Schick}, \bibinfo{person}{Jane
  Dwivedi-Yu}, \bibinfo{person}{Roberto Dess{\`\i}}, \bibinfo{person}{Roberta
  Raileanu}, \bibinfo{person}{Maria Lomeli}, \bibinfo{person}{Eric Hambro},
  \bibinfo{person}{Luke Zettlemoyer}, \bibinfo{person}{Nicola Cancedda}, {and}
  \bibinfo{person}{Thomas Scialom}.} \bibinfo{year}{2023}\natexlab{}.
\newblock \showarticletitle{Toolformer: Language models can teach themselves to
  use tools}.
\newblock \bibinfo{journal}{\emph{Advances in Neural Information Processing
  Systems}}  \bibinfo{volume}{36} (\bibinfo{year}{2023}),
  \bibinfo{pages}{68539--68551}.
\newblock


\bibitem[Schmiedmayer et~al\mbox{.}(2024)]%
        {schmiedmayer2024}
\bibfield{author}{\bibinfo{person}{Paul Schmiedmayer}, \bibinfo{person}{Vishnu
  Ravi}, {and} \bibinfo{person}{Oliver Aalami}.}
  \bibinfo{year}{2024}\natexlab{}.
\newblock \bibinfo{booktitle}{\emph{Spezi}}.
\newblock
\urldef\tempurl%
\url{https://doi.org/10.5281/zenodo.10482368}
\showDOI{\tempurl}


\bibitem[Shah et~al\mbox{.}(2022)]%
        {shah2022modeling}
\bibfield{author}{\bibinfo{person}{Raj~Sanjay Shah}, \bibinfo{person}{Faye
  Holt}, \bibinfo{person}{Shirley~Anugrah Hayati}, \bibinfo{person}{Aastha
  Agarwal}, \bibinfo{person}{Yi-Chia Wang}, \bibinfo{person}{Robert~E Kraut},
  {and} \bibinfo{person}{Diyi Yang}.} \bibinfo{year}{2022}\natexlab{}.
\newblock \showarticletitle{Modeling motivational interviewing strategies on an
  online peer-to-peer counseling platform}.
\newblock \bibinfo{journal}{\emph{Proceedings of the ACM on Human-Computer
  Interaction}} \bibinfo{volume}{6}, \bibinfo{number}{CSCW2}
  (\bibinfo{year}{2022}), \bibinfo{pages}{1--24}.
\newblock


\bibitem[Shaikh et~al\mbox{.}(2023)]%
        {shaikh2023rehearsal}
\bibfield{author}{\bibinfo{person}{Omar Shaikh}, \bibinfo{person}{Valentino
  Chai}, \bibinfo{person}{Michele~J Gelfand}, \bibinfo{person}{Diyi Yang},
  {and} \bibinfo{person}{Michael~S Bernstein}.}
  \bibinfo{year}{2023}\natexlab{}.
\newblock \showarticletitle{Rehearsal: Simulating conflict to teach conflict
  resolution}.
\newblock \bibinfo{journal}{\emph{arXiv preprint arXiv:2309.12309}}
  (\bibinfo{year}{2023}).
\newblock


\bibitem[Sharma et~al\mbox{.}(2023)]%
        {sharma2023human}
\bibfield{author}{\bibinfo{person}{Ashish Sharma}, \bibinfo{person}{Inna~W
  Lin}, \bibinfo{person}{Adam~S Miner}, \bibinfo{person}{David~C Atkins}, {and}
  \bibinfo{person}{Tim Althoff}.} \bibinfo{year}{2023}\natexlab{}.
\newblock \showarticletitle{Human--AI collaboration enables more empathic
  conversations in text-based peer-to-peer mental health support}.
\newblock \bibinfo{journal}{\emph{Nature Machine Intelligence}}
  \bibinfo{volume}{5}, \bibinfo{number}{1} (\bibinfo{year}{2023}),
  \bibinfo{pages}{46--57}.
\newblock


\bibitem[Shuster et~al\mbox{.}(2021)]%
        {shuster-etal-2021-retrieval-augmentation}
\bibfield{author}{\bibinfo{person}{Kurt Shuster}, \bibinfo{person}{Spencer
  Poff}, \bibinfo{person}{Moya Chen}, \bibinfo{person}{Douwe Kiela}, {and}
  \bibinfo{person}{Jason Weston}.} \bibinfo{year}{2021}\natexlab{}.
\newblock \showarticletitle{Retrieval Augmentation Reduces Hallucination in
  Conversation}. In \bibinfo{booktitle}{\emph{Findings of the Association for
  Computational Linguistics: EMNLP 2021}},
  \bibfield{editor}{\bibinfo{person}{Marie-Francine Moens},
  \bibinfo{person}{Xuanjing Huang}, \bibinfo{person}{Lucia Specia}, {and}
  \bibinfo{person}{Scott Wen-tau Yih}} (Eds.). \bibinfo{publisher}{Association
  for Computational Linguistics}, \bibinfo{address}{Punta Cana, Dominican
  Republic}, \bibinfo{pages}{3784--3803}.
\newblock
\urldef\tempurl%
\url{https://doi.org/10.18653/v1/2021.findings-emnlp.320}
\showDOI{\tempurl}


\bibitem[Singh et~al\mbox{.}(2023)]%
        {singh2023systematic}
\bibfield{author}{\bibinfo{person}{Ben Singh}, \bibinfo{person}{Timothy Olds},
  \bibinfo{person}{Jacinta Brinsley}, \bibinfo{person}{Dot Dumuid},
  \bibinfo{person}{Rosa Virgara}, \bibinfo{person}{Lisa Matricciani},
  \bibinfo{person}{Amanda Watson}, \bibinfo{person}{Kimberley Szeto},
  \bibinfo{person}{Emily Eglitis}, \bibinfo{person}{Aaron Miatke},
  {et~al\mbox{.}}} \bibinfo{year}{2023}\natexlab{}.
\newblock \showarticletitle{Systematic review and meta-analysis of the
  effectiveness of chatbots on lifestyle behaviours}.
\newblock \bibinfo{journal}{\emph{npj Digital Medicine}} \bibinfo{volume}{6},
  \bibinfo{number}{1} (\bibinfo{year}{2023}), \bibinfo{pages}{118}.
\newblock


\bibitem[Slov{\'a}k et~al\mbox{.}(2017)]%
        {slovak2017reflective}
\bibfield{author}{\bibinfo{person}{Petr Slov{\'a}k},
  \bibinfo{person}{Christopher Frauenberger}, {and} \bibinfo{person}{Geraldine
  Fitzpatrick}.} \bibinfo{year}{2017}\natexlab{}.
\newblock \showarticletitle{Reflective practicum: A framework of sensitising
  concepts to design for transformative reflection}. In
  \bibinfo{booktitle}{\emph{Proceedings of the 2017 CHI Conference on Human
  Factors in Computing Systems}}. \bibinfo{pages}{2696--2707}.
\newblock


\bibitem[Thudt et~al\mbox{.}(2015)]%
        {thudt2015visual}
\bibfield{author}{\bibinfo{person}{Alice Thudt}, \bibinfo{person}{Dominikus
  Baur}, \bibinfo{person}{Samuel Huron}, {and} \bibinfo{person}{Sheelagh
  Carpendale}.} \bibinfo{year}{2015}\natexlab{}.
\newblock \showarticletitle{Visual Mementos: Reflecting Memories with Personal
  Data}.
\newblock \bibinfo{journal}{\emph{IEEE Transactions on Visualization and
  Computer Graphics}} \bibinfo{volume}{22}, \bibinfo{number}{1}
  (\bibinfo{year}{2015}), \bibinfo{pages}{369--378}.
\newblock


\bibitem[Tu et~al\mbox{.}(2024)]%
        {tu2024conversational}
\bibfield{author}{\bibinfo{person}{Tao Tu}, \bibinfo{person}{Anil Palepu},
  \bibinfo{person}{Mike Schaekermann}, \bibinfo{person}{Khaled Saab},
  \bibinfo{person}{Jan Freyberg}, \bibinfo{person}{Ryutaro Tanno},
  \bibinfo{person}{Amy Wang}, \bibinfo{person}{Brenna Li},
  \bibinfo{person}{Mohamed Amin}, \bibinfo{person}{Nenad Tomasev},
  {et~al\mbox{.}}} \bibinfo{year}{2024}\natexlab{}.
\newblock \showarticletitle{Towards Conversational Diagnostic AI}.
\newblock \bibinfo{journal}{\emph{arXiv preprint arXiv:2401.05654}}
  (\bibinfo{year}{2024}).
\newblock


\bibitem[Vandelanotte et~al\mbox{.}(2023)]%
        {vandelanotte2023increasing}
\bibfield{author}{\bibinfo{person}{Corneel Vandelanotte},
  \bibinfo{person}{Stewart Trost}, \bibinfo{person}{Danya Hodgetts},
  \bibinfo{person}{Tasadduq Imam}, \bibinfo{person}{Mamunur Rashid},
  \bibinfo{person}{Quyen~G To}, {and} \bibinfo{person}{Carol Maher}.}
  \bibinfo{year}{2023}\natexlab{}.
\newblock \showarticletitle{Increasing physical activity using a just-in-time
  adaptive digital assistant supported by machine learning: A novel approach
  for hyper-personalised mHealth interventions}.
\newblock \bibinfo{journal}{\emph{Journal of Biomedical Informatics}}
  \bibinfo{volume}{144} (\bibinfo{year}{2023}), \bibinfo{pages}{104435}.
\newblock


\bibitem[Weidinger et~al\mbox{.}(2021)]%
        {weidinger2021ethical}
\bibfield{author}{\bibinfo{person}{Laura Weidinger}, \bibinfo{person}{John
  Mellor}, \bibinfo{person}{Maribeth Rauh}, \bibinfo{person}{Conor Griffin},
  \bibinfo{person}{Jonathan Uesato}, \bibinfo{person}{Po-Sen Huang},
  \bibinfo{person}{Myra Cheng}, \bibinfo{person}{Mia Glaese},
  \bibinfo{person}{Borja Balle}, \bibinfo{person}{Atoosa Kasirzadeh},
  {et~al\mbox{.}}} \bibinfo{year}{2021}\natexlab{}.
\newblock \showarticletitle{Ethical and social risks of harm from language
  models}.
\newblock \bibinfo{journal}{\emph{arXiv preprint arXiv:2112.04359}}
  (\bibinfo{year}{2021}).
\newblock


\bibitem[Welivita and Pu(2022)]%
        {welivita2022curating}
\bibfield{author}{\bibinfo{person}{Anuradha Welivita} {and}
  \bibinfo{person}{Pearl Pu}.} \bibinfo{year}{2022}\natexlab{}.
\newblock \showarticletitle{Curating a large-scale motivational interviewing
  dataset using peer support forums}. In \bibinfo{booktitle}{\emph{Proceedings
  of the 29th International Conference on Computational Linguistics}}.
  \bibinfo{pages}{3315--3330}.
\newblock


\bibitem[Welivita and Pu(2023)]%
        {welivita2023boosting}
\bibfield{author}{\bibinfo{person}{Anuradha Welivita} {and}
  \bibinfo{person}{Pearl Pu}.} \bibinfo{year}{2023}\natexlab{}.
\newblock \showarticletitle{Boosting Distress Support Dialogue Responses with
  Motivational Interviewing Strategy}. In \bibinfo{booktitle}{\emph{Findings of
  the Association for Computational Linguistics: ACL 2023}}.
  \bibinfo{pages}{5411--5432}.
\newblock


\bibitem[WHOOP(2023)]%
        {whoopcoach}
\bibfield{author}{\bibinfo{person}{WHOOP}.} \bibinfo{year}{2023}\natexlab{}.
\newblock \bibinfo{title}{Introducing WHOOP Coach, Powered By OpenAI}.
\newblock
  \bibinfo{howpublished}{\url{https://www.whoop.com/us/en/thelocker/introducing-whoop-coach-powered-by-openai/}}.
\newblock


\bibitem[Wilcox et~al\mbox{.}(2006)]%
        {wilcox2006results}
\bibfield{author}{\bibinfo{person}{Sara Wilcox}, \bibinfo{person}{Marsha
  Dowda}, \bibinfo{person}{Sarah~F Griffin}, \bibinfo{person}{Carol Rheaume},
  \bibinfo{person}{Marcia~G Ory}, \bibinfo{person}{Laura Leviton},
  \bibinfo{person}{Abby~C King}, \bibinfo{person}{Andrea Dunn},
  \bibinfo{person}{David~M Buchner}, \bibinfo{person}{Terry Bazzarre},
  {et~al\mbox{.}}} \bibinfo{year}{2006}\natexlab{}.
\newblock \showarticletitle{Results of the first year of active for life:
  Translation of 2 evidence-based physical activity programs for older adults
  into community settings}.
\newblock \bibinfo{journal}{\emph{American Journal of Public Health}}
  \bibinfo{volume}{96}, \bibinfo{number}{7} (\bibinfo{year}{2006}),
  \bibinfo{pages}{1201--1209}.
\newblock


\bibitem[Wilcox et~al\mbox{.}(2008)]%
        {wilcox2008active}
\bibfield{author}{\bibinfo{person}{Sara Wilcox}, \bibinfo{person}{Marsha
  Dowda}, \bibinfo{person}{Laura~C Leviton}, \bibinfo{person}{Jenny
  Bartlett-Prescott}, \bibinfo{person}{Terry Bazzarre},
  \bibinfo{person}{Kimberly Campbell-Voytal}, \bibinfo{person}{Ruth~Ann
  Carpenter}, \bibinfo{person}{Cynthia~M Castro}, \bibinfo{person}{Diane
  Dowdy}, \bibinfo{person}{Andrea~L Dunn}, {et~al\mbox{.}}}
  \bibinfo{year}{2008}\natexlab{}.
\newblock \showarticletitle{Active for life: Final results from the translation
  of two physical activity programs}.
\newblock \bibinfo{journal}{\emph{American Journal of Preventive Medicine}}
  \bibinfo{volume}{35}, \bibinfo{number}{4} (\bibinfo{year}{2008}),
  \bibinfo{pages}{340--351}.
\newblock


\bibitem[Wolever et~al\mbox{.}(2013)]%
        {wolever2013systematic}
\bibfield{author}{\bibinfo{person}{Ruth~Q Wolever}, \bibinfo{person}{Leigh~Ann
  Simmons}, \bibinfo{person}{Gary~A Sforzo}, \bibinfo{person}{Diana Dill},
  \bibinfo{person}{Miranda Kaye}, \bibinfo{person}{Elizabeth~M Bechard},
  \bibinfo{person}{Mary~Elaine Southard}, \bibinfo{person}{Mary Kennedy},
  \bibinfo{person}{Justine Vosloo}, {and} \bibinfo{person}{Nancy Yang}.}
  \bibinfo{year}{2013}\natexlab{}.
\newblock \showarticletitle{A Systematic Review of the Literature on Health and
  WellnessCoaching: Defining a Key Behavioral Intervention in Healthcare}.
\newblock \bibinfo{journal}{\emph{Global Advances in Health and Medicine}}
  \bibinfo{volume}{2}, \bibinfo{number}{4} (\bibinfo{year}{2013}),
  \bibinfo{pages}{38--57}.
\newblock


\bibitem[Wu et~al\mbox{.}(2022)]%
        {wu2022ai}
\bibfield{author}{\bibinfo{person}{Tongshuang Wu}, \bibinfo{person}{Michael
  Terry}, {and} \bibinfo{person}{Carrie~Jun Cai}.}
  \bibinfo{year}{2022}\natexlab{}.
\newblock \showarticletitle{AI Chains: Transparent and Controllable Human-AI
  Interaction by Chaining Large Language Model Prompts}. In
  \bibinfo{booktitle}{\emph{Proceedings of the 2022 CHI Conference on Human
  Factors in Computing Systems}}. \bibinfo{pages}{1--22}.
\newblock


\bibitem[Zamfirescu-Pereira et~al\mbox{.}(2023)]%
        {zamfirescu2023herding}
\bibfield{author}{\bibinfo{person}{JD Zamfirescu-Pereira},
  \bibinfo{person}{Heather Wei}, \bibinfo{person}{Amy Xiao},
  \bibinfo{person}{Kitty Gu}, \bibinfo{person}{Grace Jung},
  \bibinfo{person}{Matthew~G Lee}, \bibinfo{person}{Bjoern Hartmann}, {and}
  \bibinfo{person}{Qian Yang}.} \bibinfo{year}{2023}\natexlab{}.
\newblock \showarticletitle{Herding AI cats: Lessons from designing a chatbot
  by prompting GPT-3}. In \bibinfo{booktitle}{\emph{Proceedings of the 2023 ACM
  Designing Interactive Systems Conference}}. \bibinfo{pages}{2206--2220}.
\newblock


\bibitem[Zeng et~al\mbox{.}(2024)]%
        {zeng2024air}
\bibfield{author}{\bibinfo{person}{Yi Zeng}, \bibinfo{person}{Yu Yang},
  \bibinfo{person}{Andy Zhou}, \bibinfo{person}{Jeffrey~Ziwei Tan},
  \bibinfo{person}{Yuheng Tu}, \bibinfo{person}{Yifan Mai},
  \bibinfo{person}{Kevin Klyman}, \bibinfo{person}{Minzhou Pan},
  \bibinfo{person}{Ruoxi Jia}, \bibinfo{person}{Dawn Song}, {et~al\mbox{.}}}
  \bibinfo{year}{2024}\natexlab{}.
\newblock \showarticletitle{AIR-Bench 2024: A Safety Benchmark Based on Risk
  Categories from Regulations and Policies}.
\newblock \bibinfo{journal}{\emph{arXiv preprint arXiv:2407.17436}}
  (\bibinfo{year}{2024}).
\newblock


\bibitem[Zhou et~al\mbox{.}(2024)]%
        {zhou2024archer}
\bibfield{author}{\bibinfo{person}{Yifei Zhou}, \bibinfo{person}{Andrea
  Zanette}, \bibinfo{person}{Jiayi Pan}, \bibinfo{person}{Sergey Levine}, {and}
  \bibinfo{person}{Aviral Kumar}.} \bibinfo{year}{2024}\natexlab{}.
\newblock \showarticletitle{{A}r{CH}er: Training Language Model Agents via
  Hierarchical Multi-Turn {RL}}. In \bibinfo{booktitle}{\emph{Proceedings of
  the 41st International Conference on Machine Learning}}
  \emph{(\bibinfo{series}{Proceedings of Machine Learning Research},
  Vol.~\bibinfo{volume}{235})}, \bibfield{editor}{\bibinfo{person}{Ruslan
  Salakhutdinov}, \bibinfo{person}{Zico Kolter}, \bibinfo{person}{Katherine
  Heller}, \bibinfo{person}{Adrian Weller}, \bibinfo{person}{Nuria Oliver},
  \bibinfo{person}{Jonathan Scarlett}, {and} \bibinfo{person}{Felix
  Berkenkamp}} (Eds.). \bibinfo{publisher}{PMLR},
  \bibinfo{pages}{62178--62209}.
\newblock
\urldef\tempurl%
\url{https://proceedings.mlr.press/v235/zhou24t.html}
\showURL{%
\tempurl}


\end{thebibliography}
